\RequirePackage[T1]{fontenc}
\documentclass[12pt]{article}
\pdfoutput=1 
\usepackage[height=8.85in,width=6.45in]{geometry}

\usepackage[utf8]{inputenc}
\usepackage{amsmath}
\usepackage{amssymb}
\usepackage{mathtools}
\numberwithin{equation}{section}
\usepackage{slashed}
\usepackage{braket}
\usepackage[svgnames]{xcolor}
\usepackage[colorlinks,linktocpage=true,citecolor=DarkGreen,linkcolor=FireBrick]{hyperref}
\usepackage{cite}
\usepackage{graphicx}
\usepackage{tikz}
\usetikzlibrary{tqft}
\usetikzlibrary{decorations.pathmorphing}
\usetikzlibrary{decorations.markings}
\usepackage{tikz-cd}
\usepackage{times}
\usepackage[scaled]{couriers}
\usepackage{bm}
\usepackage{subfig}
\usepackage{mathrsfs}
\usepackage{amsthm}

\usepackage{xcolor}
\usepackage{mdframed}

\renewenvironment{figure}[1][]{
  \begin{originalfigure}[#1]
    \begin{mdframed}[linecolor=black!0,backgroundcolor=black!1]
}{
    \end{mdframed}
  \end{originalfigure}
}

\theoremstyle{plain}
\newtheorem{thm}{Theorem}

\newtheorem{property}[thm]{Property}

\theoremstyle{definition}

\newcommand{\vev}[1]{ \left\langle {#1} \right\rangle }

\def\d{{\rm d}}
\def\i{{\mathsf i}}
\DeclareMathOperator{\Tr}{Tr}
\DeclareMathOperator{\Det}{Det}
\DeclareMathOperator{\Index}{Index}
\DeclareMathOperator{\tr}{tr}
\def\Re{\mathop{\mathrm{Re}}}
\def\Im{\mathop{\mathrm{Im}}}

\usepackage{slashed}

\def\GS{{\mathcal W}}
\def\TMF{{\rm TMF}}
\def\SQFT{{\rm SQFT}}
\def\SQM{{\rm SQM}}
\def\SQ{({\rm SQM/SQFT})}
\def\SQwp{{\rm SQM/SQFT}}

\def\a{\alpha}

\def\cD{{\cal D}}

\def\cH{{\cal H}}
\def\cI{{\cal I}}
\def\cJ{{\cal J}}

\def\cN{{\cal N}}

\def\cS{{\cal S}}

\def\cW{{\cal W}}

\def\cZ{{\cal Z}}

\def\bA{{\mathbb A}}
\def\bB{{\mathbb B}}
\def\bC{{\mathbb C}}

\def\bI{{\mathbb I}}
\def\bJ{{\mathbb J}}

\def\bQ{{\mathbb Q}}
\def\bR{{\mathbb R}}

\def\bZ{{\mathbb Z}}

\def\sA{{\mathsf A}}
\def\sB{{\mathsf B}}
\def\sC{{\mathsf C}}

\def\sL{{\mathsf L}}
\def\sM{{\mathsf M}}
\def\sN{{\mathsf N}}

\def\sP{{\mathsf P}}

\def\sS{{\mathsf S}}
\def\sT{{\mathsf T}}
\def\sU{{\mathsf U}}

\def\sX{{\mathsf X}}
\def\sY{{\mathsf Y}}

\def\so{{\mathsf o}}

\def\su{{\mathsf u}}

\def\U{\mathrm{U}}
\def\SU{\mathrm{SU}}
\def\O{\mathrm{O}}

\def\Sp{\mathrm{Sp}}

\def\SL{\mathrm{SL}}

\def\su{\mathfrak{su}}

\def\so{\mathfrak{so}}

\def\spin{\mathrm{spin}}
\def\pin{\mathrm{pin}}
\def\e{\mathfrak{e}}

\def\beq#1\eeq{\begin{align}#1\end{align}}

\usepackage[all]{xy} 

\def\MF{\mathrm{MF}}

\def\pt{\mathrm{pt}}
\def\Nequals#1{${\mathcal{N}{=}#1}$}
\def\fiberproduct{\mathrel{\widetilde \times}}

\begin{document}

\begin{titlepage}

\begin{flushright}
TU-1271
\end{flushright}

\vskip 3cm

\begin{center}

{\Large \bfseries On invariants of two-dimensional \\[1em]
minimally supersymmetric field theories}

\vskip 1cm
Yuji Tachikawa$^1$ and Kazuya Yonekura$^2$
\vskip 1cm

\begin{tabular}{ll}
$^1$ & Kavli Institute for the Physics and Mathematics of the Universe (WPI), \\
& University of Tokyo,  Kashiwa, Chiba 277-8583, Japan\\
$^2$ & Department of Physics, Tohoku University, Sendai 980-8578, Japan
\end{tabular}

\vskip 1cm

\end{center}

\noindent
We construct a new invariant of two-dimensional $\cN{=}(0,1)$ supersymmetric quantum field theories (SQFTs), 
under a couple of assumptions on the general properties of such SQFTs motivated 
by considerations in heterotic string theory.
This new invariant is in addition to the known `primary' invariants,
i.e.~the ordinary  elliptic genus and the mod-2 elliptic genus,
and the known `secondary' invariant,
which can be considered as an SQFT analog of the Atiyah-Patodi-Singer $\eta$-invariant when its continuous variation vanishes.

We also define a spectral invariant of supercharge more generally by the integration of the expectation value of the supercharge over the fundamental region of $\mathrm{ SL}(2,\bZ)$. 
With this spectral invariant, we construct a spectral pairing which can be used to systematically detect all the known invariants, including the new one we introduce in the first half of this paper.

\end{titlepage}

\setcounter{tocdepth}{2}


\tableofcontents


\section{Introduction}

\subsection{History of invariants of supersymmetric quantum field theories}
Finding quantities associated to 
supersymmetric quantum mechanics  (SQMs) and 
supersymmetric quantum field theories (SQFTs)
which are invariant under continuous deformations
is an activity as old as the study of the supersymmetric dynamics itself.
The first among them was the Witten index of SQMs introduced in 
\cite{Witten:1982im,Witten:1982df},
which counts the number of supersymmetric vacua weighted by the fermion number parity $(-1)^F$.

Many variants of the Witten index are known by now.
For example, a $d$-dimensional SQFT can sometimes be put on  $S^{d-1}\times S^1$ preserving some supersymmetry,
and the partition function there gives us an invariant, known as the superconformal index when the theory has conformal invariance.
This equals the generating function of Witten indices of the SQMs
on eigenspaces of angular momentum operators on $S^{d-1}$.

It is then natural to ask the following question: 
are there any other invariants of $d$-dimensional SQFTs? 
In fact, in recent years, a couple of new invariants of two-dimensional (2d) \Nequals{(0,1)} SQFTs have been introduced,
and one of the aims of this paper is to discuss yet another one.
We would like to begin by recalling how these new invariants were found.

The story goes back to the mid-1980s.
Around that time, mathematicians realized that there is a class of generalizations
of the index of the Dirac operator on spin manifolds
with rather remarkable properties.
As the index of the Dirac operator was known as $\hat A$ genus,
and this generalization involved elliptic functions,
mathematicians started to call it as the \emph{elliptic genus}.
Then in \cite{Witten:1986bf},  Witten used the index (or more precisely the generating function of the indices on eigenspaces of the momentum operator on $S^1$) of  2d 
supersymmetric sigma models to give physical explanations of many of these properties.
From this reason, the index of a 2d SQFT is usually referred to 
as its elliptic genus.

It was known in mathematics that K-theory is the natural framework
in which the index of Dirac operators is to be formulated~\cite{Atiyah:1968mp}. 
It was then natural for mathematicians to ask what fits in the question mark in the schematic diagram below:
\[
\begin{array}{ccr}
\text{K-theory}  & \leadsto & \text{$\hat A$ genus of a manifold} \\
\downarrow && \multicolumn{1}{c}{\downarrow}\\
\text{???}  & \leadsto & \text{elliptic genus of a manifold} \\
\end{array}
\]
Around at the turn of this century, mathematicians finally succeeded in constructing the required object,
now known as topological modular forms (TMFs) \cite{Hopkins2002},
which have been intensively studied by mathematicians since then.
Now, K-theory has a geometric definition in terms of vector bundles.
Although TMF has a precise mathematical definition via homotopy theory, it still lacks a geometric definition.
In this direction, Stolz and Teichner introduced in \cite{Stolz:2004,Stolz:2011zj} a program to 
provide a `geometric' definition of TMF in terms of 2d \Nequals{(0,1)} SQFTs.

This program started to attract attention on the physics side since around 2018 (see e.g.~\cite{Gukov:2018iiq,Gaiotto:2018ypj,Gaiotto:2019asa,Gaiotto:2019gef,Johnson-Freyd:2020itv,Tachikawa:2021mvw,Lin:2021bcp,Tachikawa:2021mby,Yonekura:2022reu,Lin:2022wpx,Albert:2022gcs,Kaidi:2023tqo,Tachikawa:2023nne,Tachikawa:2024ucm,Saxena:2024eil,Johnson-Freyd:2024rxr,Kaidi:2024cbx}).
Although there is no proof at any level of rigor, many pieces of evidence have accumulated 
that the Stolz-Teichner program is on the right track.
In particular, many of the results within TMF, but not its algebraic-topological construction itself,
allow re-interpretation in the language of SQFTs, and can then be studied 
purely on the side of SQFTs.
The new invariants of \Nequals{(0,1)} SQFTs have been found in this manner.

One such  new invariant was studied in \cite{Tachikawa:2023nne}. 
To understand its definition, let us recall that,
in the case of SQM, 
there are situations where the ordinary Witten index vanishes,
but the number modulo two of the vacuum degeneracy is still invariant under continuous deformations.
This is the mod-2 Witten index. 
In the context of 2d SQFT, then, there are situations in which the generating function of
mod-2 Witten indices of SQMs on the eigenspaces of the momentum along $S^1$ 
provides an invariant under continuous deformations. 
This is the mod-2 elliptic genus, and fills  in the bottom-right corner of  the schematic diagram below: \[
\begin{array}{ccc}
\text{ordinary Witten index}  & \to & \text{elliptic genus} \\
\downarrow && \downarrow\\
\text{mod-2 Witten index}  & \to & \text{mod-2 elliptic genus} \\
\end{array}\quad.
\]
This invariant could have been discovered and discussed in the 1980s, 
and indeed it gives a nonzero value for the sigma model on $S^1$,
but it was studied in some detail only very recently in this historical context.
We can call the ordinary elliptic genus and the mod-2 elliptic genus
as the  `primary' invariants, since they can be defined simply 
as deformation invariants of the Hilbert space of the theory under consideration.

Another more subtle invariant was introduced on the mathematical side 
by Bunke and Naumann in \cite{Bunke}.
On the physics side, a corresponding invariant of 2d \Nequals{(0,1)} SQFTs 
was introduced in \cite{Gaiotto:2019gef} and studied further in \cite{Yonekura:2022reu}.
For a manifold $M_d$ of dimension $d$ with appropriate structure (or more precisely string structure), this new invariant can be defined
by 
using the Atiyah-Patodi-Singer (APS) $\eta$-invariant on $M_d$ which appears in the index theorem on a manifold $N_{d+1}$ such that $\partial N_{d+1}=M_d$ \cite{Atiyah:1975jf}.
As such, it can be considered as a `secondary' invariant, in contrast to the primary invariants given above.
As an example, this secondary invariant takes the value $1/24$ mod $1$ for the $S^3$ sigma model with the unit flux $\int_{S^3} H=1$ of the 3-form field strength of the $B$-field, or in other words the unit Wess-Zumino-Witten term for $\SU(2) \simeq S^3$.

The new invariant we introduce in this paper for 2d \Nequals{(0,1)} SQFTs
is subtler, and is defined for theories whose primary and secondary invariants both vanish.
It may therefore be called as a `tertiary' invariant.
As we will see in detail in this paper, for the 2d sigma model on a manifold $M_d$,
it will be defined using a manifold 
$L_{d+2}$ of dimension $d+2$ constructed in a certain way from $M_d$.
Actually, this newest invariant has already appeared in string theory contexts \cite{Kaidi:2023tqo,Tachikawa:2024ucm,Kaidi:2024cbx},
and the present paper gives its foundation for general SQFTs.
Let us discuss it in more detail.

\subsection{Our new invariant}
First, let us remind ourselves that the gravitational anomaly of a 2d quantum field theory 
is specified by an integer $\nu$, which we normalize so that a right-moving Majorana-Weyl fermion has 
the gravitational anomaly $\nu=+1$.\footnote{More precisely, this $\nu$ is the coefficient of the anomaly polynomial of the 2d theory, $ -\frac{\nu}{48}p_1(R)$, where $p_1(R)$ is the first Pontryagin class.}
We denote by $\SQFT_\nu$ the set of 2d \Nequals{(0,1)} SQFTs with gravitational anomaly $\nu$,
where our convention is that the supersymmetry is on the right-moving side.
We would like to understand functions on $\SQFT_\nu$ such that 
(i) it is invariant under continuous deformations, and
(ii) it is zero when the theory spontaneously breaks supersymmetry.\footnote{More precisely, the supersymmetry breaking in this discussion is required to persist even in the large volume limit. In the case of finite volume, the supersymmetry may be broken even for nontrivial elements of $\SQFT_{d}$. See \cite{Gaiotto:2019asa} for more discussions.}
To formalize it, introduce an equivalence relation on $\SQFT_\nu$ such that
it identifies two theories continuously connected to each other,
and it regards any theory breaking supersymmetry as trivial.
The quotient by this equivalence relation is denoted by $\SQFT_\nu(\pt)$.
The equivalence class of a theory $\sA\in\SQFT_\nu$ will be denoted by $[\sA]\in \SQFT_n(\pt)$.

In $\SQFT_\nu({\rm pt})$, we define a subgroup $\bA_\nu$ by
\beq
\bA_\nu = \{[\sA] \in \SQFT_\nu({\rm pt})~|~ I_\sA=0,  I^{\rm mod\,2}_\sA=0, I^\text{2nd}_{\sA}=0  \},
\eeq
where $I_\sA$ is the ordinary elliptic genus,
$I_\sA^{\rm mod\,2}$ is the mod-2 elliptic genus, and
$I^\text{2nd}_{\sA}$ is the secondary invariant. 
Our new invariant  is a bilinear form $\GS(\bullet, \bullet)$,
\beq
\bA_d \times \bA_{-22-d} \ni ([\sX], [\sT]) \mapsto \GS([\sX], [\sT]) \in \bQ/\bZ,
\eeq
whose existence was suggested from the mathematical theory of TMF \cite{BrunerRognes,Tachikawa:2023lwf}
and also from the consideration of heterotic string theory \cite{Kaidi:2023tqo,Tachikawa:2024ucm,Kaidi:2024cbx}.
Note that the sum of the degrees, $d+(-22-d)=-22$, is the correct gravitational anomaly
of the non-ghost part of the worldsheet of heterotic string theory.
Very roughly speaking,
this pairing computes the discrete part of the Green-Schwarz coupling 
when the internal worldsheet theory is $\sT\in \bA_{-22-d}$
and the spacetime theory is $\sX\in \bA_{d}$.
But our definition, although inspired by consideration in heterotic string theory,
does not make any direct use of heterotic string theory.
Rather, it is done by manipulating the 2d SQFTs involved,
and should be understandable for anyone interested in 2d SQFTs.

A concrete example of this pairing we discuss in this paper is for 
$(d,-22-d)=(6,-28)$.
We take the theory $\sX\in \SQFT_{6}$ to be the sigma model on $S^3\times S^3$,
with $\int_{S^3}H=1$ on both $S^3$'s involved.
For the theory $\sT\in \SQFT_{-28}$, we take the purely left-moving modular invariant theory
which has $(\e_7)_1\times (\e_7)_1$ current algebra in it,
and regard it as an \Nequals{(0,1)} theory.
Equivalently \cite{Kaidi:2023tqo,Kaidi:2024cbx}, we can take $\sT'\in \SQFT_{-28}$
obtained by fibering the purely left-moving $(\e_8)_1\times (\e_8)_1$ theory 
over the $S^4$ sigma model, by embedding the $\so(4)\simeq \su(2)\times \su(2)$ curvature of the tangent bundle of $S^4$
into $\e_8 \times \e_8$.
We will find that \begin{equation}
\GS([\sX],[\sT])=\frac12\neq0,
\end{equation}
showing that the theories $\sX$ and $\sT$ mentioned above can never be deformed continuously
to break supersymmetry spontaneously.

\subsection{Spectral invariants}
So far we have only discussed `topological invariants' of SQFTs.
They are analogous, in the case of manifolds, to the $\bZ$-valued Dirac index, the $\bZ_2$-valued Dirac index,
or the APS $\eta$-invariant when its continuous variation vanishes. 
In the case of manifolds, underlying all these is the $\eta$-invariant of manifolds which can depend continuously on geometric information such as metrics and connections. 
The APS $\eta$-invariant in dimension $d$ is connected to the Dirac index in dimension $d+1$ via the APS index theorem on manifolds with boundary.
It can also detect mod-2 indices and other invariants of manifolds, which can also be equipped with fiber bundles for gauge groups. 

The Dirac operator on manifolds and the supercharge of SQFTs have similar properties. In fact, in the case of sigma models in the context of SQM, the supercharge is really given by the Dirac operator itself, and this fact can be used to derive the Atiyah-Singer index theorem~\cite{Alvarez-Gaume:1983zxc} and also the APS index theorem~\cite{Dabholkar:2019nnc}. 
Therefore, it is natural to expect that we can define interesting quantities by using the supercharge of an SQFT, in a similar way that we define the APS $\eta$-invariant by using the Dirac operator. 

In the last part of the paper, we will introduce exactly such an analog of the APS $\eta$-invariant, defined  by using the supercharge $Q$ of an SQFT.
We call this the spectral invariant of SQFTs.
Our spectral invariant is 
defined by an integral of the expectation value of the supercharge over the fundamental region of $\SL(2,\bZ)$ transformations acting on the complex modulus $\tau$ of $T^2$,
and satisfy an analogue of APS index theorem.
 Although our definition is valid for general SQFTs, the appearance of integration over the fundamental region suggests that it might play some role in string theory.
Our spectral invariant depends on the eigenvalue spectrum of the supercharge, and hence it is a  spectral  invariant as in the case of the original APS $\eta$-invariant,
and in particular depends continuously except for integer jumps with respect to the continuous parameters of the theory.

Moreover, we define the spectral invariant not just for SQFTs but also 
for  `a formal power series of SQMs whose boundary is an SQFT'. 
This spectral invariant satisfies an SQFT-SQM analogue of the APS index theorem.
We will then see that all the topological invariants of SQFTs known so far,
i.e.~the ordinary elliptic genus,
the mod-2 elliptic genus,
the secondary invariant, and our new invariant,
have a uniform description in terms of this spectral invariant. 
Although the concept of formal power series of SQMs may look artificial, it plays a role as the detector of all the known invariants of SQFTs. If the Stolz-Teichner conjecture is correct, it does detect all the invariants. 

\subsection{Assumptions}

Before proceeding, we should mention that the construction of our new invariant
depends on a couple of (hopefully reasonable) assumptions
of the properties of SQFTs, which will be detailed later. 
Among them, the most crucial one is the assumption \begin{equation}
\SQFT_{-21}(\pt)=0,\label{eq:basic-assumption-intro}
\end{equation}i.e.~every SQFT of anomaly $\nu=-21$ can be continuously deformed to 
break supersymmetry spontaneously.
It corresponds to the mathematically proven statement $
\TMF_{-21}(\pt)=0 
$ 
in the Stolz-Teichner proposal.

The most important motivation for 
the assumption \eqref{eq:basic-assumption-intro} comes from the fact that
it leads to the absence of all global anomalies of 
heterotic string theory \cite{Tachikawa:2021mby,Yonekura:2022reu}.\footnote{%
An extremely crude explanation is as follows. For a proper treatment, consult \cite{Yonekura:2022reu} and also Section~\ref{sec:string}.
Note that the total worldsheet theory (except $bc$ and $\beta\gamma$ ghosts) 
on a heterotic worldsheet has $\nu=-22$,
since e.g.~the left-moving current algebra fermions have $\nu=-32$ 
and the ten-dimensional spacetime sigma model has $\nu=10$.
Its global anomaly may then be captured by a family of such $\nu=-22$ SQFTs.
Considering a family effectively increases  the spacetime dimensions by one,
so a single  family would correspond to an SQFT of $\nu=-21$.
If $\SQFT_{-21}(\pt)$ is zero, it would imply that all families are topologically trivial, 
leading to the absence of global anomaly in the heterotic path integral.
}
Therefore, the property \eqref{eq:basic-assumption-intro} is well-motivated by string theory, 
even without assuming Stolz-Teichner conjecture, 
and we use it for the study of general SQFTs.

The assumption \eqref{eq:basic-assumption-intro} is used e.g.~when we show that our definition of the new invariant
is independent of a number of choices we need to make during the construction.
That the well-defined-ness of our invariant relies on this assumption can also 
be viewed in a positive manner.
This is because our construction is such that, if two choices would lead to two different values of the invariant,
we would be able to construct an explicit class in $\SQFT_{-21}(\pt)$ which should be nonzero.
Therefore, the investigation of our new `invariant' in various cases could lead us to such an explicit class,
whose discovery would also tell us many things about the properties of SQFTs.

Another assumption is about $\SQFT_{-22}(\pt)$.
Namely, we assume that $\SQFT_{-22}(\pt)$ can be completely characterized by the
primary invariant, i.e.~the mod-2 elliptic genus in this case of $\nu=-22$.
This is less crucial in the sense that our constructions are still valid for a smaller subset of SQFTs if this assumption about $\SQFT_{-22}(\pt)$ is not satisfied. 

While $\SQFT_{-21}(\pt)$ is related to anomalies of heterotic string theory, $\SQFT_{-22}(\pt)$ is related to (discrete) $\theta$-angles.\footnote{%
Indeed, a heterotic worldsheet $\sM$ (including the target space as well as the internal theory) has $\nu=-22$. Then, we can include $\theta$-angles 
in the spacetime action  depending on  its class $[\sM]\in \SQFT_{-22}(\pt)$, 
just as we can include the $\theta$-angle for gauge fields depending their topological classes.
Note that 
we are \emph{not} considering each internal theory as a separate theory. 
Instead, the entire $\SQFT_{-22}$ is considered. 
For instance, although the standard $\e_8 \times \e_8$ and $\so(32)$ heterotic string theories are given by different  elements of $\SQFT_{-32}(\pt)$ \cite{Tachikawa:2024ucm},
the difference is washed out in $\SQFT_{-22}(\pt)$, which includes the ten-dimensional spacetime part.
So they are part of  ``the same theory'' in the discussions here.
The T-duality between them after $S^1$ compactification is one manifestation of this phenomenon.
}
If one wishes that string theory is ``unique'' in some sense, one might hope that there is no freedom to add $\theta$-angles.\footnote{See also \cite{Freed:2019sco} for a similar issue in M-theory, where a possible discrete $\theta$-angle in the low energy effective theory of M-theory on unorientable manifolds was identified.
The situation is not clear in the full UV theory, in that we do not yet know whether both choices are realized or not. }
Even if we assume so,
it does not mean that $\SQFT_{-22}(\pt)$ should vanish. One of the reasons is that there are physically uninteresting $\theta$-angles. 
To see it, suppose that the internal theory and the target space manifold are such that there are fermion zero modes. Then, by rotating the phases of the fermions, the path integral measure may produce additional phases which can be interpreted as the change of $\theta$-angles as is familiar in the case of the $\theta$-angle in QCD. 
Such fermion zero modes are related to the (mod-2) elliptic genus of $\SQFT_{-22}(\pt)$. 
Therefore, the best thing we could hope may be that there are no other invariants for $\SQFT_{-22}(\pt)$ than the (mod-2) elliptic  genus. 
Regardless of whether this argument is convincing or not, we will assume in the present paper 
that $\SQFT_{-22}(\pt)$ is completely characterized by the mod-2 elliptic genus.

\subsection{Organization of the paper}

The rest of the paper is organized as follows. \begin{itemize}
\item In Sec.~\ref{sec:mfds}, we start by considering the invariants of manifolds equipped with twisted string structure, i.e.~manifolds equipped with a $B$-field with the field strength 3-form $H$, a gauge field with field strength $F$, and a metric with Riemann curvature 2-form $R$, so that the condition \begin{equation}
\d H\propto \tr F^2-\tr R^2
\end{equation} is satisfied (a more precise definition will be given in the footnote \ref{foot:H} there). 
This is for us to gain some intuition about the properties of various constructions which we will pursue 
in the context of general SQFTs, in a more familiar setting of sigma models on manifolds.
It is also helpful for concrete computations of the new invariant $\cW([\sX],[\sT]) $we introduce in this paper
when one of the theories involved (say $\sX$) is indeed a sigma model.

\item In Sec.~\ref{sec:generalSQFT}, 
we give an overview of basic notions and operations on SQFTs.
In particular, we explain the holomorphic anomaly equation, which plays essential roles for our constructions of the new invariant and the spectral invariant.
We also give a review of the primary and secondary invariants. 

\item
In Sec.~\ref{sec:new}, we will construct our new invariant. 
Our definition involves various choices
in the intermediate steps, and we will spend significant efforts to show that the final result
does not depend on these auxiliary choices.
We also perform a computation of our new invariant
in one specific case, namely, 
when $\sX$ is the sigma model on $S^3\times S^3$ with unit $H$ flux on both $S^3$'s, 
and $\sT$ is the purely left-moving modular-invariant theory with left central charge $c_L=14$, containing $(\e_7)_1\times (\e_7)_1$ current algebra. 
Other examples are briefly mentioned. 

\item In Sec.~\ref{sec:differential}, we introduce our spectral invariant. Most part of this section can be read without reading Sec.~\ref{sec:mfds} and \ref{sec:new}.
For pure SQFTs, we reformulate the secondary invariant in terms of the spectral invariant. 
More generally, we define a pairing between a formal power series of SQMs and a pure SQFT,
and show that this underlies 
all the topological invariants of SQFTs, both the already known ones and the new one discussed in this paper.
We also discuss the possible ranges of the values of the primary and secondary invariants by combining various results in physics and mathematics.

\item In Sec.~\ref{sec:string}, we discuss possible applications of some of the results of Sec.~\ref{sec:differential} to partition functions in heterotic string theory.

\item At the end, we have an Appendix~\ref{sec:A}, where we give a careful derivation of the holomorphic anomaly equation. 
This formula was conjectured in \cite{Gaiotto:2019gef} and proved for sigma models in \cite{Dabholkar:2020fde}.
Here we provide the derivation for general SQFTs. We also explain a derivation of the APS index theorem by using the holomorphic anomaly equation, which may help the reader understand the discussions of Sec.~\ref{sec:differential}.
\end{itemize}


\section{Manifolds with string structure}
\label{sec:mfds}

Before going to the case of SQFTs, it is helpful to consider manifolds with twisted string structure to gain some intuition. In fact, our strategy for constructing an invariant of SQFTs is to imitate the case of manifolds. One of the main points of this section is to translate geometric quantities ($\eta$-invariant etc.) to topological quantities (APS index).

\subsection{Anomalies}
Suppose we are interested in chiral fermions coupled to gauge fields and gravity in $d$ dimensions. 
We denote a manifold with various data included (such as gauge fields, spin structure and so on) as $M$.
Let $\cD^+_d(M)$ be the chiral Dirac operator acting on the fermions. 
We expect that the partition function is roughly given by $\Det \cD^+_d(M)$.%
\footnote{For Majorana fermions, we use Pfaffians rather than determinants. The following discussion proceeds with only minor modifications in that case.} 

When there are anomalies, the determinant $\Det \cD^+_d(M)$ does not have a well-defined value in $\bC$. Instead, we proceed as follows~\cite{Witten:1999eg,Witten:2015aba,Witten:2019bou}. (See \cite{Witten:2019bou} for systematic discussions of what follows.)

We take a manifold $N$ with boundary $\partial N = M$ on which other necessary data (such as gauge fields, spin structure etc.) are extended. Let $\cD_{d+1}(N)$ be the Dirac operator on $N$ that is associated to $\cD^+_d(M)$ in an appropriate way. Then the fermion partition function is defined as
\beq
\cZ_{\rm bare}(N) = \exp ( -2\pi \i \eta(N) ) |\Det \cD^+_d(M)|, \label{eq:barefermion}
\eeq
where $\eta(N)$ is the APS $\eta$-invariant of the Dirac operator $\cD_{d+1}(N)$ with the APS boundary condition at the boundary $\partial N = M$ \cite{Atiyah:1975jf,Dai:1994kq}. The absolute value $|\Det \cD^+_d(M)|$ has no phase ambiguity and hence it is well-defined, but it may not be smooth as a function of the metric and gauge fields on $M$. The combination $ \exp ( -2\pi \i \eta(N) ) |\Det \cD^+_d(M)|$ is a smooth function. The subscript ``bare'' in $\cZ_{\rm bare}(N)$ is written because we will later add a counterterm. 

The partition function $\cZ_{\rm bare}(N)$ possibly depends on how to extend $M$ to $N$, so its argument is taken to be $N$. When $N$ has no boundary, i.e., $M = \varnothing$, it is defined as
\beq
\cZ_{\rm bare}(N) = \exp ( -2\pi \i \eta(N) ) \quad \text{when $\partial N = \varnothing$}.
\eeq
If $\cZ_{\rm bare}(N) =1$ for any closed manifold $\partial N = \varnothing$, then the partition function for general $N$ (with $\partial N = M$) can be shown to be independent of how to extend $M$ to $N$. 

For anomalous fermions, we can see obstruction for $\cZ_{\rm bare}(N) =1~(\partial N = \varnothing)$ by using the APS index theorem. Suppose $\partial N = \varnothing$ and also suppose that $N$ is the boundary of a $(d+2)$-dimensional manifold $L$ as $\partial L = N$. Let $\cD_{d+2}(L)$ be the Dirac operator on $L$ that is associated to $\cD_{d+1}(N)$ in an appropriate way. Then the APS index theorem~\cite{Atiyah:1975jf} states that the index of $\cD_{d+2}(L)$ with the APS boundary condition on $\partial L =N$ is given by
\beq
\Index \cD_{d+2}(L) = \int_L \cI_{d+2} + \eta(N), \label{eq:APSind}
\eeq
where $\cI_{d+2} = \cI_{d+2} (F, R) $ is the gauge invariant polynomial 
of the gauge field strength 2-form $F$ and the Riemann curvature 2-form $R$ 
that appears in the standard Atiyah-Singer index theorem in $(d+2)$ dimensions for the Dirac operator $\cD_{d+2}$. 
Then, for $N=\partial L$, we get
\beq
\exp ( -2\pi \i \eta(N) )  = \exp \left(2\pi \i  \int_L \cI_{d+2} \right) \quad \text{when $N=\partial L $}.
\eeq
The quantity $\cI_{d+2}$ is called the anomaly polynomial. If $\cI_{d+2}$ is nonzero, we generically have $\cZ_{\rm bare}(N) \neq 1$.

The anomaly polynomial can be cancelled by a counterterm in the following situation; this is essentially the Green-Schwarz mechanism. 
First, we assume that there is a $B$-field whose field strength 3-form $H$ satisfies, among other things,%
\footnote{\label{foot:H}The field strength $H$ of the $B$-field is required to satisfy a more precise condition. 
String worldsheet anomalies are described by a partition function $\cZ_\text{ws} : C \to \U(1)$, 
where $C$ is a spin 3-manifold equipped with a sigma model map $f: C \to M$ where $M$ is the target space. 
(We have not yet done the path integral over sigma model maps. 
Then the map $f$ is regarded as a background field.) 
For the worldsheet anomaly cancellation, the field strength $H$ of the $B$-field must satisfy $\exp(2\pi \i \int_C   H) \cZ_\text{ws}(C)=1$~\cite{Witten:1999eg}. 
One consequence of this condition is \eqref{eq:Heq}. 
Another consequence is that a certain characteristic class $ [c]-[\lambda] \in (I\Omega^\text{spin})^4(M)$ must vanish~\cite{Yonekura:2022reu}. 
Here, $ (I\Omega^\text{spin})$ is a generalized cohomology known as the Anderson dual of spin bordism, 
$[c]$ and $[\lambda]$ are characteristic classes determined by some elements of $(I\Omega^\text{spin})^4(BG)$ and $(I\Omega^\text{spin})^4(B\O(d))$ respectively, where $G$ is the internal symmetry group, $d$ is the dimension of the target space $M$,
and $BG$ and $B\O(d)$ are classifying spaces of $G$ and $\O(d)$, respectively.
Under some simplifying assumptions (e.g. that the target space $M$ is spin and $G$ is connected and simply connected), we can reduce $ [c]-[\lambda] \in (I\Omega^\text{spin})^4(M)$ to an ordinary cohomology element $ [c]-[\lambda] \in H^4(M; \bZ)$, the vanishing of which was originally conjectured and demonstrated in examples in \cite{Witten:1985mj}. Its reduction $H^4(M; \bZ) \to H^4(M; \bR)$ gives part of the information contained in \eqref{eq:Heq}, i.e. that the de~Rham cohomology class of $c(F) - \lambda(R) $ is zero. \label{footnote:ws}}  the equation
\beq
\d H = c(F) - \lambda(R) \label{eq:Heq}
\eeq
where $c(F) \sim \tr F^2$ is some characteristic class 4-form, and $\lambda(R) = p_1(R)/2$ is one half of the first Pontryagin class 4-form. We call this kind of structure of manifolds with gauge fields and the $B$-field satisfying \eqref{eq:Heq} (and other appropriate data) as twisted string structure, following the practice in mathematics.
When gauge fields (with field strength $F$) are absent and a manifold contains only gravity (i.e. metric) and the $B$-field, it is just called string structure.

Another assumption is that, under \eqref{eq:Heq}, the anomaly polynomial is written as a total derivative of a gauge invariant $(d+1)$-form $\cJ_{d+1}$ as
\beq
\cI_{d+2} = \d \cJ_{d+1}.\label{eq:totdel}
\eeq
Here $\cJ_{d+1}=\cJ_{d+1}(H, F,R)$ is a gauge invariant polynomial of $H, F$ and $R$. More explicitly, if $\cI_{d+2}$ factorizes as $\cI_{d+2} = (c(F) - \lambda(R)) \cI'_{d-2}$ for some gauge invariant polynomial $\cI'_{d-2}$, then we can take $ \cJ_{d+1}= H\cI'_{d-2}$. 

The differential form $\cJ_{d+1}$ may not be unique, depending on the dimension $d$ and the gauge group; if there is a gauge invariant closed form $\cJ'_{d+1}$ (i.e., $\d \cJ'_{d+1}=0$), we can modify $\cJ_{d+1}$ by $\cJ'_{d+1}$ as $\cJ_{d+1} \to \cJ_{d+1} + \cJ'_{d+1}$ without changing \eqref{eq:totdel}. 
By inspection, one may see that there is no such ambiguity if $d$ is even (i.e., $d+1$ is odd) since all gauge invariant polynomials of $F$ and $R$ have even degree, and $H$ is not closed (i.e., $\d H \neq 0$) by \eqref{eq:Heq}. 
On the other hand, if $d$ is odd, there can be such an ambiguity. For instance, when $d=3$, we may take 
\beq
\cJ'_{d+1=4} = \alpha c(F) + \beta \lambda(R) = \alpha \d H + (\alpha+\beta)\lambda(R), \label{eq:ambiguity}
\eeq
where $\alpha$ and $\beta$ are real numbers. 

Under the above assumptions, we modify the partition function by using $\cJ_{d+1}$ as a counterterm and define $\cZ(N)$ by
\beq
\cZ(N) &= \cZ_{\rm bare}(N)\exp \left( - 2\pi \i \int_N \cJ_{d+1} \right) \nonumber \\
&= \exp \left( -2\pi \i \eta(N) - 2\pi \i \int_N \cJ_{d+1}\right) |\Det \cD^+_d(M)|, \label{eq:totalpf}
\eeq
In particular, when $M =\varnothing$, we have
\beq
\cZ(N) =  \exp \left( -2\pi \i \eta(N) - 2\pi \i \int_N \cJ_{d+1}\right) \quad \text{when~$M=\partial N=\varnothing$}. \label{eq:Zgeometric}
\eeq
When $N = \partial L$, one can check by using the APS index theorem \eqref{eq:APSind} and the condition \eqref{eq:totdel} that $\cZ(\partial L)=1$. In other words, $\cZ$ is a bordism invariant of manifolds with twisted string structure. 
Previous papers on this approach of obtaining bordism invariants in this class of manifolds include \cite{Redden_2011,Bunke,Lee:2020ewl}.

Note that this cancellation of the effects of the anomaly polynomial $\cI_{d+2}$
by the counterterm $\cJ_{d+1}$ is the standard perturbative Green-Schwarz mechanism,
and a non-trivial value for $\cZ(N)$ would mean that there is a global anomaly
not yet cancelled by the perturbative Green-Schwarz mechanism.
We just saw that the global anomaly is a bordism invariant, i.e. $\cZ(N)=1$ when $N=\partial L$, as is generally expected.

When $N$ is not the boundary of a $(d+2)$-dimensional manifold, $\cZ(N)$ may take a nontrivial value. 
In the discussions above, this partition function is defined by using the geometric quantities $\eta(N)$ and $\cJ_{d+1}$. 
A more topological way of computing $\cZ(N)$ is as follows. Suppose that the disjoint union of $k$ copies of $N$ is realized as the boundary of a $(d+2)$-manifold $L$,
\beq
N^{\sqcup k}:=N \sqcup \cdots \sqcup N = \partial L.
\eeq
By taking an appropriate $k$, such an $L$ always exists for even $d$ since all characteristic classes of twisted string structure have even degree and hence bordism groups are torsion in odd dimensions.
Then, by using \eqref{eq:APSind} and \eqref{eq:totdel}, we get
\beq
 \eta(N) + \int_N \cJ_{d+1} = \frac{1}{k}  \Index \cD_{d+2}(L).
\eeq
Thus the partition function is given by
\beq
\cZ(N) = \exp \left( - \frac{2\pi \i}{k}  \Index \cD_{d+2}(L) \right). \label{eq:INV1}
\eeq
This formula has the advantage that the quantity $\frac{1}{k}\Index \cD_{d+2}(L) $ is a topological quantity and it is manifestly a rational number. This advantage can be used~\cite{Yonekura:2022reu} to define a corresponding invariant in SQFTs~\cite{Gaiotto:2019gef,Yonekura:2022reu}. 

The anomaly cancellation requires that this bordism invariant is 
actually trivial, i.e.~$\cZ(N) =1$ for any closed $N$ (i.e., $\partial N= \varnothing$). In the following, we assume that this anomaly cancellation condition is satisfied.

\subsection{The Green-Schwarz coupling}\label{sec:GSmanifold}

\subsubsection{Definition}
Suppose that the anomaly cancellation condition is satisfied in the sense discussed in the previous subsection. Then $\cZ(N)$ for $\partial N = M$ depends only on $M$; it is independent of the choice for $N$. Let us denote it as $\cZ(M)$. In general, it includes both the fermion contribution and the ``Green-Schwarz coupling''. Here the ``Green-Schwarz coupling'' is schematically of the form ``$\int_M B \cI'_{d-2}$'' which is closely related to the term $\int_N \cJ_{d+1}$ where $\cJ_{d+1} = H \cI'_{d-2}$.

When we take an $N$ such that $\partial N = M$, we can separate the two contributions $\cZ_{\rm bare}(N) $ and $\exp( - 2\pi \i \int_N \cJ_{d+1}) $ in $\cZ(M)$. Without choosing $N$, it is in general not possible to separate these contributions. 

However, there is a case in which we can take a separation of the two contributions without a choice of $N$. 
To be concrete, let us consider the following situation although more general situations may be possible. 
When we turn off the gauge fields, the fermions are coupled just to gravity. 
We consider the case that the fermions are non-chiral when they are coupled only to gravity, and we also assume that the dimension $d$ is even. 
In other words, these fermions are Dirac fermions when coupled only to gravity. 
The absolute value $|\Det \cD^+_d(M)|$ in this case is a well-behaved, valid partition function for non-chiral fermions in even dimensions when there is no zero mode on $M$.\footnote{%
The reason is as follows. For a non-chiral Dirac operator $\cD$, we have a chirality operator $\bar\gamma$ that anticommutes with the Dirac operator, $\{\bar\gamma, \cD\}=0$. 
Then, nonzero eigenvalues appear always in pairs as $\pm \lambda$. 
The determinant of the Dirac operator can be defined as $\Det \cD = \prod \lambda^2 \geq 0$, where the product is over each pair $\pm \lambda$. 
This is not necessarily the case for odd-dimensional fermions that have parity anomalies.} 

Then the ratio
\beq
\GS(M) &= \frac{\cZ(M)}{|\Det \cD^+_d(M)|}  \nonumber \\
&=  \exp \left( -2\pi \i \eta(N) - 2\pi \i \int_N \cJ_{d+1}\right) \quad \text{when $M$ is a string manifold} \label{eq:GScoupling}
\eeq
may be regarded as a contribution from the Green-Schwarz coupling. Assuming anomaly cancellation, both $\cZ(M)$ and $|\Det \cD^+_d(M)|$ depend only on $M$, so  $\GS(M)$ depends only on $M$ despite the fact that each of $\eta(N)$ and $\int_N \cJ_{d+1}$ depends on $N$. 

We remark that  $N$ in the formula \eqref{eq:GScoupling} need not be purely gravitational. It can be a general twisted string manifold. Only its boundary $M=\partial N$ is required to be a purely gravitational string manifold. We only consider the case that a twisted string manifold $N$ exists for a given string manifold $M$.

\subsubsection{A topological formula}
The above definition of $\GS(M)$ involves `geometric quantities', such as the $\eta$-invariant.
A more topological way to compute it, along the lines of the formula \eqref{eq:INV1}, may be as follows. 

First we assume that $M$ is a torsion element of the bordism group $\Omega^{\rm string}_d({\rm pt})$ of (untwisted) string manifolds, which implies that there exists an integer $k$ such that the disjoint union of $k$ copies of $M$ is the boundary of a manifold $N_{0}$. 
The manifold $N_0$ is a pure gravitational background equipped with the $B$-field, and
\beq
\partial N_0 = M^{\sqcup k} := \underbrace{M \sqcup \cdots \sqcup M}_\text{$k$ times}, \qquad \text{$N_0$ : (untwisted) string manifold}
\eeq
The assumption that the fermions are non-chiral when coupled only to pure gravity implies that the anomaly polynomial $\cI_{d+2}$ is zero when it is restricted to pure gravity. Then $\cJ_{d+1}$ is also zero when it is restricted to pure gravity for even $d$.\footnote{For odd $d$, $\cJ_{d+1}$ is not necessarily zero due to ambiguities mentioned around \eqref{eq:ambiguity}. }
For a purely gravitational $N_0$, the $\eta$-invariant associated to the non-chiral fermions is also zero. Thus we have
\beq
\eta(N_0) =0, \qquad \int_{N_0} \cJ_{d+1}=0. \label{eq:F4}
\eeq

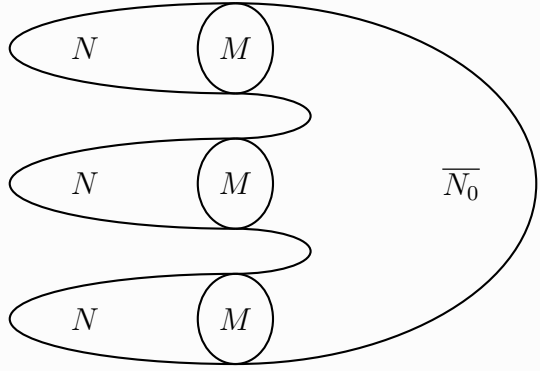
\begin{figure}
\[
\begin{tikzpicture}[yscale=.6]
\draw[thick] (0,0) ellipse (0.5 and 1) node {$M$};
\draw[thick] (0,3) ellipse (0.5 and 1) node {$M$};
\draw[thick] (0,6) ellipse (0.5 and 1) node {$M$};
\draw[thick] (0,1) arc[start angle=90, end angle=270, x radius=3, y radius= 1];
\draw[thick] (0,4) arc[start angle=90, end angle=270, x radius=3, y radius= 1];
\draw[thick] (0,7) arc[start angle=90, end angle=270, x radius=3, y radius= 1];

\draw[thick] (0,1) arc[start angle=-90, end angle=90, x radius=1, y radius= .5];
\draw[thick] (0,4) arc[start angle=-90, end angle=90, x radius=1, y radius= .5];
\draw[thick] (0,-1) arc[start angle=-90, end angle=90, x radius=4, y radius= 4];
\node at (-2,0) {$N$};
\node at (-2,3) {$N$};
\node at (-2,6) {$N$};
\node at (+3,3) {$\overline{N_0}$};
\end{tikzpicture}
\]
\caption{The gluing construction of our manifold $N^{\sqcup k} \cup \overline{N_0}$; the figure is for $k=3$. 
Note that $N$ has a twisted string structure, while $\overline{N_0}$ has an untwisted string structure.
\label{fig:first}}
\end{figure}

Notice that $M^{\sqcup k}$ is the boundary of $N^{\sqcup k}$, which is the disjoint union of $k$ copies of $N$.
Let $\overline{N_0}$ be the orientation reversal of $N_0$. By gluing $N^{\sqcup k}$ and $\overline{N_0}$ along the common boundary, we get a closed $(d+1)$-manifold
$
N^{\sqcup k} \cup \overline{N_0},
$
see Figure~\ref{fig:first}.

Now we claim the following. Suppose that $M$ is trivial as an element of the spin bordism group $\Omega^\text{spin}_d({\rm pt})$. This means that  there exists a spin manifold $N_1$, not necessarily equipped with the $B$-field or string structure, such that $\partial N_1 = M$, where in this equality we forget the $B$-field or string structure on $M$.
Under this condition, we claim that
\beq
 \eta(N) \equiv \frac{1}{k} \eta(N^{\sqcup k} \cup \overline{N_0}) \mod \bZ, \qquad \text{ if $[M] \in \Omega^\text{spin}_d({\rm pt})$ is zero.} \label{eq:claim1}
\eeq
We will later give an argument for this claim. For a while we just assume that this claim holds. The intuition for this claim is that the contribution to the $\eta$-invariant on $N^{\sqcup k} \cup \overline{N_0}$ from the pure-gravitational region $\overline{N_0}$ is zero.

Assuming \eqref{eq:claim1}, we get
\beq
\eta(N) + \int_N \cJ_{d+1} \equiv \frac{1}{k} \left( \eta( N^{\sqcup k} \cup \overline{N_0} ) + \int_{N^{\sqcup k} \cup \overline{N_0}} \cJ_{d+1}\right) \mod \bZ
\eeq
where we have used the fact that $\cJ_{d+1}$ is zero on  $\overline{N_0}$. 

\begin{figure}
\centering

 \begin{tikzpicture}[scale=2]

  \def\r{1.5}
  \def\gap{0.3}

  \draw[thick] (0,0) arc[start angle=180, end angle=270, radius=\r];

  \draw[red, very thick] (0,0) -- (\r , 0); 
 
  \draw[blue, very thick] (\r , 0) -- (\r, -\r);
 
  \node at (\r + 0.5*\gap, -0.5*\r) {$N_0$};
  \node at (\r/2, -0.5*\r) {$ {L}$};

  \node at (\r/2, 0.15) {$N^{\sqcup k}$};
  
    \node at (\r + 0.5*\gap, 0.15) {$M^{\sqcup k}$};
 
\end{tikzpicture}

\caption{The $(d+2)$-manifold $L$ and its boundary components. The $(d+1)$-manifolds $N^{\sqcup k}$ and $N_0$ are glued along their common boundary $M^{\sqcup k}$ which is a $d$-manifold. It will also be convenient to regard $L$ as a manifold with a corner $M^{\sqcup k}$, as in the figure. \label{fig:0}}
\end{figure}
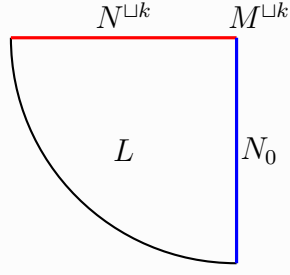

Let us now pick an $\ell$ such that 
the disjoint union of $\ell$ copies of $N^{\sqcup k} \cup \overline{N_0}$ is a boundary of some $  L$ with a twisted string structure. 
Such an $\ell$ and an $ L$ always exist for even $d$, since all characteristic classes of twisted string structure have even degree, and hence bordism groups are torsion in odd dimensions.
If $\ell \neq 1$, we can consider $k'=\ell k$ and $N_0' = N_0^{\sqcup \ell}$,
 and use $k'$ and $N'_0$ instead of $k$ and $N_0$ from the beginning. 
 Thus, without loss of generality we can assume $\ell=1$ by taking $k$ sufficiently large. Then we have 
 \beq
 \partial    L =  N^{\sqcup k} \cup \overline{N_0}. 
 \eeq
 A naive way to draw it would be as in Figure~\ref{fig:naive},
 but it turns out to be more convenient and illuminating for us later to use a schematic illustration
 as in Figure~\ref{fig:0},
 where we regard $L$ to have a boundary with two components, $N^{\sqcup k}$ and $\overline{N_0}$,
 meeting at their common boundary, $M^{\sqcup k}$.
 In this type of schematic figures, we will not carefully distinguish the orientations of manifolds.
 
 \begin{figure}
 \centering
 \begin{tikzpicture}[yscale=.3]
 \clip (-3,-8) rectangle (5,8);

\draw[thick] (-3,6) arc[start angle=180, end angle=200, x radius=3.5, y radius= 10];
\draw[thick] (-3,3) arc[start angle=180, end angle=200, x radius=3.5, y radius= 10];
\draw[thick] (-3,0) .. controls (-2,-10) and (4,-10) .. (4,3);

\draw[thick,fill=white] (0,1) arc[start angle=90, end angle=270, x radius=3, y radius= 1];
\draw[thick,fill=white] (0,4) arc[start angle=90, end angle=270, x radius=3, y radius= 1];
\draw[thick,fill=white] (0,7) arc[start angle=90, end angle=270, x radius=3, y radius= 1];

\draw[thick] (0,0) ellipse (0.5 and 1) node {$M$};
\draw[thick] (0,3) ellipse (0.5 and 1) node {$M$};
\draw[thick] (0,6) ellipse (0.5 and 1) node {$M$};

\draw[thick] (0,1) arc[start angle=-90, end angle=90, x radius=1, y radius= .5];
\draw[thick] (0,4) arc[start angle=-90, end angle=90, x radius=1, y radius= .5];
\draw[thick] (0,-1) arc[start angle=-90, end angle=90, x radius=4, y radius= 4];

\node at (0,-4) {$L$};

\node at (-2,0) {$N$};
\node at (-2,3) {$N$};
\node at (-2,6) {$N$};
\node at (+3,3) {$\overline{N_0}$};
\end{tikzpicture}

 \caption{The manifold $L$ such that $\partial L=N^{\sqcup k} \cup \overline{N_0}$.
 \label{fig:naive}}
\end{figure}
 
The APS index theorem implies that
\beq
 \eta(N^{\sqcup k} \cup \overline{N_0}) + \int_{N^{\sqcup k} \cup \overline{N_0}} \cJ_{d+1} = \Index \cD_{d+2}(  L), \qquad (\partial    L = N^{\sqcup k} \cup \overline{N_0}).
\eeq
In this way, we finally arrive at the formula
\beq
\GS(M) = \exp\left( -\frac{2\pi \i}{k} \Index \cD_{d+2}(  L)  \right) . \label{eq:topoGS}
\eeq
This formula makes the topological nature of $\GS(M) $ clearer, at the cost of making less obvious the fact that it only depends on $M$, i.e., it is independent of various choices involved in $ L$.

One of the important properties of $\GS(M) $ is that it  is a bordism invariant of $\Omega^{\rm string}_d({\rm pt})$. 
To see this, suppose that there exists a string manifold $N'_0$ such that $\partial N'_0 = M$. 
Then in the above discussion we take $N_0=(N_0')^{\sqcup k}$ and hence $N^{\sqcup k} \cup \overline{N_0} =(N \cup \overline{N_0}')^{\sqcup k} $. 
In this case one can see that $\GS(M)$ is precisely equal to the value of the anomaly $\cZ(N \cup \overline{N'_0})$, see \eqref{eq:INV1}. 
By the anomaly cancellation condition, this is trivial and hence the bordism invariance (i.e. $\GS(M)=1$ for $M=\partial N'_0$) is established. 

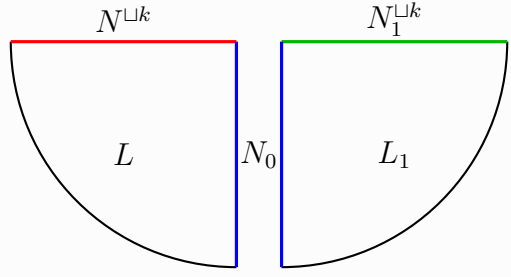
\begin{figure}
\centering

 \begin{tikzpicture}[scale=2]

  \def\r{1.5}
  \def\gap{0.3}

  \draw[thick] (0,0) arc[start angle=180, end angle=270, radius=\r];

  \draw[thick] (\r + \gap, -\r) arc[start angle=270, end angle=360, radius=\r];

  \draw[red, very thick] (0,0) -- (\r , 0); 
  \draw[green!70!black, very thick] (\r + \gap, 0) -- (2*\r + \gap, 0); 

  \draw[blue, very thick] (\r , 0) -- (\r, -\r);
    \draw[blue, very thick] (\r+\gap , 0) -- (\r+\gap, -\r);

  \node at (\r + 0.5*\gap, -0.5*\r) {$N_0$};
  \node at (\r/2, -0.5*\r) {$ {L}$};
  \node at (3*\r/2 + \gap, -0.5*\r) {$ {L}_1$};

  \node at (\r/2, 0.15) {$N^{\sqcup k}$};
  \node at (3*\r/2 + \gap, 0.15) {$N_1^{\sqcup k}$};

\end{tikzpicture}

\caption{Manifolds and boundaries appearing in \eqref{eq:mfd1} and \eqref{eq:mfd2}. \label{fig:1}}
\end{figure}

Our remaining task is to argue for the claim \eqref{eq:claim1}. The assumption was that $M$, as a spin manifold by forgetting the string structure, is the boundary of a spin manifold $N_1$. We consider the situation as in Figure~\ref{fig:1}. In the figure,  $  L$ is a manifold with a twisted string structure as before, and $  L_1$ is a spin manifold whose boundary is $N_0 \cup \overline{N_1^{\sqcup k} }$. By taking $k$ sufficiently large, such a spin manifold $  L_1$ exists since the spin bordism group $\Omega^\text{spin}_{d+1}({\rm pt})$ is torsion when $d+1$ is odd (or more precisely when it is not a multiple of 4). By definition, they have the boundaries
\beq
\partial   L = N^{\sqcup k} \cup \overline{N_0}, \qquad \partial   L_1 = N_0 \cup \overline{N_1^{\sqcup k} }. \label{eq:mfd1}
\eeq
We also consider a manifold $  L \cup_0   L_1$ which is obtained by gluing $  L $ and $  L_1$ along $N_0$. It has the boundary
\beq
\partial (  L \cup_0   L_1) = N^{\sqcup k} \cup  \overline{N_1^{\sqcup k} } = ( N \cup \overline{N_1})^{\sqcup k}. \label{eq:mfd2}
\eeq
In the figure, when there is a corner on a boundary, we may make it smooth by slightly deforming it so that we can consider the APS index theorem on $  L$, $  L_1$ and $  L \cup_0   L_1$. 

For simplicity we assume that the boundaries $ N^{\sqcup k} \cup \overline{N_0}$, $N_0 \cup \overline{N_1^{\sqcup k} }$ and $( N \cup \overline{N_1})^{\sqcup k}$ do not have zero modes. Then we can avoid subtleties of the definition of the APS indices on $  L$, $  L_1$ and $  L \cup_0   L_1$, since all zero modes are localized in the interior regions and they are normalizable even if the boundaries are extended by semi-infinite cylinders.\footnote{Such an extension will be used in the case of SQFTs, since it is difficult in SQFTs to introduce a sharp boundary. When boundaries have zero modes, we need to consider non-normalizable zero modes as well for the definition of the APS index.}

By the APS index theorem, we have
\beq
\Index \cD_{d+2}(  L) & = \eta(N^{\sqcup k} \cup \overline{N_0} )+ \int_{  L} \cI_{d+2}, \nonumber \\
\Index \cD_{d+2}(  L_1) & = \eta(N_0 \cup \overline{N_1^{\sqcup k} }) + \int_{  L_1} \cI_{d+2}, \nonumber \\
\Index \cD_{d+2}(  L \cup_0   L_1) & = k \eta(N \cup \overline{N_1} ) + \int_{   L \cup_0   L_1 } \cI_{d+2}.\label{eq:F1}
\eeq
We use the fact that the fermions are non-chiral on purely gravitational backgrounds and hence various quantities related to anomalies are zero,
\beq
\Index \cD_{d+2}(  L_1)=0, \qquad \eta(N_0 \cup \overline{N_1^{\sqcup k} })=0 , \qquad \int_{  L_1} \cI_{d+2}=0,
\qquad \eta(N_1)=0. \label{eq:F2}
\eeq
These are due to exact cancellation between ``positive and negative chiralities''.  

A crucial property of the APS index is the gluing law. As mentioned above, zero modes on $  L$ and $  L_1$ are localized in the interior regions. Moreover, when we glue $  L$ and $  L_1$ to get $  L \cup_0   L_1$, we can insert a very long cylinder of the form $\bR \times N_0$ so that the zero modes on $  L$ and $  L_1$ are separated by large distances. This fact, together with the robustness of the APS index under small perturbations, gives
\beq
\Index \cD_{d+2}(  L \cup_0   L_1) = \Index \cD_{d+2}(  L) + \Index \cD_{d+2}(  L_1) . \label{eq:F3}
\eeq
This  is the gluing law of the APS index.

From \eqref{eq:F1},  \eqref{eq:F2} and  \eqref{eq:F3}, we get
\beq
\eta(N \cup \overline{N_1} ) =\frac{1}{k}  \eta(N^{\sqcup k} \cup \overline{N_0} ).
\eeq
Finally, the Dai-Freed theorem~\cite{Dai:1994kq} (see \cite{Yonekura:2016wuc} for physical explanations) implies that 
\beq
\eta(N \cup \overline{N_1} ) \equiv \eta(N ) + \eta(  \overline{N_1} ) \mod \bZ. \label{eq:DaiFreed}
\eeq
Then the desired result \eqref{eq:claim1} can be deduced from the fact $ \eta(  {N_1} )=0$ mentioned in \eqref{eq:F2} and the fact $\eta(  \overline{N_1} ) = - \eta(  {N_1} )$.

\subsubsection{Summary of the situation and further comments}

So far, we have assumed that the dimension $d$ is even. In some cases, we can relax this condition.

One reason we restricted our attention to even $d$ was that characteristic classes in differential forms have even degrees. 
If $d+1$ is even, the relevant bordism group in degree $d+1$ may have non-torsion elements (because integration of characteristic classes may give bordism invariants), and hence the $(d+2)$-manifold $L$ in our construction may not exist. 
Aso, the differential form $\cJ_{d+1}$ is ambiguous when we can add nontrivial characteristic classes in differential forms. 
Adding a characteristic class to $\cJ_{d+1}$ corresponds to adding a Chern-Simons term in $d$-dimensional $M$. 

However, it often happens that the symmetry group $G$ is such that all the degrees of characteristic classes in less than or equal to $(d+1)$-dimensions are multiples of $4$. For instance, consider the case that $G$ is trivial. Then, we only have Pontryagin classes of manifolds and their degrees are multiples of 4. If the degrees of all characteristic classes in the  range $\leq d+1$ are multiples of $4$, we can relax the condition on $d$ and just require that $d+1$ is not a multiple of 4, $ d+1 \notin 4\bZ$. 

The condition that $d$ was even was also used when we required that the fermions coupled only to gravity were non-chiral. 
It is enough to require that the fermions are anomaly-free, and define the fermion partition function by $|\Det \cD^+_d(M)| \exp(-2\pi \i \eta(N_1))$, where $N_1$ is a purely gravitational spin manifold such that $\partial N_1=M$. When the fermions are anomaly-free, the partition function does not depend on the choice of $N_1$. In this case, the Green-Schwarz coupling can be defined by using $\eta(N \cup \overline{N_1} ) $ instead of $\eta(N)$ as
\beq
\GS(M)= \exp \left( -2\pi \i \eta(N \cup \overline{N_1}) - 2\pi \i \int_N \cJ_{d+1}\right). \label{eq:alternative}
\eeq
This definition avoids the use of the $\eta$-invariant on a manifold $N$ with boundary $\partial N \neq \varnothing$, since $N \cup \overline{N_1}$ is a closed manifold. The cost is to introduce an additional manifold $N_1$.

Our construction was complicated, so
let us recapitulate the conditions which we used to arrive at the formula \eqref{eq:topoGS}. 

\begin{enumerate}
\item The anomaly is cancelled on manifolds with twisted string structure, so that the partition function $\cZ(M)$ is defined on $d$-dimensional manifolds $M$.
That is, the definition \eqref{eq:totalpf} does not depend on the choice of $N$ and it only depends on $M = \partial N$. 
\item Our fermions are anomaly-free on purely gravitational backgrounds without gauge fields, even without the Green-Schwarz mechanism. 
Our $\GS(M)$ is to be defined for such purely gravitational $M$ with the $B$-field assigned, 
i.e.~once the string structure is chosen.
\item $d +1 \notin 4\bZ$ (or $d$ is even, depending on characteristic classes of gauge group $G$). 
\item There exists a manifold $N$ with a twisted string structure such that $\partial N = M$, where the equality is as (twisted) string manifolds. 
\item There exists a purely gravitational manifold $N_1$ such that $\partial N_1 = M$, where the equality is as spin manifolds, forgetting the $B$-field.
\item Under these conditions, the invariant $\GS$ is defined by \eqref{eq:GScoupling} or \eqref{eq:alternative}, and it has a topological formula \eqref{eq:topoGS}. 
\end{enumerate}

When we define our invariant of SQFTs, we need analogous (but not completely parallel) conditions, roughly as follows:

\begin{enumerate}
\item The anomaly cancellation is automatically guaranteed, modulo an important assumption that $\SQFT_{-21}({\rm pt}) =0$ which will be explained in Section~\ref{sec:def}.
\item The anomaly-free condition on fermions coupled to pure gravity will be replaced by a condition on the ``internal theory'' $\sT$ (in string-theoretic interpretation) that its primary invariant vanishes.
\item The restriction on $d$ is removed, but for $d +1 \in 4\bZ$ it will be required that the theory corresponding to the ``sigma model part'' $\sX$ (roughly corresponding to a sigma model with target space $M$), as well as the  ``internal theory'' $\sT$, have vanishing secondary invariants. 
\item The condition corresponding to the existence of a twisted-string manifold $N$ such that $\partial N = M$ will be automatically satisfied, modulo the assumption mentioned in Section~\ref{sec:def}.
\item The condition on the existence of a purely gravitational spin manifold $N_1$ such that $\partial N_1 = M$ will be replaced by the condition that the primary invariant of the theory $\sX$ above is zero.
\item Under these conditions, a pairing $\GS([\sX], [\sT])$ between the ``internal theory $\sT$'' and the ``sigma model part $\sX$'' is defined. 
\end{enumerate}
Here the terms the ``internal theory'' and the ``sigma model part'' are used only to aid intuition, and $\GS([\sX], [\sT])$ is defined for general SQFTs $\sX$ and $\sT$.

There is one final comment related to Section~\ref{sec:differential}. Even if fermions are not anomaly-free when they are coupled to pure gravity, the quantity on the right-hand side of \eqref{eq:alternative} still makes sense. Its values depend on $N_1$, and it is possibly not a topological invariant but depends continuously on the metric of the target space. In Section~\ref{sec:differential}, we will construct an interesting quantity in SQFTs which is analogous to \eqref{eq:alternative}.

\section{General properties of SQFTs} \label{sec:generalSQFT}

In this section, we discuss general properties of SQFTs. Most of the contents of this section are reviews, but a few points might be new, such as a general derivation of the holomorphic anomaly equation mentioned in this section and derived in Appendix~\ref{sec:A}.

\subsection{Basic notions}\label{sec:notion}

We start by recalling some basic notions in two-dimensional ${\cal N}{=}(0,1)$ SQFTs. Some backgrounds reviewed here may be found in \cite{Gaiotto:2019asa,Gaiotto:2019gef,Johnson-Freyd:2020itv,Yonekura:2022reu,Tachikawa:2023nne}.

\paragraph{Gravitational anomaly.}
In two dimensions, gravitational anomalies are specified by an integer $ \nu \in \bZ$ which is the coefficient of the anomaly polynomial.
For instance, the gravitational anomalies of $\cN{=}(0,1)$ sigma models come from the  Majorana-Weyl fermions that are superpartners of the bosonic fields. By convention, we regard these superpartner fermions as right-moving. If the dimension of the target space of a sigma model is $d$, then we take the sign convention that the gravitational anomaly is $\nu =d$. On the other hand, if we have $n$ copies of left-moving Majorana-Weyl fermions, the gravitational anomaly is $\nu = -n$.

Whenever we consider Hilbert spaces of SQFTs, they are the ones on $S^1$ which has the periodic (Ramond) spin structure. The Hamiltonian, the momentum, and the supercharge are denoted as $H$, $P$ and $Q$. We define \footnote{When an SQFT has conformal invariance, $H_L$ and $H_R$ are Virasoro generators usually denoted as $L_0-c_L/24$ and $\bar L_0-c_R/24$, where $c_L$ and $c_R$ are the left and right central charges, respectively. The gravitational anomaly is given by $\nu=2(c_R-c_L)$. However, we neither  assume nor use conformal invariance in general discussions throughout the paper, except for a few examples.} 
\beq
H_L = \frac{H+P}{2}, \qquad H_R= \frac{H-P}{2}.
\eeq
The supersymmetry algebra is simply given by
\beq
Q^2 =H_R.
\eeq

As we will consider not necessarily conformal SQFTs, 
we need to specify the circumference of $S^1$,
which we fix to be $2\pi$.
Then,  the eigenvalues of the momentum operator $P$ take values in
\beq
P \in \bZ - \frac{\nu}{24}. \label{eq:Peigenvalue}
\eeq 
Without gravitational anomalies (i.e. $\nu=0$), the momentum takes values in $\bZ$ as required by the rotational symmetry of $S^1$. However, it is shifted by gravitational anomalies. The amount of the shift $-\nu/24$ can be explicitly computed e.g. in the case of free Majorana-Weyl fermions. 

\paragraph{Compact SQFTs.}
We will consider ``compact'' SQFTs  defined as follows. Consider the Hilbert space of the theory on $S^1$. Then, a compact theory is such that the Hamiltonian $H$ has a discrete spectrum and $e^{-\beta H}$ is  trace class for any $\beta>0$, i.e. $\Tr e^{-\beta H} < +\infty$. For instance a sigma model with a compact target space $M$ without boundary (i.e., $M$ is closed) is a compact theory. 

\paragraph{Orientation reversal.}
For an SQFT $\sM$, we define its orientation reversal\footnote{%
Here the orientation refers to the target space orientation in the case of sigma models.
} $\overline{\sM}$ by adding a discrete theta term associated to ${\rm Hom}(\Omega^\text{spin}_2({\rm pt}), \U(1)) \simeq \bZ_2$.
This discrete theta term is also known as the Arf theory,
or the topologically nontrivial phase of the Kitaev chain.
See \cite{Kapustin:2014dxa,Freed:2016rqq,Yonekura:2018ufj} for the importance of spin bordism groups for discrete theta angles.
 
 For instance, consider the sigma model with target space $\bR$, and change the orientation of it. The fermion $\psi$ which is the superpartner of the position operator $X$ of $\bR$ also changes sign under the orientation reversal  as $\psi \to -\psi$. Then the theta term is generated from the path integral measure of $\psi$, because the path integral measure changes sign under $\psi \to -\psi$ when there are odd number of zero modes of $\psi$. 

\paragraph{Product of SQFTs.}
When we have two theories $\sM_1$ and $\sM_2$,
we define their product $\sM_1\times \sM_2$ to be the combined theory
such that its Hilbert space is the tensor product 
of the Hilbert spaces of $\sM_1$ and $\sM_2$. 
When $\sM_1$ and $\sM_2$ have the gravitational anomalies $\nu_1$ and $\nu_2$,
the product $\sM_1\times\sM_2$ has the gravitational anomaly $\nu_1+\nu_2$.
It is graded-commutative (up to isomorphism): $\sM_1 \times  \sM_2 \simeq (-1)^{\nu_1\nu_2} \sM_2 \times \sM_1$, where the sign factor means orientation reversal, $-\sM = \overline{\sM}$.\footnote{%
The use of the sign for orientation reversal is slightly misleading because the sum of $\sM$ and $-\sM$ is not zero. 
However, the notation $\overline{\sM}$ using the bar is inconvenient when we consider equations such as $\sM_1 \times \sM_2 \simeq (-1)^{\nu_1\nu_2} \sM_2 \times \sM_1$, 
so we opt for using the minus sign for orientation reversal. 
Later we will introduce bordism classes of SQFTs $[\sM]$. Then we indeed have $[\sM] +[-\sM]=0$. }

In the case of sigma models with target spaces $M_1$ and $M_2$,
their product is the sigma model with target space $M_1\times M_2$.
In this case, the graded-commutativity comes from reordering of the fermionic superpartners of the coordinate degrees of freedom.

\paragraph{Sum of SQFTs.}
When we have two theories $\sM_1$ and $\sM_2$, their disjoint union $\sM_1 \sqcup \sM_2$ (which may also be denoted by the sum $\sM_1 + \sM_2$) is the theory in which we have two different vacua corresponding to $\sM_1$ and $\sM_2$, respectively, and we have the theories $\sM_1$ and $\sM_2$ on each of these vacua. For instance, in the case of sigma models with target spaces $M_1$ and $M_2$, we have a sigma model with a target space  $M_1 \sqcup M_2$ that is the disjoint union of $M_1$ and $M_2$.

\paragraph{Mildly noncompact SQFTs.}
In addition to compact theories, we also need to consider some noncompact theories. We do not try to discuss a precise definition, but let us give an intuitive explanation. Let us start from the case of sigma models. Consider a manifold $N$ with boundary $\partial N = M$. In this case, we attach a semi-infinite cylinder $\bR_{\geq 0} \times M$ to $N$, where $\bR_{\geq 0}$ is the space of real nonnegative numbers. We glue $N$ and $\bR_{\geq 0} \times M$ along $M$ and the result is a noncompact manifold without boundary. By abuse of notation, we denote the noncompact manifold also as $N$, and we say that ``$M$ is the boundary of $N$, $\partial N=M$'' although this is mathematically an imprecise statement. A little more precisely, one may say that $M$ is a boundary of $N$ at infinity. Whenever we consider a sigma  model whose target space has boundary, we always consider the version made noncompact in this way, so hopefully there is no confusion. 

More generally, an SQFT $\sN$ with boundary $\sM=\partial \sN$ is considered in the following way. The theory $\sN$ is noncompact in the sense that the Hamiltonian $H$ has a continuous spectrum. But the continuous spectrum is controlled in the sense that it is the same as the continuous spectrum of $\bR_{\geq 0} \times \sM$, where $\bR_{\geq 0}$ is regarded as a sigma model with the target space $\bR_{\geq 0}$; 
here we do not care what happens in the ``negative region'' of $\bR_{\geq 0}$ as far as it is made compact in the sense of energy spectrum. 
The part $\sM$ is compact. Notice that $\bR_{\geq 0}$ has the gravitational anomaly $\nu=1$, so if the gravitational anomaly of $\sM$ is given by $\nu_\sM$, then the gravitational anomaly of $\sN$ is $\nu_\sN=\nu_\sM +1$, in accord with the case of manifolds. 

\paragraph{Position operator.}
We may rephrase the situation of the previous paragraph as follows. On $\sN$, we assume that there exists an operator $X$, which is intuitively regarded as ``a position operator'', with the following properties. Let $\ket{x}$ be an eigenstate of $X$ with 
\beq
X\ket{x} = x\ket{x}. \label{eq:position}
\eeq
(We suppress other quantum numbers that are necessary to specify states.) In the region $x\geq 0$, the operator $X$ is the position operator for $\bR_{\geq 0} $. The wavefunction $\langle x | \Psi \rangle$ of a quantum state $\ket{\Psi}$ behaves as a quantum mechanical wavefunction on $\bR_{\geq 0}$. In the region $x<0$, the operator $X$ does not necessarily have an interpretation as a simple position operator. For instance, in the case of a sigma model with a target space $N$, we may take $X$ to be a smooth function on $N$ which coincides with the coordinate of $\bR_{\geq 0}$ in the cylinder region $\bR_{\geq 0} \times M$. In the interior region, the function is more complicated, and it has a lower bound $X \geq x_{\rm low}$ for some $x_{\rm low}<0$. 

In the formalism of using $X$, the boundary theory $\sM$ is recovered as follows. Let us add a left moving Majorana-Weyl fermion $\lambda$ to the theory $\sN$. We regard $X$ and $\lambda$ as the bottom components of the corresponding $\cN=(0,1)$ supermultiplets (i.e., a chiral multiplet and a fermi multiplet, respectively). Then we introduce a superpotential 
\beq
\int \d \theta\, m\lambda (X- x_*) \label{eq:superpot}
\eeq
where $m$ and $x_*$ are parameters, and $\theta$ is the supercoordinate of $\cN=(0,1)$ superspace. If $m \to \infty$ and $x_*>0$, we see that both $X$ and $\lambda$ become very massive and decouple in the low energy limit, leaving the theory $\sM$ at $X=x_*$. 
On the other hand, the compactness of the theory $\sN$ in the region $x_* < 0$ implies that the energy becomes large in the limit $x_* \to -\infty$ (or it may be that $x_* < x_{\rm low}$ is not in the spectrum of the operator $X$ as in the case of a smooth function in a sigma model). Thus, in the limit $x_* \to - \infty$, the above superpotential breaks supersymmetry. We conclude that $\sM = \partial \sN$ can be deformed to a theory in which supersymmetry is spontaneously broken,
by just varying $x_*$. 
Conversely, if a theory $\sM$ can be deformed by a parameter $x_*$ to a theory in which supersymmetry is spontaneously broken, then it is possible to realize it as the boundary of some $\sN$, since we can promote the parameter $x$ to a dynamical field to get a theory $\sN$. See \cite{Gaiotto:2019asa,Gaiotto:2019gef,Johnson-Freyd:2020itv,Yonekura:2022reu} for more discussions.

\begin{figure}
\centering

\begin{tikzpicture}[scale=1.2, thick]

  \def\a{2}

  \draw[red, very thick] (-3,0.5) -- (-2,0.5);
  \draw[green!70!black, very thick] (-2,0.5) -- (2,0.5);
  \draw[blue, very thick] (2,0.5) -- (3,0.5);

  \fill[black] (-3,0.5) circle (5pt);
  \fill[black] (3,0.5) circle (5pt);

  \node at (-2.5,0.8) {$\mathsf{N}_1$};
  \node at (0,0.9) {$[-a,a] \times \mathsf{M}$};
  \node at (2.5,0.8) {$\mathsf{N}_2$};

  \draw[->] (-3.5,-0.2) -- (3.5,-0.2) node[anchor=west] {$x$};

  \foreach \x/\label in {-2/{$-a$}, 0/{$0$}, 2/{$a$}} {
    \draw (\x,-0.25) -- (\x,-0.15);
    \node at (\x,-0.5) {\label};
  }

\end{tikzpicture}

\caption{Gluing of two theories $\sN_1$ an $\sN_2$. The black blobs are compact regions.  \label{fig:6}}
\end{figure}
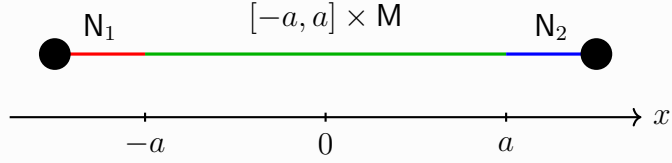

\paragraph{Gluing of SQFTs.}
When we have two theories $\sN_1$ and $\sN_2$ with boundaries $\partial \sN_1 = \partial \sN_2 = \sM$, then we may glue them along $\sM$ to obtain a compact theory which we denote as $\sN_1 \cup \overline{\sN_2}$. See Figure~\ref{fig:6}. We do not try to give a precise definition, but the intuition is as follows. We assume the existence of an operator $X$ with the following properties. Let $\ket{x}$ be the eigenstate of $X$ with eigenvalue $X \ket{x} = x\ket{x}$ as before. In the region $x \leq -a$, the theory is described by degrees of freedom of $ \sN_1$. In the region  $x \geq a$, the theory is described by degrees of freedom of $ \overline{\sN_2}$. In the region $-a \leq x \leq a$, the theory is described by $[-a,a] \times \sM$ and $X$ is the position operator of $[-a,a]$. Whenever we glue two theories $\sN_1$ and $\sN_2$, we implicitly assume that $a$ is taken very large, $ a \gg 1$. Then $\sN_1$ and $\sN_2$ are separated by a large distance, and what happens in compact regions (i.e., black blobs in Figure~\ref{fig:6}) will not be very much affected by the gluing.

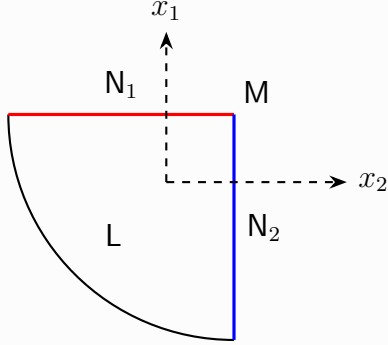
\begin{figure}
\centering

\begin{tikzpicture}[scale=2]

  \def\r{1.5}

  \draw[thick] (-\r,0) arc[start angle=180, end angle=270, radius=\r];

  \draw[red, very thick] (0,0) -- (-\r,0);

  \draw[blue, very thick] (0,0) -- (0,-\r);

  \draw[thick,dashed,-Stealth] (0-0.3*\r ,0-0.3*\r) -- (0-0.3*\r,1.0-0.3*\r) node[above] {$x_1$};
  \draw[thick,dashed,-Stealth] (0-0.3*\r,0-0.3*\r) -- (1.2-0.3*\r,0-0.3*\r) node[right] {$x_2$};

  \node at (-0.8,-0.8) {$\mathsf{L}$};
  \node at (-0.75,0.2) {$\mathsf{N}_1$};
  \node at (0.2,-0.75) {$\mathsf{N}_2$};
  \node at (0.15,0.15) {$\mathsf{M}$};

\end{tikzpicture}

\caption{An SQFT $\sL$ with a corner. \label{fig:2}}
\end{figure}

\paragraph{SQFTs with a corner.}
We will also need to deal with ``SQFTs with a corner'' in the following sense. If a theory $\sL$ has a corner, it has two noncompact regions which have overlap: see Figure~\ref{fig:2}. One noncompact region is $\bR_{\geq 0} \times \sN_1$ and the other is $\overline{\bR'_{\geq 0} \times \sN_2}$. We parametrize $\bR_{\geq 0} $ and $\bR'_{\geq 0} $ by coordinates $x_1$ and $x_2$,
\beq
\bR_{\geq 0}=\{ x_1 \geq 0\}, \qquad \bR'_{\geq 0}=\{ x_2 \geq 0\}.
\eeq
 $\sN_1$ and $\sN_2$ are SQFTs with boundaries $\partial \sN_1 =  {\partial \sN_2} = \sM$. In more detail, $\sN_1$ has noncompact region $\bR'_{\geq 0} \times \sM$, and $\sN_2$ has noncompact region $\bR_{\geq 0} \times {\sM}$. The noncompact regions $\bR_{\geq 0} \times \sN_1$ and $\overline{\bR'_{\geq 0} \times \sN_2}$ are overlapping in the region $\bR_{\geq 0} \times \bR'_{\geq 0} \times \sM$, where we have used the fact that $\overline{ \bR'_{\geq 0} \times \bR_{\geq 0} \times \sM } = \bR_{\geq 0} \times \bR'_{\geq 0} \times \sM$.

After some deformation of the behavior at infinity, the theory $\sL$ with a corner can also be regarded as an SQFT with boundary $\sN_1 \cup \overline{\sN_2}$. 
Our main interest on $\sL$ will be the APS index discussed below. 
As far as the theory $\sN_1 \cup \overline{\sN_2}$ does not have zero modes, the modification at infinity does not affect the APS index on $\sL$.

\paragraph{CPT or time-reversal symmetry.}
As mentioned before, gravitational anomalies are classified by $\nu \in \bZ$. When we consider the structure of Hilbert spaces, the mod-8 reduction $\nu$ mod $8$ often plays an important role.

The CPT symmetry in two dimensions reduces to a time reversal symmetry in one dimension after compactification on  $S^1$. The following discussions apply to both two-dimensional CPT and one-dimensional time-reversal symmetry, so we denote the symmetry as $T$.  
In one-dimensional quantum mechanics, what we discuss is the case of $\pin^{-}$, and anomalies are classified by $\nu \in \bZ_8$ \cite{Fidkowski:2009dba,Kapustin:2014dxa,Witten:2015aba}.

The time-reversal (or CPT) symmetry $T$, the fermion parity $(-1)^F$ (for even $\nu)$, and the supercharge $Q$ satisfy the algebraic relations given as follows, see e.g. the Appendix of \cite{Tachikawa:2023nne} for details.
Namely, when $\nu=2m$, we have
\begin{equation}
\begin{aligned}
  T^2&=(-1)^{\frac12 m(m-1)}, & (-1)^F  T &= (-1)^m T(-1)^F,\\
   TQ &= (-1)^m QT,&  (-1)^FQ &= -Q(-1)^F  
\end{aligned} \label{eq:time1}
\end{equation}
 and when $\nu=2m-1$, we have
 \beq
 T^2=(-1)^{\frac12 m(m-1)}, \qquad   TQ = (-1)^m QT.  \label{eq:time2}
\eeq
Consequences of these algebras will be used at several points in this paper.

\paragraph{APS index.}
Now we are going to define the APS index of a theory $\sL$ with boundary $\partial \sL = \sN$ (or a corner). When the gravitational anomaly $\nu_\sL$ of $\sL$ is even, the fermion parity operator $(-1)^{F}$ is well-defined. Then we can consider the index by using the supercharge $Q$ as an analog of the Dirac operator and the fermion parity $(-1)^F$ as an analog of the chirality operator. This is the standard Witten index~\cite{Witten:1982df}. 

For simplicity, let us explain the case that there are no zero modes on the boundary $\sN$ in the sense that there are no states in $\sN$ that are annihilated by the supercharge $Q$. (More general case will be briefly mentioned in footnote~\ref{footnote:generalAPS}.) 
Then we expect that the states of $\sL$ that are annihilated by $Q$ are localized in the interior region, which means that these states are normalizable states in the sense of quantum mechanics.

Now we decompose the Hilbert space $\cH^\sL$ of $\sL$ as
\beq
\cH^L = \bigoplus_{n \in \bZ} \cH^{\sL}_{n - \nu_\sL/24},
\eeq
where $\nu_\sL$ is the gravitational anomaly of $\sL$, and $\cH^{\sL}_{n - \nu_\sL/24}$ is the eigenspace of $P$ with eigenvalue $P=n- \nu_\sL/24$. On each subspace $\cH^{\sL}_{n - \nu_\sL/24}$, we define  $\Index_{n-\nu_\sL/24}$ to be the number of states annihilated by $Q$ with $(-1)^F=+1$ minus the number of states annihilated by $Q$ with $(-1)^F=-1$. This is the APS index for $\cH^{\sL}_{n - \nu_\sL/24}$. We define
\beq
I_\sL = \kappa  \sum_{n \in \bZ} q^{n - \nu_\sL/24} \Index_{n-\nu_\sL/24}, \qquad q=e^{2\pi \i \tau}, \label{eq:2dAPS}
\eeq
where $\tau$ is a complex variable with $\Im \tau>0$, $q=e^{2\pi \i \tau}$, and $\kappa$ is defined by
\beq
\kappa = \left\{ \begin{array}{lcl} 
1, & &\nu_\sL  \equiv 0 \mod 8, \\
  \frac{1}{2}, &&  \nu_\sL  \equiv 4 \mod 8.
\end{array} \right. \label{eq:kappadef}
\eeq
The reason for introducing $\kappa$ is as follows. When $  \nu_\sL  \equiv 4 \mod 8$, the CPT symmetry in two dimensions, denoted by $T$ in \eqref{eq:time1}, satisfies $T^2=-1$ and $[T, Q]=[T,(-1)^F]=0$. 
Then all states appear in pairs related by $T$; this is the Kramers degeneracy when $T^2=-1$. 
In particular all the indices $ \Index_{n-\nu_\sL/24}$ are even, i.e.~$ \Index_{n-\nu_\sL/24} \in 2\bZ$. 
Thus we multiply them by $\frac12$ to get integers. 
For $\nu_\sL = 2 \mod 4$, the index is automatically zero because \eqref{eq:time1} implies $\{T, (-1)^F\}=0$ and hence states with $(-1)^F = \pm 1$ appear in pairs.\footnote{In more general cases in which the boundary $\sN$ has zero modes, the boundary condition breaks symmetry between $(-1)^F=1$ and $(-1)^F=-1$ and hence the APS index is not necessarily zero. See footnote~\ref{footnote:generalAPS}.} When $\nu_\sL$ is odd, $(-1)^F$ is ill-defined. 

We call the quantity $I_{\sL}$ given in \eqref{eq:2dAPS} the APS index for the whole theory $\sL$.
This is defined as a Laurent series in $q$, up to the overall factor $q^{-\nu_\sL/24}$. 

\paragraph{Gluing law of APS indices.}

Suppose that $\sL$ and $\sL'$ have the same boundary, $\partial \sL = \partial \sL' = \sN$. We can glue them and get a new theory, $\sL'' = \sL' \cup \overline{\sL}$. The APS index has the gluing law
\beq
I_{\sL''} = I_{\sL'}- I_\sL . \label{eq:APSgluinglaw}
\eeq
This follows from the same argument as in the case of manifolds. Zero modes are localized in the interior region of $ \sL$ and $ \sL'$. We always assume that $ \sL$ and $ \sL'$ are separated by a large distance when they are glued (similarly to Figure~\ref{fig:6} with $a \gg 1$).
The APS index is robust under small perturbations, so the gluing law should hold. 

\paragraph{Partition function.}

If $\sL$ is a compact theory, $I_\sL$ can also be computed as
\beq
I_\sL = \kappa \Tr_{\sL} (-1)^F q^{H_L} \bar q^{H_R} \quad \text{if $\sL$ is compact}, \label{eq:EG}
\eeq
where $q=e^{2\pi \i \tau}$.
For compact $\sL$, this is the standard elliptic genus of $\sL$ introduced in \cite{Witten:1986bf}, up to the factor $\kappa$. 
It then only depends on $\tau$, i.e., $\partial_{\bar \tau} I_\sL=0$ due to cancellation 
between bosonic states with $(-1)^F=+1$ and fermionic states with $(-1)^F=-1$ when $Q^2 \neq 0$. 
Notice that for states with $H_R=Q^2=0$, we have $H_L=P$. 
Notice also that when the right hand side is computed by the path integral on a torus $T^2$,
the variable $\tau$ is the complex modulus of $T^2$.

When $\sL$ is noncompact, the formula \eqref{eq:EG} is not valid, due to various reasons.
This is the subject of the next subsection.

\subsection{Partition functions of noncompact SQFTs}\label{sec:pt}

Let us consider an SQFT $\sL$ with boundary $\partial \sL = \sN$ such that the gravitational anomaly $\nu_\sL$ is even. As mentioned in the previous subsection, an SQFT with a corner can also be regarded as an SQFT with a boundary and hence the following discussions are expected to apply.

The gravitational anomaly $\nu_\sN$ of the boundary $\sN=\partial \sL$ is odd, $\nu_\sN=\nu_\sL-1$. We use the convention that the Hilbert space of $\sN$ is small and $(-1)^F$ is not defined.\footnote{More concretely, suppose we have a single Majorana zero mode. The partition function $\Tr e^{-\beta H}$ of the single Majorana zero mode is $\sqrt{2}$, 
since two Majorana fermion zero modes can be realized by Pauli matrices and has the partition function $2$.
Therefore  ``the Hilbert space dimension'' of a single Majorana fermion zero mode is $\sqrt{2}$. 
The actual Hilbert space dimension is either 1 or 2 depending on one's definition. We use the convention that the dimension is 1. \label{footnote1} }

Now let us consider the partition function
\beq
Z_\sL = \text{``~}  \Tr_{\sL} (-1)^F q^{H_L} \bar q^{H_R} \text{~''} , \quad (q=e^{2\pi \i \tau}).
\eeq
However, for noncompact theories, the Hamiltonian has  continuous spectrum and it is not immediately clear whether $Z_\sL$ can be defined. 
If there is no insertion of $(-1)^F$ inside, the trace $\Tr q^{H_L} \bar q^{H_R}$ is divergent because there are infinitely many states with finite $H_L$ and $H_R$ in the Hilbert space. 
When we insert $(-1)^F$, there is huge cancellation among bosonic states with $(-1)^F=+1$ and fermionic states with $(-1)^F=-1$ related by supersymmetry $Q$. 
This cancellation was the reason that the right hand side of \eqref{eq:EG} was independent of $\bar \tau$ for compact $\sL$. 
Because of this cancellation, we may hope that $Z_\sL$ is defined even if $\sL$ is noncompact. 

For noncompact $\sL$, we claim the following three properties of $Z_\sL$ listed below,
which were originally conjectured in  \cite{Gaiotto:2019gef} and 
proved for sigma models in \cite{Dabholkar:2020fde}.
We give a full derivation in Appendix~\ref{sec:A}.
\begin{enumerate}
\item $Z_\sL$ is defined and has the expected behavior under modular transformations (i.e., $\SL(2,\bZ)$ transformations) of $(\tau, \bar\tau)$.\footnote{%
More details about $\SL(2,\bZ)$ transformations will be discussed below \eqref{eq:Ztransf1}.} 
\item By regarding $\tau$ and $\bar \tau$ as independent variables and taking $\bar \tau \to -\i  \infty$ while $\tau$ is fixed, the quantity $Z_\sL$ becomes the APS index $I_\sL$ 
in the following manner:
\beq
I_\sL(\tau) =\kappa \lim_{\substack{\bar\tau \to -\i  \infty, \\ \tau\text{:\,fixed}}} Z_\sL(\tau, \bar\tau), \label{eq:ZlimitI}
\eeq
if there are no zero modes on the boundary $\sN=\partial \sL$.\footnote{The formula with boundary zero modes will be given in \eqref{eq:correctiontoAPS}.  }
\item The derivative of $Z_\sL$ with respect to $\bar \tau$ is given by
\beq
\frac{\partial Z_\sL}{\partial \bar\tau} =\frac{\i}{2 (\Im \tau)^{1/2} \eta(\tau)}  \Tr_\sN  q^{H_L} \bar q^{H_R} Q_{\sN} , \label{eq:holanomeq}
\eeq
where the trace on the right hand side is taken in the Hilbert space of the boundary theory $\sN$, and $\eta(\tau)=q^{1/24}\prod_{n \geq 1}(1-q^n)$ is the Dedekind $\eta$-function.
\end{enumerate}
Let us comment on each of these points. 

About the first point, even though ``the trace in the Hilbert space'' is difficult to define directly, we may compute it by using the path integral on $T^2$, at least when $\tau$ and $\bar\tau$ are complex conjugates of each other. Then the claim is that the path integral gives a finite answer. 

About the second point, due to noncompactness of $\sL$, the cancellation among bosonic and fermionic states are not perfect and $Z_\sL$ may depend on $\bar\tau$. However, if we take the limit $\bar \tau \to -\i  \infty$, the states contributing to the trace are restricted to $H_R=Q^2=0$. Thus the partition function is expected to coincide with the APS index $I_\sL$ in this limit.

About the third point, this is the most nontrivial claim,
and is the holomorphic anomaly equation, conjectured in \cite{Gaiotto:2019gef} and proved for sigma models in \cite{Dabholkar:2020fde}. 
We give a general derivation in Appendix~\ref{sec:A}.


One of the important consequences of the above properties for the purposes of the present paper is as follows. Suppose that $\sN$ satisfies the condition that
\beq
 \Tr_{\sN} Q q^{H_L} \bar q^{H_R}=0. \label{eq:vanishingQ}
\eeq
In this case, $Z_{\sL}$ is independent of $\bar \tau$ by the third property, and then the second property implies 
\beq
I_\sL = \kappa Z_\sL \quad \text{under the condition \eqref{eq:vanishingQ}} , \label{eq:APSpartition}
\eeq
where $\kappa$ is defined in \eqref{eq:kappadef}.
Then the first property implies that $I_\sL$ has the expected behavior under modular transformations, just as if $\sL$ were compact.

A class of theories satisfying the condition \eqref{eq:vanishingQ} is as follows; this class plays an important role later. Suppose that $\sN$ is a product of two compact theories $\sA$ and $\sB$ with gravitational anomalies $\nu_\sA$ and $\nu_\sB$, respectively; $\sN = \sA \times \sB$. 
As we assume $\nu_\sN$ is odd, one of  $\nu_\sA$ or $\nu_\sB$ is even and the other is odd. 
Without loss of generality, we can assume that $\nu_\sA$ is even and $\nu_\sB$ is odd. 
In the product $\sA \times \sB$, the supercharge $Q$ is given in terms of the supercharges $Q_\sA$ and $Q_\sB$ of $\sA$ and $\sB$ as
\beq
Q = Q_{\sA} \otimes 1 + (-1)^F \otimes Q_{\sB},
\eeq
where the factor $(-1)^F$ in the second term is the usual rule for operators of odd fermion parity.\footnote{In general, if $\psi_\sA$ and $\psi_\sB$ are operators of $\sA$ and $\sB$ with odd fermion parity $\{ (-1)^F, \psi_{\sA,\sB}\}=0$, then they are embedded in $\sA \times \sB$ as $\psi_\sA \otimes 1$ and $(-1)^F \otimes \psi_B$. This rule is necessary since we want these two operators to anticommute with each other. 
The same $(-1)^F$ appears in the Jordan-Wigner transformations because of the same reason.}
 Then we compute
\beq
 &\Tr_{\sN} Q q^{H_L} \bar q^{H_R} \nonumber \\
& = (  \Tr_{\sA}Q_{\sA}  q^{H_L} \bar q^{H_R} ) ( \Tr_{\sB}  q^{H_L} \bar q^{H_R} )
+ ( \Tr_{\sA}  (-1)^F  q^{H_L} \bar q^{H_R} )(  \Tr_{\sB}  Q_{\sB}  q^{H_L} \bar q^{H_R} ). 
\eeq
The first term is zero because $(-1)^F$ is well-defined in $\sA$ and we are computing the one-point function of the operator $Q_\sA$ which is odd under $(-1)^F$. (For the theory $\sB$, the fermion parity $(-1)^F$ is not defined since $\nu_\sB$ is odd, and hence the one-point function of $Q_\sB$ need not vanish.) 
Therefore, we get
\beq
 \Tr_{\sN} Q  q^{H_L} \bar q^{H_R}   =  ( \Tr_{\sA}  (-1)^F   q^{H_L} \bar q^{H_R}   )(  \Tr_{\sB}  Q_{\sB}  q^{H_L} \bar q^{H_R} ).\label{eq:prodv}
\eeq
This vanishes in either of the following two cases. One case is that the elliptic genus of $\sA$ is zero, 
\beq
\Tr_\sA (-1)^F  q^{H_L} \bar q^{H_R} =0. \label{eq:Avanish}
\eeq
Another case is that the gravitational anomaly $\nu_\sB$ of $\sB$ is of the form
\beq
\nu_\sB \in 1+ 4\bZ.   \label{eq:Bvanish}
\eeq
In this case, the CPT symmetry, denoted by $T$ in \eqref{eq:time2}, is such that it anticommutes with the supercharge $Q_\sB$, i.e., $\{T, Q_\sB\}=0$.
Then, positive and negative eigenvalues of the supercharge $Q_\sB$ appear in pairs, and there is exact cancellation. This implies $ \Tr_{\sB}  Q_\sB  q^{H_L} \bar q^{H_R}=0$.\footnote{%
This is analogous to the fact that the APS $\eta$-invariant for pure gravitational backgrounds is identically zero for $d \in 1+4\bZ$ if there is no zero mode. For the case of manifolds, there is a kind of charge conjugation symmetry for the Dirac operator which plays the role of the time reversal symmetry for quantum mechanics. } 
We conclude that  the relation \eqref{eq:vanishingQ} holds under the condition of either \eqref{eq:Avanish} or \eqref{eq:Bvanish}.

Let us explain in a little more detail about the $\SL(2,\bZ)$ transformation properties
of $Z_\sL$. 
We claim that $Z_\sL $ satisfies
\beq
Z_\sL (\tau+1, \bar\tau+1) &= e^{- 2\pi \i \nu/24} Z_\sL (\tau, \bar\tau) , \label{eq:Ztransf1} \\
Z_\sL ( -1/\tau, -1/\tau) &=  e^{-2\pi \i \nu/8} Z_\sL (\tau, \bar\tau) \quad \text{if} \quad |\tau|=1. \label{eq:Ztransf}
\eeq
The first equation \eqref{eq:Ztransf1} is an immediate consequence from \eqref{eq:Peigenvalue}. 
The second equation \eqref{eq:Ztransf} follows from the fact that in a homomorphism $\sigma : \SL(2,\bZ)\to \U(1)$, the transformation $T \in \SL(2,\bZ)$ and the transformation $S \in \SL(2,\bZ)$ are related by $\sigma(S) = \sigma(T)^3$ because of the relation $(ST)^3$ and $S^4=1$. 
The restriction $|\tau|=1$ in the second equation is by the following reason. We are considering SQFTs which do not necessarily have conformal invariance, and hence the partition function depends on the area of the torus $T^2$ as well as the complex modulus $\tau$. We have fixed the length of the $S^1$ to be $2\pi$, so the area of the torus is $
(2\pi)^2 \Im \tau$. 
The $\SL(2,\bZ)$ transformation is a symmetry (modulo gravitational anomalies) as far as the area of the torus $T^2$ is fixed. 
When $|\tau|=1$, one can check that the area is preserved under the transformation $S \in \SL(2,\bZ)$.

The Dedekind $\eta$-function satisfies $\eta(\tau+1) = e^{2\pi \i  /24} \eta(\tau) $ and $\eta(-1/\tau) = \sqrt{-\i \tau} \eta(\tau)$. Then, $\eta(\tau)^{\nu}Z_\sL$ has the same transformation law (on $|\tau|=1$ for $S$) as modular forms of weight $\nu/2$.

For the purpose of later use, we record the following crucial fact which follows from \eqref{eq:APSpartition}. Consider the case that $\nu_\sL=-20$. 
The modular transformation property of $Z_\sL$ and the equality \eqref{eq:APSpartition} imply that $F(\tau)=\eta(\tau)^4 I_\sL(\tau)$ is a modular form of weight 2,
\beq
F(\tau+1) = F(\tau), \quad F(-1/\tau) = \tau^2 F(\tau) \quad \text{where $F(\tau)=\eta(\tau)^4 I_\sL(\tau)$},
\eeq
where the transformation under $\tau \to -1/\tau$ follows from analytic continuation from that on $|\tau|=1$.
Let us expand $F(\tau)$ as
\beq
F(\tau) = \sum_{n \in \bZ} b_n q^n, \qquad (b_n \in \bZ).
\eeq
For a modular form of weight 2, it is known that\footnote{For a modular form $F(\tau)$ of weight 2, the 1-form $F(\tau) \d \tau$ is modular invariant. One has $b_0 = \int_{-1/2+ \i \infty}^{1/2+ \i \infty} F(\tau) \d \tau$. By deforming the integration contour to the boundary of the fundamental region of $\tau$, one can show that this integral vanishes. We will explain it in a more general context in Section~\ref{sec:differential}. } 
\beq
\left[ \eta(\tau)^4 I_{ \sL}(\tau)  \right]_{q^0}  : = b_0=0, \label{eq:weight2vanish}
\eeq
where $[ \bullet ]_{q^0}$ means that we take the coefficient of $q^0$. 
The fact that the $q^0$ term of a weight 2 modular form vanishes have played an important role in the proof of the absence of perturbative anomalies of general heterotic string theories~\cite{Schellekens:1986xh,Lerche:1987qk,Lerche:1988np}. We will use this fact for noncompact theories $\sL$ whose boundary $\sN = \partial \sL$ satisfies the condition \eqref{eq:vanishingQ}.

\subsection{Primary and secondary invariants of SQFTs}\label{sec:primarysecondary}

We want to consider invariants of SQFTs. 
For that, we first need to discuss what we mean by invariance.
For this purpose, let us define $\SQFT_\nu$ as the `set of compact SQFTs' whose gravitational anomaly is specified by $\nu$.\footnote{%
$ \SQFT_\nu$ is actually not precisely ``the set of all SQFTs'' because there is no such thing, in the same way that there is no such thing as ``the set of all sets''. 
It is also problematic to define $ \SQFT_\nu$ to be ``the set of isomorphism classes of SQFTs'', such as the space of coupling constants, because $\sM \in \SQFT_\nu$ may have automorphisms which can be used for nontrivial gluing such as in Figure~\ref{fig:6}, but the information of automorphisms is lost once we take ``isomorphism classes''. 
The situation is analogous to the bordism category of manifolds. 
For the purposes of the present paper, it is enough to regard $\SQFT_\nu$ as ``the class of objects in the bordism category of SQFTs''.
}

We introduce the following equivalence relation in the set $\SQFT_\nu$. Two SQFTs $ \sA, \sB \in \SQFT_\nu$ are equivalent, $\sA \sim \sB$, if there exists a noncompact theory $\sC$ with gravitational anomaly $\nu_\sC = \nu+1$ such that
\beq
\sA \sqcup \overline{\sB}  = \partial \sC.
\eeq
We denote the equivalence class of theories under the equivalence relation as $[\sA]$. We also denote the set of equivalence classes of theories as $\SQFT_\nu({\rm pt})$.\footnote{%
This notation is motivated by some relation with algebraic topology.
There, an $\Omega$-spectrum is a sequence of spaces $E_n$ such that 
$E_n \simeq \Omega E_{n+1}$, from which 
we can define a generalized cohomology theory $E^n(X):=[X,E_n]$,
and a corresponding generalized homology theory $E_n(X)$ by an analogous formula.
In particular, $E^n(\pt)=\pi_0 E_n=E_{-n}(\pt)$.
Here we are using a different convention where $\pi_0 E_\nu = E_{+\nu}(\pt)$ 
without the minus sign.

More importantly, in this paper, we avoid assuming that $\SQFT_\nu$ forms a spectrum. 
To regard $\SQFT_\nu$ as a spectrum, we need to make it a sequence of topological spaces rather than a sequence of classes of objects in a category.
It is not yet clear what is the proper  definition of points of $ \SQFT_\nu$, and how to introduce an appropriate topology. 
We need a concrete model, analogous to ``the space of Fredholm operators in a (fixed) Hilbert space'' for SQMs.
 Once an appropriate model of ``the space of SQFTs'' is found, it is likely that they form a spectrum satisfying $\SQFT_n \simeq \Omega \SQFT_{n-1}$,
 along the lines of arguments in \cite{Johnson-Freyd:2020itv}.
\label{footnote:spectrum}
}

One can check that $\SQFT_\nu({\rm pt})$ has the structure of an abelian group. For instance, the sum is given by $[\sA] + [\sB]=[\sA \sqcup \sB]$, the unit $0 \in \SQFT_\nu({\rm pt})$ is given by $0=[\partial \sC]$ for any noncompact theory $\sC$, and the inverse $-[\sA] \in \SQFT_\nu({\rm pt})$ is given by $[\overline{\sA}]$. 
Not only that, $\bigoplus_\nu \SQFT_\nu(\pt)$ has the structure of a graded ring. The product is given by $[\sA][\sB] = [\sA \times \sB]$. 

Suppose we have some function $F(\sA)$ of $\sA \in \SQFT_\nu$. We say that $F$ is a bordism invariant if $F$ only depends on the equivalence class $[\sA]$, i.e. $F(A) = \widetilde F ([A])$ for some $\widetilde F$. 

The above  concept of bordism for SQFTs is a natural generalization of the one for manifolds. In string theory context, it is intermediate between bordism of manifolds and that of full quantum gravity. The bordism of SQFTs includes full effects of $\alpha'$, but nonperturbative effects of string coupling are missing. The bordism group of full quantum gravity is conjectured to be trivial~\cite{McNamara:2019rup}. For more discussions and some applications, see~\cite{Kaidi:2024cbx}.

\subsubsection{Primary invariants}
The elliptic genus $I_\sA$ is an important example of bordism invariants. The reason is that the elliptic genus is invariant under continuous deformations of a theory (as is generally true for Witten index), and when a theory $\sA$ is realized as the boundary of some $\sC$, then it can be deformed to another theory in which supersymmetry is spontaneously broken, as discussed around \eqref{eq:superpot}. Thus we get $I_\sA=0$ if $\sA = \partial \sC$.

In addition to the elliptic genus, there is also the ``mod-2 elliptic genus'' studied in \cite{Tachikawa:2023nne}. The elliptic genus can be nonzero for $\nu \equiv 0, 4 \mod 8$, while the mod-2 elliptic genus can be nonzero for $\nu \equiv 1, 2 \mod 8$. 

For $\nu \equiv 1 \mod 8$, we count the number of zero modes (in the sense of the supercharge $Q$) modulo 2,
at each eigenspace of $P$, the momentum along the spatial $S^1$. This is invariant under continuous deformations by the following reason. The algebra \eqref{eq:time2} in this case is $TQ = - QT$, and it implies that nonzero modes of $Q$ always appear in pairs: if a state $\ket{+}$ has eigenvalue $Q\ket{+} =\lambda \ket{+}$ with $\lambda \neq 0$, then the state $\ket{-}:=T\ket{+}$ has eigenvalue $Q\ket{-} = - \lambda \ket{-}$. The two states $\ket{+}$ and $\ket{-}$ are different because $\lambda \neq -\lambda$ for $\lambda \neq 0$. Then, even if $\lambda$ accidentally goes to zero under continuous deformations, the number of zero modes modulo 2 does not change. 

For $\nu \equiv 2 \mod 8$, the algebra \eqref{eq:time1} has $T(-1)^F=- (-1)^F T$. Then all states are degenerate, and the number of zero modes is always even. 
One can again check by using the algebra \eqref{eq:time1} that the number of nonzero modes are multiples of 4, and hence the number of zero modes modulo $4$ is invariant. In this case, it is convenient to divide the number by 2, so the mod-2 index at each eigenspace of $P$ is either $0$ or $1$ modulo 2. 
 
We denote the mod-2 elliptic genus as $I^{\rm mod\,2}_\sA$. It is of the form
\beq
I^{\rm mod\,2}_\sA = \sum_{n \in \bZ} q^{n-\nu_\sA/24}  \Index^{\rm mod\,2}_{n-\nu_\sA/24} 
\eeq
where $ \Index^{\rm mod\,2}_{n-\nu_\sA/24} \in \bZ_2$, and $q$ is regarded as a formal variable rather than a complex number.

The usual and mod-2 elliptic genera are KO-theoretic quantities of the Hilbert space of the theory; for this viewpoint, see \cite{Tachikawa:2023nne} and its Appendix in particular.
They may be regarded as primary invariants of SQFTs. 

\subsubsection{Secondary invariant} \label{app:BN}
There is also another, more subtle invariant introduced mathematically in the context of TMF by Bunke and Naumann \cite{Bunke} and studied in the physics context of SQFTs in \cite{Gaiotto:2019gef,Yonekura:2022reu}. 

It is possible to define it for general $\nu$ (at least in physics),\footnote{Results in mathematics suggest that the invariant will be interesting only when $\nu \in 3+24\bZ$. For example, the invariant $I^\text{2nd}_\sA $ defined in \eqref{eq:2ndary} is just a reduction of the mod-2 elliptic genus $I^{\rm mod\,2}_\sA$ when $\nu=1,2 \mod 8$. In such a case,  $I^\text{2nd}_\sA $ may not give new information. This is analogous to the fact that the APS $\eta$-invariant for purely gravitational backgrounds is reduced to the mod-2 index in dimensions $d=1,2 \mod 8$.} 
but in the present paper we only use it in the case
\beq
\nu = 4m-1 . \label{eq:secondarydim}
\eeq
In this case, $I_\sA$ and $I^{\rm mod\,2}_\sA$ are automatically zero.
A conjecture about topological modular forms (mentioned later) suggests that any element $[\sA] \in \SQFT_{4m-1}({\rm pt})$ is torsion. 
Assuming it, there is a positive integer $k$ and a theory $\sB$ whose boundary is given by
\beq
\partial\sB = \sA^{\sqcup k}: = \underbrace{\sA \sqcup \cdots \sqcup \sA}_\text{$k$ times} .
\eeq

In general, for any abelian group $\bA$, we denote by $\bA((q))$ the set of formal Laurent series with coefficients in $\bA$ (i.e., elements of the form $\sum_{n =-p}^\infty a_n q^n$ for some $p \in \bZ$ and $a_n \in \bA$).
We also denote by $\MF_{2m}$ the set of (weakly holomorphic\footnote{``Weakly holomorphic'' means that the only possible pole is at $q=0$.}) modular forms of weight $2m$,
\beq
\MF_{2m}=\bigl\{f(\tau) \mid  f(-1/\tau)= \tau^{2m} f(\tau),~~ f(\tau)=\sum a_n q^n~(a_n \in \bZ)\bigr\}
\eeq
We then let \begin{equation}
I^\text{2nd}_\sA := \frac 1k [I_\sB] \in \eta(\tau)^{-4m} \frac{\bQ((q))}{ \bZ((q)) +  \MF_{2m}\otimes \bQ} \label{eq:2ndary}
\end{equation}
where $I_{\sB}$ is the APS index of the non-compact theory $\sB$ whose gravitational anomaly is $\nu_{\sB} = 4m$.

If we use the spectral invariant which will be introduced in Sec.~\ref{sec:differential}, which is an analog of the APS $\eta$-invariant, we can give a more direct definition of $I^\text{2nd}_\sA$ without taking $\sB$. This point will be explained in Sec.~\ref{sec:secondrevisit}.

Let us discuss three properties of $I^\text{2nd}_\sA$ for $\sA\in \SQFT_{4m-1}$
which we will need later.
\begin{property}
\label{prop:1}
If $I^\text{\upshape 2nd}_\sA$ vanishes, then we can take an appropriate integer $k$  and $\sB$ such that
$\partial \sB=\sA^{\sqcup k}$ and $\frac 1k I_\sB(\tau) \in \eta(\tau)^{-4m} \bZ((q))$.
\end{property}
\noindent 
To see this, let us first take a $\sB'$ such that $\partial \sB' = \sA^{\sqcup k'}$.
That $I^\text{2nd}_\sA$ vanishes means that \begin{equation}
\frac{1}{k'} \eta (\tau)^{4m} I_{\sB'} =  f(\tau)   \mod \bZ((q)),
\end{equation} where $f(\tau) \in \MF_{2m}\otimes \bQ $. 
Although $\SQFT_{4m}({\rm pt}) \to \MF_{2m}$ is not a surjection, it is a surjection after $\otimes \bQ$ (see Sec.~\ref{sec:const}).
This means that there is a compact SQFT $\sC$ such that $f(\tau) = \frac{1}{\ell} \eta (\tau)^{4m} I_{\sC}(\tau) $, for some $\ell$.
We now take $\sB:=(\sB')^{\sqcup \ell} - \sC^{\sqcup k'}$, and $k:= k'\ell$, and we are done.

\begin{property}
\label{prop:2}
Assume $\SQFT_{-21}({\rm pt})=0$. 
Then, for any $\sA$ and $\sB$ of $\nu_\sA=4m-1$ and $\nu_{\sB}=-20-4m$ respectively, $I^\text{\upshape 2nd}_{\sA\times \sB}=0$.
\end{property} 
\noindent 
Indeed, if it is nonzero, it implies $\SQFT_{-21}({\rm pt})\neq 0$.

\begin{property}
\label{prop:3}
Let 
\beq
I : \SQFT_{-20-4m}({\rm pt}) \to \eta(\tau)^{20+4m} \cdot \MF_{(-20-4m)/2} 
\eeq
be given by $I([\sB]):=I_\sB$. 
Then there is a well-defined pairing  \begin{equation}
\frac{ \eta(\tau)^{20+4m} \cdot  \MF_{(-20-4m)/2}}{    I (\SQFT_{-20-4m}({\rm pt})  ) } \times   \SQFT_{4m-1}({\rm pt}) \to \bQ/\bZ
\end{equation} 
given by \begin{equation}
(f,[\sA]) \mapsto [\eta(\tau)^4  \cdot f(\tau)  \cdot  I_\sA^\text{\upshape 2nd} ]_{q^0} \mod \bZ.
\end{equation}
\end{property} 
\noindent
Let us check that it does not depend on the choices of representatives of $f(q)$ and $ I_\sA^\text{2nd}$.  
First, suppose that we shift $ I_\sA^\text{2nd}$ by an element $g(\tau) \in  \eta(\tau)^{-4m} \cdot  \MF_{2m} \otimes \bQ$. Then, $\eta(\tau)^4 f(\tau) g(\tau)$ is a modular form of weight 2. As mentioned in \eqref{eq:weight2vanish}, the coefficient of $q^0$ in a weight 2 modular form is zero. Next, suppose that we shift $ I_\sA^\text{2nd}$ by an element $h(\tau) \in   \eta(\tau)^{-4m} \cdot \bZ((q))$. Then the coefficients of $\eta(\tau)^4 f(\tau) h(\tau)$ are integers and hence they are zero modulo $\bZ$. Finally, suppose we shift $f(\tau)$ by $I_\sB$ for an arbitrary $\sB$. Then we have $I_\sB  I_\sA^\text{2nd} =  I_{\sA \times \sB}^\text{2nd}$, which is zero by the second property mentioned above. 

This allows us to constrain $I(\SQFT_{-20-4m}({\rm pt}) )$ if we know $\SQFT_{4m-1}$, or vice versa.
For example, consider the sigma model whose target space is a round $S^3$ with unit 3-form flux $\int_{S^3}H=1$. We denote this theory as $S^3_{H=1}$.
The pairing of $f(\tau)=1 \in  \eta(\tau)^{24} \cdot \MF_{(-24)/2} (=\MF_0) $ and $\sA= S^3_{H=1}$  is $1/24$,
implying that if $a   \in I(\SQFT_{-24}({\rm pt}) )$ then $a$ is a multiple of $24$.
This consideration will be generalized extensively in Sec.~\ref{sec:const}.

\section{The new invariant}\label{sec:new}

In this section, we are going to construct a new invariant of SQFTs. On manifolds, we started from geometric and spectral quantities and then derived a topological formula \eqref{eq:topoGS}. 
In this section, we will define the invariant by a topological formula analogous to \eqref{eq:topoGS}, and then check that it is independent of various choices. 
We will come back to the construction of the spectral invariants in Sec.~\ref{sec:differential}.

\subsection{Definition of the new invariant}\label{sec:def}
In $\SQFT_\nu({\rm pt})$, we define a subgroup $\bA_\nu$ by 
\beq
\bA_\nu = \{[\sA] \in \SQFT_\nu({\rm pt})~|~ I_\sA=0,  I^{\rm mod\,2}_\sA=0, I^\text{2nd}_{\sA}=0  \}. \label{eq:nontrivialgroup}
\eeq
This is the subgroup in which all the primary and secondary invariants are zero.

We would like to define an invariant that can detect this subgroup.
The invariant we are going to define is a bilinear form $\GS(\bullet, \bullet)$,
\beq
\bA_d \times \bA_{-22-d} \ni ([\sX], [\sT]) \mapsto \GS([\sX], [\sT]) \in \bQ/\bZ. 
\eeq

The number $-22$ can be motivated by the worldsheet construction of heterotic string theories. It is the correct gravitational anomaly for the worldsheet theories, without the ghost contributions. For instance, in ten-dimensional supersymmetric heterotic string theories, $-22=-32+10$ where $-32$ comes from the left-moving current algebra and $10$ comes from the ten-dimensional target space. 
More generally, the non-ghost part of the heterotic string worldsheet has to cancel the gravitational anomaly of the $bc$ and $\beta\gamma$ ghost systems, which have gravitational anomaly $+22$.

Our construction of the invariant $\cW$ depends on the following conjectures:
\begin{enumerate}
\item $\SQFT_{-21}({\rm pt}) =0$.
\item $\bA_\nu$ are torsion groups, i.e., for any  $[\sA] \in \bA_\nu$ there exists an integer $k$ such that $k[\sA]=0$.
\item For $ ([\sX], [\sT]) \in \bA_d \times \bA_{-22-d}$, the product vanishes, $[\sX][\sT] =0 \in \SQFT_{-22}({\rm pt}) $.
\end{enumerate}
These are suggested by the theory of topological modular forms (TMF) and the Stolz-Teichner conjecture~\cite{Stolz:2004,Stolz:2011zj}.

The first conjecture $\SQFT_{-21}({\rm pt}) =0$ is suggested by the fact known in topological modular forms, ${\rm TMF}_{-21}({\rm pt}) =0$, as pointed out in \cite{Tachikawa:2021mby}. It plays an important role in the argument that general heterotic string theories do not have global anomalies~\cite{Tachikawa:2021mby,Yonekura:2022reu}. 

The second conjecture comes from the fact that the corresponding subgroup $\bA^\TMF_\nu$ of $\TMF_{\nu}(\pt)$, defined by vanishing of primary and secondary invariants, is indeed torsion \cite{Tachikawa:2023lwf}.\footnote{%
We warn the reader that, in \cite{Tachikawa:2023lwf},  
the subgroup $A_\nu$ of $\TMF_{\nu}(\pt)$ was defined by the vanishing of primary invariants only.
This group $A_\nu$ was already torsion. As $\bA_\nu^\TMF$ is a further subgroup of $A_\nu$.
that this is also torsion follows.
}

The third conjecture is also supported by the corresponding property of $\TMF$.
By looking at Table~2 and 3 of \cite{Tachikawa:2021mby}, we see that ${\rm TMF}_{-22}({\rm pt}) $ is completely characterized by primary  invariants. 
By the definition of $\bA_{d}$, the primary invariants of $[\sX][\sT] $ are zero. 
Therefore $[\sX][\sT]$ is zero in the case of $\TMF$.

 If the second and third conjectures are not valid in the case of $\SQFT$, we can just take a subset $\bB_{d,-22-d} \subset \bA_d \times \bA_{-22-d}$ in which the statements are valid, and define the invariant $\GS$ only on $\bB_{d,-22-d}$. We can find explicit examples of nontrivial elements of $\bB_{d,-22-d}$, so the second and third conjectures are not crucial for applications to concrete examples, such as Section~7 of \cite{Kaidi:2024cbx}. On the other hand, the first conjecture $\SQFT_{-21}({\rm pt}) =0$ will play a crucial role for the definition and well-definedness of $\GS([\sX], [\sT]) $.

In the following discussions, it may help the reader's understanding to imagine that $\sX$ is a sigma model with a target manifold $X$ of dimension $d$, and $\sT$ is an internal theory of a heterotic string theory. The internal theory $\sT$ produces some massless fermions on $X$, and then we can compare the discussions in this section to that in Section~\ref{sec:GSmanifold}. 
Our definition of $\GS([\sX], [\sT])$ is parallel to the formula \eqref{eq:topoGS}.

Now we define $\cW$. In the definition, we make several choices at intermediate steps. The independence of the final quantity from those choices will be discussed later in Sec.~\ref{sec:well-defined-ness}.

Given $([\sX], [\sT]) \in \bA_d \times \bA_{-22-d}$, we take representatives $\sX $ and $\sT$ of $[\sX]$ and $ [\sT]$, respectively. We denote 
\beq
\sM = \sX \times \sT.
\eeq
The equation $[\sM]=[\sX][\sT] =0$ implies that there exists a noncompact theory $\sN$ such that 
\beq
\partial \sN = \sM. 
\eeq
Also, $k[\sX]=0$ for some integer $k$ implies that there exists a noncompact theory $\sY$ such that the disjoint union of $k$ copies of $\sX$, denoted $\sX^{\sqcup k} = X \sqcup \cdots \sqcup X$, is a boundary of $\sY$, i.e. $\partial \sY = \sX^{\sqcup k} $. 

When $d\in -1 + 4\bZ$, we choose the above $k$ and $\sY$ such that 
\beq
\frac1k  I_{\sY}(\tau) \in \eta(\tau)^{-(d+1)}\bZ((q)) \qquad (\text{when $d\in -1 + 4\bZ$}).  \label{eq:BNcondition}
\eeq
The existence of such $(k, \sY)$ is shown by the assumption that $I^\text{2nd}_{\sX}=0$ and then
using the Property~\ref{prop:1} given in Sec.~\ref{app:BN}.

\begin{figure}
\centering

 \begin{tikzpicture}[scale=2]

  \def\r{1.5}
 
  \draw[thick] (0,0) arc[start angle=180, end angle=270, radius=\r];

  \draw[red, very thick] (0,0) -- (\r , 0); 
 
  \draw[blue, very thick] (\r , 0) -- (\r, -\r);
 
  \node at (\r + 0.3 , -0.5*\r) {$\mathsf{N}_0$};
  \node at (\r/2, -0.5*\r) {$ \mathsf{L}$};
 
  \node at (\r/2, 0.15) {$\mathsf{N}^{\sqcup k}$};
 
\end{tikzpicture}

\caption{The non-compact theory $ {\sL}$ used in our construction of $\GS([\sX],[\sT])$. 
Here, $\partial \sN=\sX\times \sT$, $\sN_0=\sY\times \sT$ and $\partial\sY=\sX^{\sqcup k}$.
\label{fig:3}}
\end{figure}
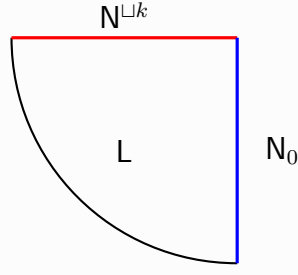

Denoting $\sN_0 = \sY \times \sT$, we have 
\beq
\partial \sN_0 = \sM^{\sqcup k} , \qquad (\sN_0 = \sY \times \sT, \quad \partial \sY = \sX^{\sqcup k}).
\eeq
By gluing $\sN^{\sqcup k}$ and $\overline{\sN_0}$ along $\sM^{\sqcup k}$, we get a compact theory  $\sN^{\sqcup k} \cup \overline{\sN_0}$. Then $\SQFT_{-21}({\rm pt}) =0$ implies that there exists a noncompact theory $ {\sL}$ such that
\beq
\partial  {\sL} = \sN^{\sqcup k} \cup \overline{\sN_0}.
\eeq
It will be convenient to think $ {\sL}$ as a theory with a corner, as in Figure~\ref{fig:3}.

We consider the APS index $I_{  \sL}=I_{  \sL}(\tau)$ of $  \sL$. 
The gravitational anomaly of $  \sL$ is $\nu_{  \sL} = -20$ and hence $\eta(\tau)^4 I_{  \sL}(\tau)$ is a Laurent series in $q$, which we write as
\beq
\eta(\tau)^4 I_{  \sL}(\tau) =  \sum_{ n \in \bZ} a_n q^n \qquad (a_n \in \bZ).
\eeq
We define
\beq
\GS([\sX], [\sT]) =  \frac{1}{k}\left[ \eta(\tau)^4 I_{  \sL}(\tau)  \right]_{q^0} = \frac{a_0}{k}  \in \bQ/\bZ. \label{eq:defofGS}
\eeq
This is the definition of our invariant $\GS([\sX], [\sT]) \in \bQ/\bZ$. The construction here should be compared with that of \eqref{eq:topoGS}.\footnote{The physical meaning of $\eta(\tau)^4$ in \eqref{eq:defofGS} in the context of heterotic string theories is as follows. Suppose that $\sX$ is a sigma model, and suppose also we go to the theory $  \sL$ by extending $\sX$ to a higher-dimensional manifold whose dimension is  $\nu_\sX + 2$. Then, the worldsheet fields corresponding to the additional two dimensions have oscillation modes whose contribution to the partition function is $\eta(\tau)^{-2}$. Thus we need to include a factor of $\eta(\tau)^2$ to cancel them since they are unphysical. Another factor of $\eta(\tau)^2$ comes from the $bc$-ghosts. \label{foot:ghost} }

We remark that the vanishing of ordinary and mod-2  indices in the subgroup $\bA_\nu$ defined in \eqref{eq:nontrivialgroup} suggests that the constituents of the boundary, i.e., $\sN $ and $\sN_0$, do not generically have zero modes. For the sake of simplicity of arguments, in this section we will always assume that various boundaries do not have zero modes. 

We have made several choices in the above definition, so it is not obvious that $\GS([\sX], [\sT]) $ depends only on $[\sX]$ and $[\sT]$. We discuss this issue in the next subsection.

\subsection{Well-definedness of the invariant} 
\label{sec:well-defined-ness}

Here we argue that $\GS([\sX], [\sT]) $ is independent of the choices we made during its definition. 
Recall that the choices made in the definition are the following:
\begin{enumerate}
\item We first took explicit representatives $\sX$ and $\sT$ from the equivalence classes $[\sX]$ and $[\sT]$.
\item We then chose an $\sN$ satisfying $\partial\sN=\sM$.
\item We also chose a $k$ and $\sY$ such that $\partial \sY=\sX^{\sqcup k}$.
\item We then took an $\sL$ such that $\partial \sL=\sN^{\sqcup k} \cup \overline{\sN_0}$,
where $\sN_0=\sY\times \sT$.
\end{enumerate}
We study the dependence on these choices in the following order: 
on the choices of $\sL$ and $\sY$, on the choice of $\sN$, 
on the choice of $k$, 
and on the choices of representatives $\sX$ and $\sT$.

\paragraph{Dependence on $  \sL$ and $\sY$.}

\begin{figure}
\centering

\begin{tikzpicture}[scale=2]
  \def\r{1.5}

  \draw[thick] (0,2*\r -0.7*\r ) arc(90:180:\r);
  \draw[red,very thick] (-\r,\r -0.7*\r ) -- (0,\r -0.7*\r );
  \draw[blue,very thick] (0,\r -0.7*\r ) -- (0,2*\r -0.7*\r );
  \node at (-0.5*\r,1.5*\r -0.7*\r )           {$ \mathsf{L}'$};
  \node at (0.2,   1.5*\r -0.7*\r )           {$\mathsf{N}_0'$};

  \draw[thick] (0,-\r) arc(270:180:\r);
  \draw[red,very thick] (-\r,0) -- (0,0);
  \draw[blue,very thick] (0,0) -- (0,-\r);
  \node at (-0.5*\r, -0.5*\r)        {$\mathsf{ L}$};
  \node at ( -\r/2,    0.2)         {$\mathsf N^{\sqcup k}$};
  \node at ( 0.2,    -0.5*\r)        {$\mathsf N_0$};
\end{tikzpicture}

\caption{The configuration to be used in the proof that $\GS([\sX], [\sT]) $ is independent of $  \sL$ and $\sY$. Here $\sN_0 = \sY \times \sT$. \label{fig:4}}
\end{figure}
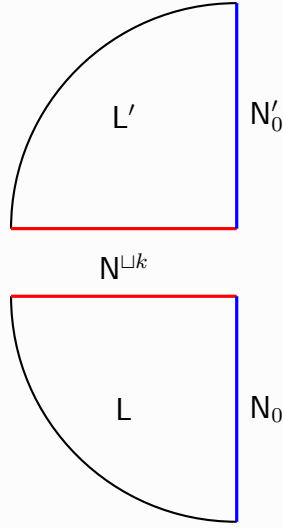

First, we fix $k$ and $\sN$, and show that $\GS([\sX], [\sT]) $ does not depend on the choices of $ {\sL}$ and also of  $\sY$ used in $\sN_0=\sY \times \sT$. 
To see the independence, we take another $\sN'_0= \sY' \times \sT$ and $  \sL'$, and consider the situation as in Figure~\ref{fig:4}. 
We glue $  \sL'$ and the orientation reversal of $  \sL$ along $\sN^{\sqcup k}$ to get a theory $  \sL''$ whose boundary is $\sN'_0 \cup \overline{\sN_0}$.
Let $\sY''= \sY' \cup \overline{\sY}$ be the compact theory obtained by gluing $\sY'$ and $\overline \sY$. Then we get
\beq
\partial (  \sL'') = \sN'_0 \cup \overline{\sN_0} = \sY'' \times \sT \qquad (\sY''= \sY' \cup \overline{\sY}).
\eeq

We will establish the relation
\beq
\frac1k[\eta(\tau)^4 I_{  \sL''}(\tau) ]_{q^0} \equiv 0 \mod \bZ \label{eq:vanishL''}
\eeq
below.
Once this is achieved,
we  can use the gluing law of the APS index
\beq
I_{  \sL''} = I_{  \sL'} - I_{   \sL}  
\eeq
to show 
\beq
\frac1k\left[ \eta(\tau)^4 I_{  \sL'}(\tau)  \right]_{q^0} -\frac1k \left[ \eta(\tau)^4 I_{  \sL}(\tau)  \right]_{q^0} 
=
\frac1k \left[ \eta(\tau)^4 I_{  \sL''}(\tau)  \right]_{q^0}  \equiv 0 \mod \bZ,
\eeq
meaning that $\GS([\sX], [\sT]) $ is independent of the choices of $  \sL$ and $ \sY $.

What remains then is to establish \eqref{eq:vanishL''}.
We do this by considering the three cases $d\in 2\bZ$, $d\in 1+4\bZ$, $d\in -1+4\bZ$ in turn.
The first case and the second case are easier. The third case is more difficult.

First, when $d$ is even, the elliptic genus of $\sT$ is zero by the condition $[\sT] \in \bA_{-22-d}$. 
Then we have \begin{equation}
\Tr_\sT (-1)^F  q^{H_L} \bar q^{H_R} =0.
\label{eq:Tcond1}
\end{equation}
Second, when $d\in 1+4\bZ$, which also implies $-22-d \in 1+4\bZ$, we have 
\beq
\Tr_{\sT}  Q_\sT  q^{H_L} \bar q^{H_R} =0 
\label{eq:Tcond2}
\eeq
because of the algebra $TQ=-QT$ as mentioned in the paragraph containing \eqref{eq:Bvanish}.
In either case, we have shown in the paragraph containing \eqref{eq:prodv} that the product $ \sY'' \times \sT$ satisfies the condition \eqref{eq:vanishingQ}, 
guaranteeing that $I_{  \sL''}$ is independent of $\bar\tau$.
Then, because of the property of modular forms of weight 2 given in \eqref{eq:weight2vanish}, we can conclude \eqref{eq:vanishL''},
even without taking modulo $\bZ$.

The third case when $d\in -1+4\bZ$ is somewhat trickier. 
In this case we choose $k'$ and $\sU$ such that  $\partial \sU=\sT^{\sqcup k'}$
and $\frac{1}{k'}I_{\sU}(q) \in \eta(\tau)^{-(-22-d+1)} \bZ((q))$, using the assumption that $I^\text{2nd}_\sT=0$ and the Property~\ref{prop:1} in Sec.~\ref{app:BN}.
We have \begin{equation}
 \frac 1{k'} \frac{\partial Z_{\sU}}{\partial\bar\tau}
=\frac{\i}{2 (\Im \tau)^{1/2}\eta(\tau)}\Tr_{\sT}  Q_\sT  q^{H_L} \bar q^{H_R} .
\end{equation}
Using this, we have \begin{equation}
\begin{aligned}
\frac1k \frac{\partial}{\partial \bar\tau} Z_{  \sL''}(\tau,\bar \tau)
&=\frac1{k} \frac{\i}{2 (\Im \tau)^{1/2}\eta(\tau)}
\left( \frac1{\kappa_{\sY''}} I_{\sY''} (\tau) \Tr_{\sT}  Q_\sT  q^{H_L} \bar q^{H_R}\right)\\
&=
\frac1k \frac{\partial}{\partial \bar\tau}
\left(\frac1{\kappa_{\sY''}} I_{\sY''} (\tau) \frac1{k'}Z_\sU(\tau,\bar \tau)\right),
\end{aligned}
\end{equation}
where, in general, $\kappa_{\sA}$ is $\kappa$ introduced in \eqref{eq:kappadef} for a theory $\sA$.
This means that  the combination \begin{equation}
\frac1k Z_{  \sL''}(\tau,\bar \tau)- \frac1{k} \frac1{\kappa_{\sY''}} I_{\sY''} (\tau) \frac1{k'}Z_\sU(\tau,\bar \tau)
\end{equation} is independent of $\bar \tau$.
This combination transforms in a natural way under the modular group.
In particular, by taking $\bar\tau \to -\i \infty$ and multiplying with $\kappa_{\sL''}$, 
we find \begin{equation}
\frac1{k} I_{  \sL''}(\tau)-  \frac{\kappa_{\sL''}}{\kappa_{\sY''} \kappa_{\sU}} \frac1{k} I_{\sY''}(\tau) \frac1{k'}  I_\sU(\tau)
=
\frac1{k} I_{  \sL''}(\tau)- \frac1{k} I_{\sY''}(\tau) \frac1{k'}  I_\sU(\tau)
\label{III}
\end{equation} transforms naturally under the modular group,
where we used $\kappa_{\sL''}=\kappa_{\sY''}\kappa_{\sU}=\frac12$.
Therefore, the right hand side of \eqref{III} becomes 
a modular form of weight 2 if we multiply it by $\eta(\tau)^4$,
from which we find that  \begin{equation}
\frac1{k}[\eta(\tau)^4 I_{  \sL''}(\tau) ]_{q^0} 
= [\eta(\tau)^4\frac1{k} I_{\sY''} (\tau)  \frac1{k'} I_{\sU}(\tau) ]_{q^0} .
\end{equation}  
Now, from the gluing property of the APS index \eqref{eq:APSgluinglaw}, we have
\beq
\frac{1}{k} I_{\sY''} (\tau) = \frac1k I_{\sY'}(\tau)-\frac 1kI_{\sY}(\tau),
\eeq
which is in $\eta(\tau)^{-(d+1)}\bZ((q))$ by our condition \eqref{eq:BNcondition} on $\sY$ and $\sY'$.
Similarly, we arranged so that $ \frac1{k'} I_{\sU}(\tau) \in  \eta(\tau)^{-(-22-d+1)} \bZ((q))$.
Combined, we find \eqref{eq:vanishL''}, this time modulo $\bZ$.

\paragraph{Dependence on $\sN$.}

\begin{figure}
\centering

\begin{tikzpicture}[scale=2]
  \def\r{1.5}    
  \def\gap{0.3}  

  \draw[thick] (-\r,0) arc(180:270:\r);
  \draw[red,very thick]    (-\r,0)      -- (0,0);    
  \draw[blue,very thick]   (0,0)        -- (0,-\r); 

  \draw[thick] ({\gap+\r},0) arc(0:-90:\r);
  \draw[red,very thick]    (\gap,0)     -- ({\gap+\r},0);    
  \draw[blue,very thick]   (\gap,0)     -- (\gap,-\r);       

  \node at (-0.5*\r,   -0.5*\r)          {$\mathsf{ L}$};
  \node at ({\gap+0.5*\r}, -0.5*\r)      {$\mathsf{L} '$};

  \node at (-0.5*\r,   0.15)             {$\mathsf{N}^{\sqcup k}$};
  \node at ({\gap+0.5*\r},   0.15)       {$\mathsf{N}'^{\sqcup k}$};

  \node at ({\gap/2},   -0.5*\r)         {$\mathsf{N}_0$};
\end{tikzpicture}

\caption{The configuration to be used in the proof that $\GS([\sX], [\sT]) $ is independent of $\sN $. \label{fig:5}}
\end{figure}
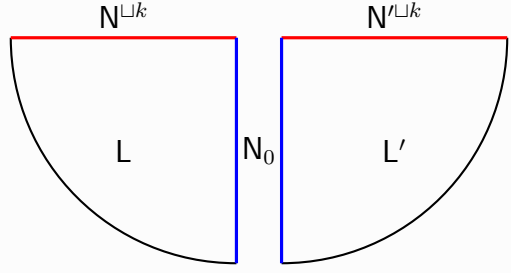

Next we show that $\GS([\sX], [\sT]) $ is independent of the choice of $\sN$ such that $\partial \sN = \sX \times \sT$. To see this, we take another $\sN'$ and $  \sL'$ and consider the situation as in Figure~\ref{fig:5}. We glue $  \sL'$ and the orientation reversal of ${  \sL}$ along $\sN_0$ to get $  \sL''$. Its boundary is given by
\beq
\partial   \sL'' = \sN'^{\sqcup k} \cup \overline{\sN}^{\sqcup k} = (\sN' \cup \overline \sN)^{\sqcup k}.
\eeq
By the gluing law of the APS index, we have
\beq
I_{  \sL''} = I_{  \sL'} - I_{   \sL} .\label{LLL}
\eeq

Now, we take $\sL'''$ such that 
\beq
\partial \sL''' = \sN' \cup \overline{\sN}.
\eeq
Such an $\sL'''$ is guaranteed to exist by $\SQFT_{-21}({\rm pt}) =0$.
Then the disjoint union of $k$ copies of $ \sL'''$, which we denote as $\sL'''^{\sqcup k}$, has the boundary $ (\sN' \cup \overline \sN)^{\sqcup k}$. By gluing $\sL'''^{\sqcup k}$ and $\overline{\sL'' }$ along $ (\sN' \cup \overline \sN)^{\sqcup k}$, the result $\sL'''^{\sqcup k} \cup \overline{\sL''}$ is a compact theory and hence by the property of a modular form of weight 2,
\beq
0=\left[  \eta(\tau)^4 I_{ \sL'''^{\sqcup k} \cup \overline{\sL''} }(\tau) \right]_{q^0} = - \left[ \eta(\tau)^4 I_{  \sL''}(\tau)  \right]_{q^0} + k \left[ \eta(\tau)^4 I_{ \sL'''}(\tau)  \right]_{q^0}.
\eeq
Combined with \eqref{LLL}, we have
\beq
\frac{1}{k} \left[ \eta(\tau)^4 I_{  \sL'}(\tau)  \right]_{q^0} - \frac{1}{k} \left[ \eta(\tau)^4 I_{  \sL}(\tau)  \right]_{q^0} = \left[ \eta(\tau)^4 I_{ \sL'''}(\tau)  \right]_{q^0} \in \bZ.
\eeq
This shows that the value of $\GS([\sX], [\sT]) $ in $\bQ/\bZ$ 
 is independent of the choice of $\sN$. We remark that the fact $ \left[ \eta(\tau)^4 I_{ \sL'''}(\tau)  \right]_{q^0} \in \bZ$ is essentially the global anomaly cancellation condition~\cite{Yonekura:2022reu}.

 \paragraph{Dependence on $ k$ and the linearity.}
We now  move on to the dependence on $k$.
Suppose we chose $k_1$, $\sY_1$ and $\sL_1$ on one hand
and $k_2$, $\sY_2$ and $\sL_2$ on the other.
We now consider choosing $k':=k_1 k_2$ instead.
In this case, we can just take $\sY'_1:=\sY_1^{\sqcup k_2}$ and $\sL'_1:= {\sL_1}^{\sqcup k_2}$ in the above construction and then we immediately have
\beq
\frac{1}{k_1}   I_{  \sL_1}(\tau) = \frac{1}{k_1 k_2}   I_{  \sL_1^{\sqcup k_2} }(\tau).
\eeq
Similarly, we can also take $\sY'_2:=\sY_2^{\sqcup k_1}$ and $\sL'_2:=\sL_2^{\sqcup k_1}$ in the construction and then
we have \begin{equation}
\frac{1}{k_2}   I_{  \sL_2}(\tau) = \frac{1}{ k_1 k_2}   I_{ \sL_2^{\sqcup k_1} }(\tau).
\end{equation}  
Therefore the independence of the result from the choice of 
$k_1$, $\sY_1$ and $\sL_1$  versus $k_2$, $\sY_2$ and $\sL_2$
can be subsumed into 
the independence from the choices between $\sY'_1$, $\sL'_1$ and $\sY'_2$, $\sL'_2$ at fixed $k'=k_1k_2$,
which was already demonstrated.
 
We also note that the linearity 
\beq
\GS([\sX_1]+[\sX_2], [\sT]) &= \GS([\sX_1], [\sT]) +\GS([\sX_2], [\sT]) , \nonumber \\
\GS([\sX], [\sT_1]+ [\sT_2]) &=  \GS([\sX], [\sT_1])  + \GS( [\sX],  [\sT_2]) \label{eq:linearity}
\eeq
can be shown using a similar technique. 
To define $\GS([\sX_1]+[\sX_2], [\sT]) $, we take $k$ sufficiently large so that both $k[\sX_1]=0$ and $k[\sX_2]=0$ hold. Then the configuration for $\GS([\sX_1]+[\sX_2], [\sT]) $ can be taken to be just the disjoint union of the configurations for $\GS([\sX_1], [\sT]) $ and $\GS([\sX_2], [\sT])$. The linearity for $\GS([\sX], [\sT_1]+ [\sT_2]) $ is shown in the same way. 

\paragraph{Dependence on representative $\sX$.}

Our remaining tasks are to show that $\GS([\sX], [\sT]) $ is independent of the representatives $\sX $ and $\sT$ of $[\sX] $ and $[\sT]$, respectively. 
We have already established that $\GS([\sX], [\sT]) $ is independent of $  \sL, \sY$ and $\sN$. Then the linearity \eqref{eq:linearity} implies that
it is sufficient to show that  $\GS([\sX], [\sT]) $ vanishes when $\sX = \partial \sY_1$ or $\sT = \partial \sU_1$ for some $\sY_1$ or $\sU_1$. 

Let us consider the case $\sX = \partial \sY_1$. In this case, we can choose the integer $k$ and the SQFT $\sY$ in the definition of $\GS([\sX], [\sT]) $ to be 
\beq
k=1, \quad \sY=\sY_1.
\eeq
We remark that the condition \eqref{eq:BNcondition} is trivially satisfied for $k=1$.
Because $k=1$, the value of $\GS([\sX], [\sT]) $ in $\bQ/\bZ$ 
is manifestly zero.

\paragraph{Dependence on representative $\sT$.}

Finally we consider the case $\sT = \partial \sU_1$.
We choose $\sN$ to be 
$\sN =(-1)^{\nu_\sX} \cdot \sX \times \sU_1$, where the sign $(-1)^{\nu_\sX} $ means
the choice of orientation. Recall also that $\sN_0 = \sY \times \sT$.

We need to choose an $  \sL$ whose boundary is $\sN^{\sqcup k} \cup \overline{\sN_0}$. 
In this case, we may take $  \sL$ to be a theory with a corner, given by the product
\beq
  \sL = (-1)^{\nu_\sX} \cdot  \sY \times \sU_1 . \label{eq:indepT}
\eeq
When boundaries do not have zero modes (and hence bulk zero modes are normalizable), the APS index has the simple multiplicative property that
\beq
I_{\sL} = (-1)^{\nu_\sX} I_\sY I_{\sU_1}.
\eeq
Unless $\nu_\sY =d+1 \in 4\bZ$, the index $I_\sY$ on the right-hand side is automatically zero by the algebra of time-reversal symmetry \eqref{eq:time1} or \eqref{eq:time2}.
Therefore, we only need to consider the case $d \in -1 +4\bZ$. 

When $d \in -1+ 4\bZ$, we have $\frac1k I_\sY(\tau) \in \eta(\tau)^{-(d+1)} \bZ((q))$ by the condition \eqref{eq:BNcondition}
and therefore the right hand side of
 \begin{equation}
\frac1k\left[ \eta(\tau)^4 I_{  \sL}(\tau)  \right]_{q^0} 
= (-1)^{\nu_\sX}  \frac1k \left[\eta(\tau)^4  I_\sY(\tau)   I_{\sU_1}(\tau) \right]_{q^0}
\end{equation}
is in $\bZ$.

This concludes our check of well-definedness of $\GS([\sX], [\sT]) $.

\subsection{Properties of the invariant}

We have established that our pairing 
\beq
\GS :  \bA_d \times \bA_{-d-22} \to \bQ/\bZ
\eeq 
is well-defined as a bilinear form on $\bA_{d}\times \bA_{-22-d}$,
\beq
\GS([\sX_1]+[\sX_2], [\sT]) &= \GS([\sX_1], [\sT]) +\GS([\sX_2], [\sT]) , \nonumber \\
\GS([\sX], [\sT_1]+ [\sT_2]) &=  \GS([\sX], [\sT_1])  + \GS( [\sX],  [\sT_2])  .
\eeq
It has two additional nice properties we discuss below.

\paragraph{Symmetry.}

We claim
\beq
\GS([\sX],[\sT]) = (-1)^{\nu_\sX \nu_\sT} \GS([\sT],[\sX]). \label{eq:gcomm}
\eeq
Before showing it, recall that
there is an apparent asymmetry in the definition of $\GS([\sX],[\sT])$ in the treatment of $\sX$ and $\sT$,
in that we took a $\sY$ such that $\partial \sY = \sX^{\sqcup k}$,
and considered a noncompact theory $  \sL$ with corners such that 
$\partial   \sL= \sN^{\sqcup k} \cup \overline{\sY \times \sT}$.
We want to study what happens if we take $\ell$, $\sU$ and $\sL'$ such that 
\beq
\partial \sU= \sT^{\sqcup \ell}, \qquad \partial   \sL'=\sN^{\sqcup \ell} \cup \overline{(-1)^{\nu_\sX} \sX \times \sU}
\eeq
instead. We will get the same pairing, as we now see.

Let us consider $\sL^{\sqcup \ell}$ and $\sL'^{\sqcup k}$. Their boundaries have a common part $\sN^{\sqcup k\ell}$. Let $\sL''$ be the theory obtained by gluing $\sL^{\sqcup \ell}$ and $\sL'^{\sqcup k}$ along $\sN^{\sqcup k\ell}$. The situation is similar to Figure~\ref{fig:4}, but now we replace $k \to k\ell$, $\sL \to \sL^{\sqcup \ell}$, $\sL' \to \sL'^{\sqcup k}$, $\sN_0 \to \sY \times \sT^{\ell}$, and $ \sN'_0 \to (-1)^{\nu_\sX} \sX^{\sqcup k} \times \sU$ in Figure~\ref{fig:4}. 
The boundary of $\sL''$ is given by
\beq
\partial  \sL''=(\sY \times \sT^{\sqcup \ell} ) \cup \overline{(-1)^{\nu_\sX}  \sX^{\sqcup k} \times \sU}.
\eeq
Notice that the product $(-1)^{\nu_\sY} \sY \times \sU$ has the same boundary, so we can glue $\sL''$ and (the orientation reversal of) $(-1)^{\nu_\sY} \sY \times \sU$ to get a compact theory $\sL'''$,
\beq
\sL''' = \sL'' \cup \overline{(-1)^{\nu_\sY} \sY \times \sU}.
\eeq
As usual, we have
\beq
[\eta(\tau)^4 I_{\sL'''}]_{q^0}=0. 
\eeq
Also, the APS index (in the absence of boundary zero modes) has the properties that
\beq
I_{\sL''' } &= I_{\sL''} -   (-1)^{\nu_\sY} I_{\sY \times \sU} = ( \ell I_{\sL}  - k I_{\sL'}) -  (-1)^{\nu_\sY}  I_{\sY} I_{\sU} .
\eeq
Combining them gives
\beq
\frac1k [\eta(\tau)^4 I_{\sL}(\tau) ]_{q^0} -  \frac1\ell  [\eta(\tau)^4 I_{\sL'}(\tau) ]_{q^0}   =  (-1)^{\nu_\sY} \left[ \eta(\tau)^4  \frac{1}{k} I_{\sY} \cdot \frac{1}{\ell} I_{ \sU}\right]_{q^0}.\label{eq:diff-k-l}
\eeq
Unless $d \in -1+4\bZ$, the right-hand side is zero by the algebra of time-reversal symmetry \eqref{eq:time1} or \eqref{eq:time2}.
When $d \in -1+4\bZ$, both $ \frac{1}{k} I_{\sY} $ and $ \frac{1}{\ell} I_{ \sU}$ have integer coefficients because of Property~\ref{prop:1} of Sec.~\ref{app:BN}.
We conclude that
\beq
\frac1k [\eta(\tau)^4 I_{\sL}(\tau) ]_{q^0} \equiv  \frac1\ell  [\eta(\tau)^4 I_{\sL'}(\tau) ]_{q^0}  \mod \bZ.
\eeq

Now the asymmetry between $\sX$ and $\sT$ has disappeared. Then the relation \eqref{eq:gcomm} simply follows from $\sX \times \sT = (-1)^{\nu_\sX \nu_\sT} \sT \times \sX$.

\paragraph{Multiplicative structure.}

Let us define graded Abelian groups
\beq
\SQFT_\bullet(\pt)=\bigoplus_\nu \SQFT_\nu(\pt), \qquad \bA_\bullet = \bigoplus_\nu \bA_\nu.
\eeq
The graded Abelian group $\SQFT_\bullet (\pt)$ is a (graded) ring.
The graded Abelian group $\bA_\bullet$ is actually an ideal of $\SQFT_\bullet (\pt)$, i.e., if $[\sX] \in \bA_\bullet$ and $[\sA] \in  \SQFT_\bullet(\pt)$, then $[\sX][\sA] \in \bA_\bullet$. 
This can be seen as follows. The primary invariants are KO-theoretic invariants and hence they are multiplicative, so if $\sX$ has vanishing primary invariants, then so does $\sX \times \sA$ for any $\sA$. Next, suppose that the secondary invariant of $\sX$ also vanishes. We take $(k,\sY)$ such that $\partial \sY = \sX^{\sqcup k}$ and then the secondary invariant is determined by $\frac1k I_{\sY}$. The secondary invariant of $\sX \times \sA$ is determined by $\frac1k I_{\sY \times \sA}$ because $\partial (\sY \times \sA) = (\sX \times \sA)^{\sqcup k}$. If $\sX$ has vanishing primary invariants, then the boundary of $\sY$ does not generically have zero modes. In this case the APS index has the property that $ I_{\sY \times \sA} = I_\sY I_{\sA}$, and hence the secondary invariant of $\sX \times \sA$ also vanishes for any $\sA$. 

For notational simplicity we regard $\GS$ as a bilinear form 
\beq
\bA_\bullet \times \bA_\bullet \to \bQ/\bZ,
\eeq
such that $\GS([\sX], [\sT])=0$ if $\nu_\sX + \nu_\sT \neq -22$. 

We claim that if $[\sX], [\sT] \in \bA_\bullet$ and $[\sA] \in \SQFT_\bullet (\pt)$, then
\beq
\GS([\sX][\sA], [\sT])=\GS([\sX], [\sA][\sT]). \label{eq:multiplicability}
\eeq
This is because we can take the intermediate theories $\sN$ and $\sN_0$ in the definition of $\GS$
to be the same for the computations of both 
the left-hand side and the right-hand side. We take $\sN$ such that $\partial \sN = \sX \times \sA \times \sT$. We also take $\sY$ such that $\partial \sY = \sX^{\sqcup k}$, and then $\sN_0=\sY \times \sA \times \sT$. 

\subsection{Examples}
\label{sec:examples}

Now that we have defined the new invariant and discussed its basic properties,
we would like to have a look at concrete examples.
We discuss four cases listed in Table~\ref{tab:examples}  in this section.
There, for $\sX$, we use $S^3_{H=1}$ which is the sigma model on $S^3$ with the unit $H$ flux,
or equivalently the \Nequals{(0,1)} $\SU(2)$ Wess-Zumino-Witten model at level $1$.
Similarly, $\SU(3)_{H=1}$ and $\Sp(2)_{H=1}$ denote \Nequals{(0,1)} Wess-Zumino-Witten model
at level one for each group indicated.
For $\sT$, we use purely left-moving spin modular-invariant CFTs
where we indicate the maximal chiral algebra contained in each of them;
according to \cite{BoyleSmith:2023xkd} this uniquely fixes the theory in the range $0\le c_L\le 16$.

The first three pairings were discussed in \cite{Kaidi:2024cbx},
and the last one was discussed in \cite{Tachikawa:2024ucm},
using a different technique and formulation of computation
using string theory.
Here we perform the computation using the definition 
and the properties of the new invariant established in our discussions so far in this paper.

\begin{table}
\[
\begin{array}{c@{\ \in\ }l|c@{\ \in\ }l|c|c}
\sX & \bA_{d} & \sT & \bA_{-d-22} & \GS([\sX],[\sT]) & \text{in Section}\\
\hline
\hline
(S^3_{H=1} )^2 & \bA_{6} & (\e_7)_1\times (\e_7)_1 & \bA_{-28} & 1/2 & \text{Sec.~\ref{subsec:6}}\\
\SU(3)_{H=1} & \bA_{8} & \su(16)_1 & \bA_{-30} & 1/2& \text{Sec.~\ref{subsec:8}}\\
(S^3_{H=1} )^3 & \bA_{9} & (\e_8)_2 & \bA_{-31} & 1/2& \text{Sec.~\ref{subsec:8}}\\
\Sp(2)_{H=1} & \bA_{10} & \so(16)_1\times \so(16)_1 & \bA_{-32} & 1/3& \text{Sec.~\ref{subsec:10}}
\end{array}
\]
\caption{The examples of the pairing $\GS([\sX],[\sT])$ we discuss in this paper.
For the notations, see the main text.
\label{tab:examples} }
\end{table}

In the discussions of the examples, we will implicitly use the following string-theoretic\footnote{It is not necessary to require that SQFTs have conformal invariance as far as the computations of the invariant is concerned.} interpretation. The theory $\sT$ is regarded as an internal theory and it produces massless fermions in the target space of $\sX$. Then the invariant for SQFTs is the same as the one discussed in Sec.~\ref{sec:mfds} for fermions and manifolds.

\subsubsection{The first example}
\label{subsec:6}

Our first and main example is when  $\sX=S^3_{H=1}\times S^3_{H=1} \in \SQFT_6$
and $\sT=(\e_7)_1\times (\e_7)_1\in \SQFT_{-28}$.
Before beginning the computation, we need to check that $[\sX] \in \bA_{6}$
and $[\sT] \in \bA_{-28}$.
For this, recall that the ordinary elliptic genus can only be nonzero when $\nu \equiv 0 \mod 4$,
the mod-2 elliptic genus only when $\nu \equiv 1,2 \mod 8$,
and we have required the vanishing of the secondary invariant  only when $\nu \equiv 3 \mod 4$.

For $\sX$, all the relevant invariants vanish automatically. 
For $\sT$, the only nonzero possibility is the ordinary elliptic genus,
which in this case is simply the trace in the R-sector 
weighted by $(-1)^F$.
According to the data tabulated in \cite{BoyleSmith:2023xkd}, 
the R-sector consists of 
one irreducible representation $\mathbf{56}$ of one $(\e_7)_1$ with $(-1)^F=+1$,
and 
one irreducible representation $\mathbf{56}'$ of another $(\e_7)_1$ with $(-1)^F=-1$.\footnote{Modulo excited modes, the appearance of these representations can be seen in string theory interpretation as follows. We start from $(\e_8)_1 \times (\e_8)_1$ heterotic string theory, which produces fermions (gauginos) in the representation $({\bf 248} \otimes {\bf 1} )\oplus ({\bf 1} \otimes {\bf 248})$ with (say) $(-1)^F=1$.
Then we compactify it on $S^4$ with an instanton of one $\e_8$ and an anti-instanton of another $\e_8$ by using $\su(2) \times \e_7 \subset \e_8$ and putting the (anti)-instanton inside $\su(2)$. After the compactification, we get massless fermions in ${\bf 56} \otimes {\bf 1}$ with $(-1)^F=+1$ and ${\bf 1} \otimes {\bf 56}  $ with $(-1)^F=-1$. The worldsheet theory flows to $
 (\e_7)_1\times (\e_7)_1$ \cite{Kaidi:2023tqo,Kaidi:2024cbx}. This will also be used below \eqref{eq:e7e8}. \label{footnote:e8}}
Therefore, the trace $\Tr (-1)^F q^{H_L} \bar{q}^{H_R}$ is simply zero.

Now that we checked that $[\sX] \in \bA_{6}$ and $[\sT] \in \bA_{-28}$,
we proceed to the computation of the pairing. 
We first need to pick $\sN$ such that 
\begin{equation}
\partial \sN=\sX \times \sT = S^3_{H=1} \times (S^3_{H=1})' \times (\e_7)_1\times (\e_7)_1,
\end{equation}
where we distinguished two factors of $S^3_{H=1}$ by putting a prime on one of them.
To construct such an $\sN$, we note that $(S^3_{H=1})'$ can be filled in 
with the four-dimensional ball $B^4$, equipped with an $\e_7$ instanton gauge field with field strength $F$, such that 
$\d H=c(F)-\lambda(R)$ is satisfied:
\beq
\int_{S^3} H = \int_{B^4} c(F)=1,
\eeq
where we have used the fact that $\lambda(R)$ is trivial on $B^4$.
Then we can fiber the $(\e_7)_1\times (\e_7)_1$ theory over $B^4$, using one of the $\e_7$ factor.
Let us denote the resulting noncompact theory by \begin{equation}
\sA = (B^4 \fiberproduct (\e_7)_1)\times (\e_7)_1,
\end{equation}
It satisfies
\begin{equation}
\partial\sA=(S^3_{H=1})' \times (\e_7)_1\times (\e_7)_1.
\end{equation}
We can then take 
\begin{equation}
\sN=-S^3_{H=1} \times \sA.
\end{equation}

Next, we need to pick a $\sY$ such that $\partial \sY=\sX^{\sqcup k}$.
For this, we take $\sB:=\text{K3}\setminus (B^4)^{\sqcup 24}$ which is the K3 surface with 24 balls removed.
Recall that the K3 surface has
$
\int_{\text{K3}} \lambda(R) =-24.
$
We equip $\sB$ with an $H$ field strength such that $\d H=-\lambda(R)$, so that 
\begin{equation}
\partial \sB= (S^3_{H=1})^{\sqcup 24},
\end{equation}
which is consistent because
\beq
24 \int_{S^3} H = -\int_{\text{K3}\setminus (B^4)^{\sqcup 24}} \lambda(R)=24.
\eeq
We then set $\sY=\sB \times (S^3_{H=1})'$ with $k=24$ and \begin{equation}
\sN_0=\sB \times (S^3_{H=1})' \times (\e_7)_1\times (\e_7)_1.
\end{equation} 
We now need to find an $  \sL$ such that $\partial  \sL=\sN^{\sqcup k}\cup \overline{\sN_0}$.
This is easily done as a noncompact theory with a corner, \begin{equation}
  \sL = -\sB \times \sA.
\end{equation}
We then have \begin{equation}
I_{ \sL}=-I_{\sB}\times I_{\sA}.
\end{equation} 
The coefficient of the leading term in $I_{\sB}$ is the Atiyah-(Patodi-)Singer index of the Dirac operator on K3 (with 24 copies of $B^4$ removed), multiplied by $\kappa=1/2$. The index theorem states that this index is given by $- \frac{1}{24} \int p_1(R) = - \frac{1}{24} \int 2\lambda(R) = 2$, and hence by multiplying $\kappa=1/2$ we get
\begin{equation}
I_{\sB} = \eta(q)^{-4} (1+O(q)).
\end{equation} 
The coefficient of the leading term in $I_{\sA}$ is the index of the Dirac operator on $B^4$ coupled to the $\e_7$ instanton in the representation ${\bf 56}$. This index is given by $12$\footnote{This may be seen e.g. from the fact that ${\bf 56}$ decomposes under $\su(2) \times \so(12) \subset \e_7$ as $({\bf 2} \otimes {\bf 12}) \oplus {\bf 2^5}$, and then putting the instanton inside $\su(2)$.} and hence
\begin{equation}
I_{\sA}= 12 +O(q)
\end{equation} 
where the exponent $a$ of the leading term $q^a$ is determined from the fact that ${\bf 56}$ has scaling dimension $3/4$ compared with the NS ground state, so $a=3/4-(4+7+7)/24=0$.\footnote{Alternatively, as in footnote~\ref{footnote:e8}, we may go to  $(\e_8)_1 \times (\e_8)_1$ on $S^4$ instead of $(\e_7)_1 \times (\e_7)_1$. Then ${\bf 248}$ has scaling dimension $1$ as it should be for massless gauginos in the ten-dimensional target space. Then $a=1 - (4+4+8+8)/24=0$.}
Therefore we get \begin{equation}
\GS([ S^3_{H=1} \times  S^3_{H=1}], [ (\e_7)_1\times (\e_7)_1]) = \frac1{24}[\eta(q)^4 I_{ \sL}]_{q^0} = \frac12.
\end{equation}

\subsubsection{The second and the third examples}
\label{subsec:8}

For other examples, a direct evaluation of the invariant will be discussed elsewhere.
In the present paper, we give short discussions of other methods only as a guide for the reader: interested readers should consult the original work \cite{Kaidi:2024cbx,Tachikawa:2024ucm}.

For the second and third examples, we utilize the bordism invariance of the pairing
to evaluate them. The discussion here is completely parallel to the discussion in \cite[Sec.~7.7]{Kaidi:2024cbx},
so we will be brief.

First, the bordism relation
\begin{equation}
(\e_7)_1\times (\e_7)_1 \sim S^4 \fiberproduct  \bigl( (\e_8)_1\times(\e_8)'_1\bigr) \label{eq:e7e8}
\end{equation} 
was demonstrated in \cite[Sec.~5.2]{Kaidi:2024cbx},
where $S^4$ is equipped with an $\e_8$ instanton of instanton number $+1$
and an $\e_8'$ instanton of instanton number $-1$ (see also footnote~\ref{footnote:e8} for a brief discussion.)
It was then shown in \cite[Sec.~7.6]{Kaidi:2024cbx} that it is further bordant to \begin{equation}
\sim S^3_{H=1} \times (S^1 \fiberproduct \bigl((\e_8)_1\times(\e_8)'_1\bigr))
\end{equation} where $S^1$ is now equipped with a holonomy flipping two $\e_8$ factors.
Finally, in \cite[Sec.~5.3]{Kaidi:2024cbx}, it was shown that \begin{equation}
S^1 \fiberproduct \bigl((\e_8)_1\times(\e_8)'_1\bigr) \sim (\e_8)_2.
\end{equation}
Combining these facts, we get
\beq
[(\e_7)_1\times (\e_7)_1] = [S^3_{H=1} ][ (\e_8)_2].
\eeq
By using the result from the previous subsection, we find \begin{equation}
\begin{aligned}
 \GS([S^3_{H=1}]^3, [(\e_8)_2])
&=\GS([S^3_{H=1}]^2, [S^3_{H=1} ][ (\e_8)_2])\\
&= \GS([S^3_{H=1}]^2, [(\e_7)_1\times (\e_7)_1]) \\
&=\frac12
\end{aligned}
\end{equation}
where in the first equality we used the property \eqref{eq:multiplicability}. This is one of the results listed in Table~\ref{tab:examples}.

Second, the equivalence \begin{equation}
S^1_\text{periodic} \times (\e_8)_2 \sim \su(16)_1
\end{equation} was shown in \cite[Sec.~7.6]{Kaidi:2024cbx}.
Furthermore, the string bordism relation \begin{equation}
(S^3_{H=1})^3 \sim \SU(3)_{H=1} \times S^1_\text{periodic}
\end{equation} is known in mathematics \cite{Hopkins2002}.
Then we find \begin{equation}
\begin{aligned}
 \GS([\SU(3)_{H=1}] , [ \su(16)_1]) 
&= \GS([\SU(3)_{H=1} ], [S^1_\text{periodic}] [ (\e_8)_2] )\\
&=\GS([\SU(3)_{H=1} ][ S^1_\text{periodic}], [(\e_8)_2])\\
&=\GS([S^3_{H=1}]^3, [(\e_8)_2]) \\
&=\frac12 .
\end{aligned}
\end{equation}
This is another one of the results listed in Table~\ref{tab:examples}.

\subsubsection{The fourth example}
\label{subsec:10}

Our fourth example is to take $\sX=\Sp(1)_{H=1} \in \SQFT_{10}$
and $\sT=\so(16)_1\times \so(16)_1 \in \SQFT_{-32}$.
The only possibly non-zero primary or secondary invariants 
are the mod-2 elliptic genus of $\sX$ and the ordinary elliptic genus of $\sT$.
The former vanishes since it is known that 
\begin{equation}
3[\Sp(1)_{H=1} ]=0 \label{3torsion}
\end{equation}
as a string bordism class \cite{Hopkins2002},
whereas the latter vanishes by a direct computation using the data given in \cite{BoyleSmith:2023xkd}.
Therefore our new invariant $\GS([\sX],[\sT])$  is defined without problem, and 
the condition \eqref{3torsion} means that it should be given by \begin{equation}
\GS([\Sp(1)_{H=1} ],[\so(16)_1\times \so(16)_1])=\frac a3 
\end{equation}  where $a$ is an integer.

A direct evaluation of the invariant will be discussed elsewhere.
We can indirectly evaluate it using the relation of our pairing to the discrete part of 
the Green-Schwarz coupling in the heterotic string,
and then using various string dualities to evaluate the coupling.
This was performed in \cite{Tachikawa:2024ucm},\footnote{%
The rough argument is as follows. 
$\Sp(2)$ is an $S^3$ fiberation over $S^7$. 
We have a unit $H$ flux in $S^3$, so let us regard it as the angular direction of an NS5-brane.
Then, the Green-Schwarz phase associated to 
$S^7$ with a nontrivial $S^3$ bundle can be thought of as measuring the anomaly
of the NS5-brane, under an $\SU(2)$ transformation 
specified by elements of $\pi_6(\SU(2))\simeq \bZ_{12}$.
As the 10d anomaly of the $\so(16)_1\times \so(16)_1$ heterotic string
is the difference of those of the $(\e_8)_1\times (\e_8)_1$ heterotic string 
and the $\so(32)_1$ heterotic string,
we can evaluate this NS5-brane anomaly by using the known anomaly
of the M5-brane in M-theory and that of the D5-brane in the Type I theory,
leading to the result \eqref{frac13}.
}
and the result was  \begin{equation}
\GS([\Sp(1)_{H=1} ],[\so(16)_1\times \so(16)_1])=\pm \frac13,\label{frac13}
\end{equation} where the sign depends on the choice of the orientations of $\sX$ 
and $\sT$.

\section{Spectral invariants}
\label{sec:differential}

Our strategy so far was to use topological quantities. In this section, we define a spectral invariant of supercharges which is analogous to the APS $\eta$-invariant for Dirac operators, and discuss their relation to the invariants discussed in Section~\ref{sec:primarysecondary} and \ref{sec:new}. Here, a spectral invariant of $Q$ means that it only depends on the eigenvalue spectrum of $Q$. 

The invariant defined in this section is not always a topological invariant, in the same way that the APS $\eta$-invariant is not always a topological invariant. 
Recall that the APS $\eta$-invariant of Dirac operators on $d$-manifolds
is related to the APS index and a characteristic differential form (such as the $\hat A$ polynomial) in dimension $d+1$.
Furthermore, when the characteristic differential form vanishes, 
the $\eta$-invariant gives a bordism invariant in $d$ dimensions.
Our spectral invariant of supercharges of degree $\nu$ has similar properties:
it is related to APS indices in degree $\nu+1$, 
and bordism invariants in degree $\nu$ in some situations. 

Our version of the $\eta$-invariant is  defined by integrating $\vev{Q} = \Tr q^{H_L} \bar q^{H_R} Q$ over the fundamental region on the $\tau$-plane. In this sense, it might be seen as a ``stringy'' version of the $\eta$-invariant. An application to the partition functions of heterotic string theory will be discussed in Section~\ref{sec:string}.
In this section, we consider more general class of theories than in string theory.

By using our $\eta$-invariant, 
we will define a pairing $\eta(\sX, \sS)$, which can detect in a systematic way
various bordism invariants we have described so far in this paper. 
Here, $\sX \in \SQFT_{d}$, and $\sS$ has a boundary $\partial \sS = \sT$ such that $\sT \in \SQFT_{-22-d}$. However,
to define the set in which $\sS$ takes values, we need to consider not only SQFTs but also supersymmetric quantum mechanics (SQM) with time-reversal symmetry. 

The introduction of SQMs in our context may look somewhat artificial at first. The main motivation for introducing them is that there is an interesting relation between TMF and KO which corresponds to the relation between SQFT and SQM discussed in this section~\cite{Tachikawa:2023lwf}. 
On the other hand, SQMs are not necessary in heterotic string theory since string worldsheet is described by 2d superconformal field theory. Still, the $\eta$-invariant defined in this section is important as will be discussed in Section~\ref{sec:string}. 

This section is organized as follows. 
In Sec.~\ref{sec:formal}, we first discuss formal Laurent series of SQMs,
which play a central role in our construction of the pairing $\eta(\sX,\sS)$.
Then in Sec.~\ref{sec:the-eta}, we define our spectral invariant of supercharges,
and discuss the analogue of the APS index theorem in our setup. 
In Sec.~\ref{sec:secondrevisit}, a direct definition of the secondary invariant by using the spectral invariant is given.
In Sec.~\ref{sec:the-pairing}, we use the results obtained so far to define
the spectral pairing $\eta(\sX,\sS)$, which will be the focus of the rest of the section.
After discussing in Sec.~\ref{sec:generalstr} the general structure of the spectral pairing,
we show in Sec.~\ref{sec:inv1} how various known bordism invariants can be uniformly
understood in terms of the spectral pairing. Then, in Sec.~\ref{sec:const} we discuss interesting constraints on the primary and secondary invariants in SQFTs.
Finally in Sec.~\ref{sec:inv2}, we will rephrase the new invariant we introduced in this paper
using our spectral pairing.

\subsection{Formal Laurent series of SQMs}\label{sec:formal}
Most of the notions discussed in Section~\ref{sec:notion} are valid in SQMs. For example, we can define compact and mildly noncompact SQMs, and consider their APS indices. 
On the other hand, worldsheet gravitational anomalies $\nu$ are not defined as integers in SQMs. Time-reversal anomalies are classified by $\bZ_8$~\cite{Fidkowski:2009dba,Kapustin:2014dxa,Witten:2015aba} which is manifested in the algebras \eqref{eq:time1} and \eqref{eq:time2}.

An SQFT can also be seen as a formal Laurent series of SQMs in the following way. In an SQFT, we rewrite
\beq
q^{H_L} \bar q^{H_R} = q^P |q|^{2Q^2} \label{eq:rewrite}
\eeq
where recall that $H_L=\frac12 (H+P) $ and $H_R=\frac12 (H-P)=Q^2$. 
For each eigenvalue $P=n -\nu/24$, we have the corresponding eigenspace $\cH_{n  - \nu/24}$ on which the supercharge $Q$ acts. Therefore, each $\cH_{n  - \nu/24}$ can be regarded as the Hilbert space of an SQM.
As mentioned in Section~\ref{sec:notion}, the CPT symmetry in a 2d SQFT
gives a time-reversal symmetry in the SQM defined in each eigenspace of $P$.

We can consider a more general formal power series of SQMs (not necessarily related to an SQFT) of the form,
\beq
\sS= \eta(\tau)^{-\nu_\sS } \sum_{n } q^{n} \sS_{n}. \label{eq:formalLaurent}
\eeq
In this expression, each $\sS_n$ is an SQM. The sum is over $n \geq n_0$ for some lower bound $n_0\in \bZ$, or in other words $\sS_n$ are empty for $n <n_0$. The motivation for the overall factor $\eta(\tau)^{-\nu_\sS } $ (and how to choose the value of $\nu_\sS$) will be discussed later. The expression \eqref{eq:formalLaurent} is only a formal expression, and it only means that we have infinitely many SQMs, $\sS_n$.\footnote{%
Note that for any mathematical concept $R$ for which an addition and a multiplication are consistently defined,
there is no problem in considering the ring $R((q))$ of formal Laurent series whose coefficients are in $R$.
Here we are simply using $R=\sS$, where the addition and the multiplication are the direct sum and the tensor product
of the underlying Hilbert spaces involved.
}

The reader may wonder why we have to consider such an artificially-looking object
as a formal Laurent series of SQMs. One natural source of such a formal Laurent series
is to consider a 2d \Nequals{(0,1)} sigma model on a spin manifold $M$
which does \emph{not} necessarily satisfy the condition $\d H=c(F)-\lambda(R)$ (or the more precise condition mentioned in footnote~\ref{footnote:ws}).
As it does not satisfy the worldsheet anomaly cancellation condition,
it does not define a genuine 2d SQFT. There is, however, no problem in 
quantizing the space of maps $S^1\to M$ equipped with fermionic directions,
and writing down the Hilbert space, the supercharge and the Hamiltonian.
This gives a formal Laurent series of SQMs.
There is of course no guarantee that the resulting theory is modular invariant,
even when $M$ is compact.
In this case, it is convenient to set $\nu_M=\dim M$
and include a factor of $\eta(\tau)^{-\nu_M}$,
to reproduce the same shift of the background $S^1$ momentum
as in the case when the worldsheet anomaly vanishes and the theory is an SQFT.

Let $Q$ represents supercharge for any theory. By comparing with the case of SQFTs and looking at \eqref{eq:rewrite}, it is clear that we should define 
the partition function, the APS index, and the expectation value of $Q$ in the theory $\sS$ as
\beq
Z_\sS &=  \eta(\tau)^{-\nu_\sS }  \sum_n q^n  \Tr_{\sS_n} (-1)^F  |q|^{2Q^2}, \nonumber \\
I_\sS & = \eta(\tau)^{-\nu_\sS } \sum_n q^n  \Index(\sS_n), \nonumber \\
\vev{Q}_{\sS}&=\eta(\tau)^{-\nu_\sS }  \sum_n q^n  \Tr_{\sS_n}|q|^{2Q^2} Q \label{eq:formalQ}
\eeq
where the notation is hopefully obvious. 

When all of $\sS_n$ is compact, we call the formal Laurent series itself as compact,
and denote by $\SQM_\nu$ the set of all such formal Laurent series of compact SQMs
with a fixed gravitational anomaly $\nu$.\footnote{The fact that the time reversal anomalies in 1d are classified by $\bZ_8$ implies that $\SQM_\nu$ has periodicity $\SQM_{\nu+8} \simeq \SQM_\nu$. This is essentially the same as the periodicity of KO-theory. }
Now, suppose that the theory $\sS$ has boundary. More precisely, we define the boundary of $\sS$ in terms of the boundaries of each $\sS_n$ by
\beq
\partial\sS= \eta(\tau)^{-\nu_\sS +1 } \sum_{n} q^{n} \partial \sS_{n}
\eeq
Let us denote this boundary as $\sT=\eta(\tau)^{-\nu_\sT } \sum_{n} q^{n} \sT_{n}$ where $\nu_\sT = \nu_\sS-1$. 
The derivation of the holomorphic anomaly equation in Appendix~\ref{sec:A} is essentially the same, and the result is
\beq
\frac{\partial Z_\sS}{\partial \bar\tau} =\frac{\i}{2 (\Im \tau)^{1/2} \eta(\tau)} \vev{Q}_\sT, \label{eq:hol3}
\eeq
where the factor $1/\eta(\tau)$ just comes from the difference of the overall factors $ \eta(\tau)^{-\nu_\sS }$ and $ \eta(\tau)^{-\nu_\sT }$.

We will consider formal power series of SQMs, $\sS$, whose boundary is given by an SQFT, $\sT$.
An SQFT has a definite value $\nu_\sT$ of the gravitational anomaly, and we want to choose $\nu_\sS$ as $\nu_\sT+1$. This is the reason that we have introduced the factor $\eta(\tau)^{-\nu_\sS }$ in \eqref{eq:formalLaurent}. We denote the set of formal Laurent series of SQMs with $\nu_\sS=\nu$ whose boundary is an SQFT as
\beq
\SQ_{\nu} = \{ \sS~|~\partial \sS  \in \SQFT_{\nu -1} \}. \label{eq:SQM/SQFT}
\eeq
This strange-looking notation is motivated by the following standard practice in algebraic topology.
The rest of this subsection is motivational, so can be skipped by the reader.

Let $M\cS$ be the `classifying space of manifolds with $\cS$ structure', such that 
$M\cS_d(\pt)$ is the bordism group of manifolds with structure $\cS$.
More precisely, $M\cS$ is known as the Thom spectrum of manifolds with $\cS$ structure.
Consider two structures $\cS$ and $\cS'$, such that there is a forgetful map $\cS\to \cS'$.
One example is to take $(\cS,\cS')=(\text{spin structure}, \text{orientation})$
and another is to take $(\cS,\cS')=(\text{string structure},\text{spin structure})$.
In such a case, we can consider a map $M\cS\to M\cS'$.
In algebraic topology, when a map from a spectrum $A$ to another $B$ is given as $A\to B$,
the `quotient of $B$ by (the image of) $A$' can be constructed, and is denoted by $B/A$,
provided with the natural quotient map $A\to B\to A/B$, which comes
with a connecting map:\footnote{As remarked in footnote~\ref{footnote:spectrum}, we are using a non-standard convention that $A_\nu^\text{here} = A_{-\nu}^\text{standard}$
so that $A_\nu(\pt)=\pi_0A_\nu^\text{here}$ rather than $A_\nu(\pt)=\pi_0 A_{-\nu}^\text{standard}$.} 
\begin{equation}
A_\nu \to B_\nu \to (A/B)_\nu \to A_{\nu-1}.
\end{equation} 
Applying it to $M\cS$ and $M\cS'$, we have \begin{equation}
(M\cS)_d \to (M\cS')_d \to (M\cS'/M\cS)_d \to M\cS_{d-1}.
\end{equation} In this case, $(M\cS'/M\cS)_d$ is known to be the classifying
space of $d$-dimensional manifolds $M$ with $\cS'$ structure whose boundary 
$\partial M$ is equipped with a compatible $\cS$ structure.
To see why this property is sensible, note that a 
compact $d$-dimensional manifold $M$ with $\cS'$ structure 
naturally gives an element in $(M\cS'/M\cS)_d(\pt)$ by regarding its boundary to be the empty set $\varnothing$.
A $d$-dimensional manifold equipped with $\cS$ structure in the bulk (as well as the boundary) is regarded as trivial in $(M\cS'/M\cS)_d(\pt)$.
The map $(M\cS'/M\cS)_d(\pt) \to M\cS_{d-1}(\pt)$ simply extracts the manifold at the boundary.

With this background explained, let us come back to our case.
Namely, $\SQFT_\nu$ was the `set' of SQFTs with gravitational anomaly $\nu$.
We can partially forget the structure of an SQFT and regard it as a formal Laurent series of SQMs,
defining the map $\SQFT_\nu \to \SQM_\nu$.
Motivated by the above construction about $M\cS'/M\cS$, we come to the definition \eqref{eq:SQM/SQFT}.
Finally, note that the operation of  associating an \Nequals{(0,1)} sigma model to a manifold
gives maps $M\text{String}_d(\pt) \to \SQFT_d(\pt) $ and $M\text{Spin}_d(\pt)  \to \SQM_d(\pt)  $,\footnote{%
The spaces like $M\text{String}_d$ are not ``the set of manifolds''. They are constructed as Thom spaces. On the other hand, in the present paper, we are loosely regarding $\SQFT_d$ as the set of SQFTs and we are not assuming them to form an $\Omega$-spectrum as remarked in footnote~\ref{footnote:spectrum}. By these reasons, we avoided to write $M\text{String}_d \to \SQFT_d$. }
which fit into the following commutative diagrams together with their quotients: \begin{equation}
\begin{array}{ccccccc}
M\text{String}_d(\pt)  & \to & M\text{Spin}_d (\pt) & \to & M\text{Spin}/M\text{String}_d (\pt) & \to & M\text{String}_{d-1}(\pt)  \\
\downarrow & & \downarrow& & \downarrow & & \downarrow  \\
\SQFT_d(\pt)  & \to &\SQM_d(\pt)  &\to & \SQM/\SQFT_d(\pt) & \to & \SQFT_{d-1}(\pt).
\end{array}
\end{equation}
The third vertical map, $(M\text{Spin}/M\text{String})_d(\pt) \to (\SQM/\SQFT)_d(\pt)$
gives us a class of concrete elements of this somewhat abstract concept $(\SQM/\SQFT)_d(\pt)$.
Namely,
we consider a $d$-dimensional spin manifold $M$ whose boundary has a string structure on
its boundary $N=\partial M$.
We can then consider an \Nequals{(0,1)} sigma model on it.
Then we have a formal Laurent series of SQMs from the bulk $M$,
whose boundary is given by a compact SQFT on $N=\partial M$.

\subsection{The $\eta$-invariant and the index theorem} \label{sec:the-eta}

Consider any compact theory $\sN$. It may be either an SQFT or a formal Laurent series of SQMs. 
We are going to define a spectral invariant of its supercharge $Q$, which we call as the $\eta$-invariant $\eta_\sN$. This is an analog of the APS $\eta$-invariant in the case of spin manifolds.

For later purpose, we introduce $\kappa$ given by
\beq
\kappa = \left\{ \begin{array}{ll} 
  1 &   (\nu_\sN \equiv -1, 0,1,5,6 \mod 8),\\
  \frac{1}{2} &  (\nu_\sN  \equiv 2,3,4 \mod 8)
\end{array} \right. \label{eq:kappadef1}
\eeq
due to the following reasons. The time reversal symmetry as well as other symmetries satisfies the algebra \eqref{eq:time1} or \eqref{eq:time2} which we reproduce here for the convenience of the reader; when $\nu_{\sN}=2m$, we have
\begin{equation}
\begin{aligned}
  T^2&=(-1)^{\frac12 m(m-1)}, & (-1)^F  T &= (-1)^m T(-1)^F,\\
   TQ &= (-1)^m QT,&  (-1)^FQ &= -Q(-1)^F 
\end{aligned}
\end{equation}
 and when $\nu_{\sN}=2m-1$, we have
 \beq
 T^2=(-1)^{\frac12 m(m-1)}, \qquad   TQ = (-1)^m QT. \label{eq:timealgebra}
\eeq
When $\nu \equiv 2,3,4 \mod 8$, the algebra implies that the dimension of each eigenspace of $Q$ (with eigenvalue $Q=Q_0  $) is a multiple of 2. This is because $T(-1)^F$, $T$, and $T$ for $\nu \equiv 2,3$ and $4$, respectively, commute with $Q$ and their squares are $-1$. For $\nu_{\sN}\equiv -1,0,1,5,6 \mod 8$, the dimension of each eigenspace of $Q$ is a multiple of 1. These facts will be used later. 

For $\nu_{\sN}\equiv 5,6 \mod 8$, the $\eta$-invariant modulo $\bZ$ will be zero. For $\nu_{\sN}\equiv 0,4 \mod 8$ and $\nu_{\sN}\equiv 1,2 \mod 8$, the $\eta$-invariant modulo $\bZ$ will be determined by the usual elliptic genus and the mod-2 elliptic genus, respectively, as we will come back to later.
Thus, potentially non-topological cases are $\nu \equiv -1, 3 \mod 8$. We remark that when we have a theory $\sL$ with $\partial \sL = \sN$ and hence $\nu_\sL=\nu_\sN+1$,  $\kappa$ defined here is consistent with what was introduced in \eqref{eq:kappadef} for $\nu_\sL \in 4\bZ$ in the sense that $\kappa_\sL=\kappa_\sN$.

\paragraph{The case of no zero modes of $Q$.}
Let us first consider the case that $Q$ does not have zero modes in the theory $\sN$. 
Let $F$ be the fundamental region of $\tau$ for the $\SL(2,\bZ)$ action,
\beq
F=\{ \tau \in \bC ~|~\Im \tau>0,  -1/2 \leq \Re \tau \leq 1/2, ~|\tau| \geq 1 \}.
\eeq
Let $\vev{Q}_\sN$ be the expectation value of the supercharge $Q$ as defined in \eqref{eq:formalQ}. 
Notice that this is a function of $\tau, \bar\tau$. 
Then, we define the $\eta$-invariant\footnote{Unfortunately, the symbol $\eta$ is used both for the spectral $\eta$-invariant and also the Dedekind $\eta$ function. 
The Dedekind $\eta$ function always has $\tau$ or $q$ as its argument, so hopefully there is no confusion. 
}
 of $\sN$ by
\beq
\eta_\sN  = \kappa \int_F \d \tau \wedge \d \bar\tau \frac{\i  \eta(\tau)^3}{2 (\Im \tau)^{1/2}} \vev{Q}_\sN.\label{eq:eta-inv1}
\eeq
This quantity $\eta_\sN$ is a spectral invariant of $Q$, meaning that it is determined by the eigenvalue spectrum of $Q$. 

More generally, let $f(q)$ be any Laurent series of $q$ (possibly multiplied by some power of $\eta(\tau)$) whose coefficients are integers.  Then we define
\beq
\eta_\sN(f)  = \kappa \int_F \d \tau \wedge \d \bar\tau \frac{\i  \eta(\tau)^3}{2 (\Im \tau)^{1/2}} \vev{Q}_\sN f(q).
\eeq
We will always assume below that $f(q)$ is such a Laurent series of $q$ with integer coefficients,
possibly multiplied by some power of $\eta(\tau)$,
and we will not repeat this comment.

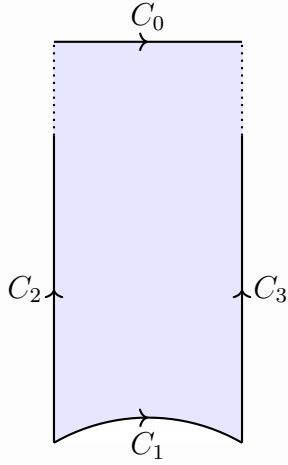
\begin{figure}
\[
\begin{tikzpicture}[scale=2.5]


  \begin{scope}
    \clip (-0.5,{sin(60)}) rectangle (0.5,3);
    \fill[blue!10] (-1,{sin(60)}) rectangle (1,3);
    \fill[white] (0,0) circle (1);
  \end{scope}

  \draw[thick, postaction={decorate}, decoration={markings, mark=at position 0.5 with {\arrow{>}}}] 
  (-0.5,{sin(60)}) -- (-0.5,2.5) node [midway, left] {$C_2$};
\draw[thick, dotted] (-0.5,2.5)--(-0.5,3);
  \draw[thick,postaction={decorate}, decoration={markings, mark=at position 0.5 with {\arrow{>}}}] (0.5,{sin(60)}) -- (0.5,2.5) node [midway, right] {$C_3$};;
\draw[thick, dotted] (0.5,2.5)--(0.5,3);
  \draw[thick, postaction={decorate}, decoration={markings, mark=at position 0.5 with {\arrow{>}}}] (-0.5,3)--(0.5,3) node[midway,above] {$C_0$};

  \draw[thick, domain=120:60, postaction={decorate}, decoration={markings, mark=at position 0.5 with {\arrow{>}}}] plot ({cos(\x)}, {sin(\x)});

\node (A) at (0,.85) {$C_1$};

\end{tikzpicture}
\]
\caption{The fundamental region $F$ and the four components of its boundary.
\label{fig:fundamental-region}}
\end{figure}

The definition of $\eta_\sN$ is motivated by the following index theorem. Consider a theory $\sL$ whose boundary is $\sN$. Let us consider a 1-form
\beq
\omega_\sL = \kappa \eta(\tau)^4 Z_\sL \d \tau \label{eq:1-form}
\eeq
where $ Z_\sL$ is the partition function of $\sL$. By the holomorphic anomaly equation \eqref{eq:hol3}, the exterior derivative is given by
\beq
\d  (f  \omega_\sL ) =  - \kappa \frac{\i  \eta(\tau)^3}{2 (\Im \tau)^{1/2}} \vev{Q}_\sN f(q) \d \tau \wedge \d \bar\tau
\eeq
Also notice that the boundary of the fundamental region (including the boundary at infinity) is given by
\beq
\partial F = -C_0+C_1 -C_2 + C_3, \label{eq:bdryF}
\eeq
where
\beq
C_0 &=\{ \tau ~|~ \tau= \tau_1 + \i \infty, ~ -1/2 \leq \tau_1 \leq 1/2\}, \nonumber \\
C_1 &=\{ \tau ~|~ |\tau|=1, ~ -1/2 \leq \Re \tau \leq 1/2\}, \nonumber \\
C_2 &= \{ \tau ~|~ \tau= -1/2 + \i \tau_2, ~ \sqrt{3}/2 \leq \tau_1 \leq \infty \},   \nonumber \\
C_3 &= \{ \tau ~|~ \tau= 1/2 + \i \tau_2, ~ \sqrt{3}/2 \leq \tau_1 \leq \infty \} , \label{eq:contours}
\eeq
and the orientations are given to them so that \eqref{eq:bdryF} holds (i.e.~the direction of increasing $\Re \tau$ for $C_0$ and $C_1$, and increasing $\Im \tau$ for $C_2$ and $C_3$),
see Fig.~\ref{fig:fundamental-region}.

Then, by the Stokes theorem, we get
\beq
- \int_{C_0} f \omega_\sL + \int_{C_1} f \omega_\sL -  \int_{C_2} f \omega_\sL + \int_{C_3} f \omega_\sL = - \kappa \int_F  \frac{\i  \eta(\tau)^3}{2 (\Im \tau)^{1/2}} \vev{Q}_\sN f(q) \d \tau \wedge \d \bar\tau.
\eeq
We only consider the case that $f \omega_\sL$ is invariant under $\tau \to \tau+1$ (i.e. only integral powers of $q$ appear in $f(q) \eta(\tau)^4 Z_\sL$). Then,
we have $\int_{C_2} f \omega_\sL = \int_{C_3} f \omega_\sL$. Moreover, from the property \eqref{eq:ZlimitI} of the APS index, we have
\beq
 \int_{C_0} f \omega_\sL = [f(q) \eta(\tau)^4 I_\sL]_{q^0}.
\eeq
Let us denote
\beq
\hat{A}_\sL  =  \int_{C_1} \omega_\sL ,  \qquad  \hat{A}_\sL(f)  =  \int_{C_1} f \omega_\sL  \label{eq:Ahatgenus}
\eeq
This is an analog of the $\hat A$-genus in the usual APS index theorem.
Then we get the index theorem
\beq
 [\eta(\tau)^4 I_\sL]_{q^0} &= \hat{A}_\sL + \eta_\sN, \nonumber \\
   [f(q) \eta(\tau)^4 I_\sL]_{q^0} &= \hat{A}_\sL(f) + \eta_\sN(f) .\label{eq:indextheorem}
\eeq
This is valid when $f\omega_\sL$ is invariant under $\tau \to \tau+1$.

We note that our definition of the spectral invariant \eqref{eq:eta-inv1} and the derivation of the associated index theorem~\eqref{eq:indextheorem}
can be considered as  a natural SQFT generalization of an analogous manipulation in SQM
to derive the standard APS index theorem, see Appendix~\ref{sec:SQMAPS}.

\paragraph{Corrections by zero modes.}
Let us next consider the more general case that $Q$ possibly has zero modes. For a theory $\sN= \eta(\tau)^{-\nu_\sN } \sum_{n} q^{n} \sN_{n}$, we denote
\beq
\dim \ker Q = \eta(\tau)^{-\nu_\sN } \sum_{n} q^{n} \dim \ker_{\sN_n} Q,
\eeq
where $\dim \ker_{\sN_n} Q$ is the dimension of the kernel of $Q$ on $\sN_n$. Then, we define the $\eta$-invariant by
\beq
\eta_\sN&= \kappa \int_F \d \tau \wedge \d \bar\tau \frac{\i  \eta(\tau)^3}{2 (\Im \tau)^{1/2}} \vev{Q}_\sN + \frac{1}{2} \kappa [\eta(\tau)^3 \dim \ker Q]_{q^0} ,  \nonumber \\ 
\eta_\sN(f) &= \kappa \int_F \d \tau \wedge \d \bar\tau \frac{\i  \eta(\tau)^3}{2 (\Im \tau)^{1/2}} \vev{Q}_\sN f(q) + \frac{1}{2} \kappa [f(q) \eta(\tau)^3 \dim \ker Q]_{q^0} .  
\label{eq:eta-inv}
\eeq
The correction term may be motivated as follows. Suppose that some nonzero eigenvalue $Q_0$ of $Q$ approaches zero when we vary parameters of $\sN$. The contribution of such a small eigenvalue $Q_0$ to $\vev{Q}_\sN$ is given by $q^{P_0} Q_0 |q|^{2Q_0^2}$, where $P_0$ is the eigenvalue of $P$ for the mode under consideration. For $|Q_0| \ll 1$, the integral in the definition of the $\eta$-invariant is dominated by the contribution from the region $\Im \tau \to \infty$ and we get
\beq
\kappa \int_F \d \tau \wedge \d \bar\tau \frac{\i  \eta(\tau)^3}{2 (\Im \tau)^{1/2}}  q^{P_0} Q_0 |q|^{2Q_0^2} f(q) &\sim \kappa  [\eta(\tau)^3 q^{P_0} f(q)]_{q^0} \int^{\infty} \frac{\d \tau_2}{\sqrt{\tau_2}} Q_0 e^{-4\pi \tau_2 Q_0^2} \nonumber \\
& \sim  \kappa  [\eta(\tau)^3 q^{P_0} f(q)]_{q^0}  \frac{Q_0}{2|Q_0|}, \label{eq:smallengenmode}
\eeq
where $\tau = \tau_1 + \i \tau_2$, and in the first equality we performed the integral over $\tau_1$ for somewhat large $\tau_2$. (Small $\tau_2$ region does not contribute in the limit $Q_0 \to 0$.)
Therefore, when $Q_0 \to +0$, this term becomes a contribution to the second term of \eqref{eq:eta-inv}. Notice that when $Q_0$ is exactly zero, it does not contribute to $\vev{Q}_\sN$ and hence its contribution is not included in the first term of \eqref{eq:eta-inv}. 

The discussion above also shows that when $Q_0$ goes from positive infinitesimal $+0$ to negative infinitesimal $-0$, the $\eta$-invariant jumps discontinuously. 
This jump is given by 
\beq
 \kappa  [\eta(\tau)^3 q^{P_0} f(q)]_{q^0} .
\eeq
$\kappa$ is chosen precisely so that this jump is an integer; for the values of $\nu_\sN$ with $\kappa=1/2$, the dimension of the eigenspace of $Q$ with eigenvalue $Q=Q_0$ is a multiple of 2.
Therefore, $\eta_\sN$ modulo $\bZ$ is a smooth function of the parameters of $\sN$. 

Now, let us consider how the index theorem is modified. 
When boundary $\partial \sL$ has zero modes, we claim that the APS index is modified as
\beq
I_\sL(\tau) =\kappa \left( \lim_{\substack{\bar\tau \to -\i  \infty \\ \tau\text{:\,fixed}}} Z_\sL(\tau, \bar\tau) +  \frac{1}{2} \eta(\tau)^{-1} \dim \ker Q\right). \label{eq:correctiontoAPS}
\eeq
A sketch of the reasoning is as follows. 
As discussed above, the $\eta$-invariant is continuous when a zero eigenvalue $Q_0=0$ is deformed to a small positive value $0< Q_0 \ll 1 $. On the other hand, it can be shown that the APS index is also continuous under such deformation, if it is appropriately defined.\footnote{The appropriate definition is as follows. We count states with $(-1)^F=+1$ which are not necessarily square normalizable but the wavefunctions are bounded. On the other hand, we count states with $(-1)^F=-1$ which are square normalizable. Under this definition, the index is unchanged if $Q$ is slightly deformed to $Q + \epsilon$ for positive infinitesimal $\epsilon>0$ as discussed around eq.~(3.17) of \cite{Yonekura:2022reu}. \label{footnote:generalAPS}
} In the index theorem \eqref{eq:indextheorem}, the $ \hat{A}_\sL(f) $ is defined by the integral over $C_1$ and hence $\tau$ moves only in a bounded region (in particular, it does not go to the region $\Im \tau \to \infty$). Therefore, the $ \hat{A}_\sL(f) $ is also continuous when $Q_0$ is changed. Then, by continuity, the index theorem \eqref{eq:indextheorem} should hold even when $Q_0=0$. For this equation to hold, we need the correction term in \eqref{eq:correctiontoAPS}. 

It would be interesting to derive \eqref{eq:correctiontoAPS} by a more direct analysis of $ Z_\sL(\tau, \bar\tau) $. 
At any rate, we claim that the index theorem \eqref{eq:indextheorem} is still valid even if $Q$ has zero modes. 
The subtleties about the zero modes of $Q$ are related to ``IR modes'' in the sense of the target space, and the corresponding points are well-known in the original APS index theorem for Dirac operators.

\paragraph{Continuity.}
Contrary to the case of the usual APS index theorem, we do not have a simple formula for the ``$\hat A$-genus'' $ \hat{A}_\sL$ for general $\sL$, and hence one might think that the index theorem \eqref{eq:indextheorem} is not so useful. However, $ \hat{A}_\sL$ actually has a few nice properties. One of them is continuity. As mentioned above,
it is defined by the integral of the partition function $Z_\sL$ on $C_1$, on which $\tau$ is finite. Hence, it is expected to be a smooth function of the parameters of the theory. 
On the other hand, the spectral quantity $\eta_\sN$ is not a smooth function in general as the result \eqref{eq:smallengenmode} clearly shows.
The APS index $I_\sL$ is also not smooth, but both $\eta_\sN$ and $I_\sL$ jump only by integers, so they are smooth modulo $\bZ$.
In summary,
\begin{itemize}
\item $\hat{A}_\sL$ is smooth but it depends on the entire bulk $\sL$.
\item $\eta_\sN$ depends only on the boundary $\sN$, but it is not smooth in general. It becomes smooth only after reduction modulo $\bZ$.
\end{itemize}

\subsection{The secondary invariant revisited}\label{sec:secondrevisit}

For the APS $\eta$ invariants of Dirac operators,
there are cases when the $\hat{A}$ contribution simply vanishes
and the APS $\eta$ invariants give rise to bordism invariants.
Similarly, there are cases when $\hat{A}_\sL$ vanishes from various reasons,
so that our spectral invariant $\eta_\sN$ gives rise to a bordism invariant when it is applied to SQFTs.
This allows us to give a direct definition of the secondary invariant using the spectral invariant.
Let us see how it goes.

When the theory $\sL$ is an SQFT rather than a formal Laurent series of SQMs, the partition function $ Z_\sL(\tau, \bar\tau) $ has a natural transformation property under  $S \in \SL(2,\bZ)$, at least on the circle $|\tau|=1$ as given in \eqref{eq:Ztransf}. 
Suppose also that we choose $f(q)$ such that the product $f(q) \eta(\tau)^4 Z_\sL(\tau, \bar\tau)$ has the same transformation law as modular forms of weight 2, at least on $|\tau|=1$. Then, the 1-form $f\omega_\sL = f(q) \eta(\tau)^4 Z_\sL \d \tau$ is invariant under the transformation $\tau \to -1/\tau$ on $|\tau|=1$.
On the other hand, the orientation of $C_1$ is reversed under $\tau \to -1/\tau$ and hence the integral of $f\omega$ over $C_1$ vanishes. Therefore, we get 
\beq
\hat{A}_\sL(f)=0. \label{eq:vanishingAhat}
\eeq

It is possible that $\hat{A}_\sL(f)$ vanishes by other reasons than modular invariance.  Recall that the partition function is given by
$Z_{\sL}= \Tr_{\sL} (-1)^F q^{H_L} \bar q^{H_R}$ (up to the regularization discussed in Appendix~\ref{sec:A}). 
Let us list various situations under which we get $\hat{A}_\sL(f)=0$:
\begin{itemize}
\item If $\nu_\sL \equiv 0 \mod 4$, we take $f(q)$ such that the product $f(q) \eta(\tau)^4 Z_\sL(\tau, \bar\tau)$ has the same transformation law as modular forms of weight 2 on $|\tau|=1$, as already mentioned.

\item If $\nu_\sL \equiv 2 \mod 4$, the time reversal symmetry satisfies $(-1)^FT = - T(-1)^F$ (see \eqref{eq:time1}) and hence the partition function is zero by the exact cancellation between $(-1)^F= +1$ and $(-1)^F=-1$.

\item If $\nu_\sL$ is odd, the fermion parity $(-1)^F$ can be defined only if we double the Hilbert space of $\sL$. This is just the direct sum of two copies of the smaller Hilbert space in which $(-1)^F$ is not defined, and the partition function is zero by the exact cancellation between $(-1)^F= +1$ and $(-1)^F=-1$ in the large Hilbert space.

\end{itemize}

In general, when  $\sN$ is an SQFT (rather than a formal Laurent series of SQMs) with $\nu_\sN=\nu$ and $f(q)$ is chosen such that \eqref{eq:vanishingAhat} is satisfied for any $\sL$ with $\nu_\sL=\nu+1$, the quantity $\eta_\sN(f) \mod \bZ$ is a bordism invariant:
\beq
\eta(f) : \SQFT_{\nu}({\rm pt}) \ni [\sN] \to  \eta_\sN(f)  \in \bR/\bZ. \label{eq:geometricsecondary}
\eeq
This can be seen as follows. Suppose we have two theories $\sN$ and $\sN'$ which are bordant, $\overline{\sN} \sqcup \sN' = \partial \sL$ where $\sL$ is an SQFT (rather than a formal Laurent series of SQMs). Then by the index theorem \eqref{eq:indextheorem} and the vanishing of $\hat{A}_\sL(f)$, we get
\beq
\eta_{\sN'}(f) - \eta_{\sN}(f) =  [f(q)\eta(\tau)^4 I_\sL]_{q^0} \in \bZ.
\eeq
This establishes the bordism invariance of $\eta_N(f)$ modulo $\bZ$.  

It is  straightforward to compute $\eta_{\sN}(f)$ for $\nu_\sN \notin -1+4\bZ$, with the result
\beq
\eta_{\sN}(f) \equiv \left\{ \begin{array}{ll}
\frac12 [f(q) \eta(\tau)^3 I_\sN]_{q^0} & \nu_\sN = 0, 4 \mod 8 \\
\frac12 [f(q) \eta(\tau)^3 I^{\rm mod\,2}_\sN ]_{q^0}  &  \nu_\sN = 1, 2 \mod 8 \\
0 & \nu_\sN = 5, 6 \mod 8 
\end{array}\right. \mod \bZ.
\eeq
In these cases the nonzero contributions come only from the second summand of \eqref{eq:eta-inv}.
Therefore, $\eta_{\sN}(f)$ can give new information in addition to  the primary invariant only in the case $\nu_\sN \in -1+4\bZ$.

The property \eqref{eq:vanishingAhat} may also be used for the description of the secondary invariant as follows. 
Consider an SQFT $\sN$, and suppose that $k$ copies of $\sN$ is a boundary of an SQFT $\sL$,
\beq
\sN^{\sqcup k} = \partial \sL.
\eeq
Then, the index theorem \eqref{eq:indextheorem} gives
\beq
\eta_\sN(f) = \frac{1}{k}  [f(q) \eta(\tau)^4 I_\sL]_{q^0} \equiv [f(q) \eta(\tau)^4 I^\text{2nd}_\sN]_{q^0}, \mod \bZ\label{eq:eta-to-2nd}
\eeq
where the second equality follows from the definition of the secondary invariant given in \eqref{eq:2ndary}. 

Let us check that the right hand side of \eqref{eq:eta-to-2nd} is  determined by the secondary invariant $ I^\text{2nd}_\sN$ as an equivalence class, rather than by its representative.
Recall that the combination $\eta(\tau)^{\nu_\sL} I^\text{2nd}_\sN$ is defined modulo $ \bZ((q)) $ and $  \MF_{\nu_\sL/2}\otimes \bQ $. 
If we change $ \eta(\tau)^{\nu_\sL} I^\text{2nd}_\sN$ by elements of  $ \bZ((q)) $, then $ [f(q) \eta(\tau)^4 I^\text{2nd}_\sN]_{q^0}$ changes by integers. 
On the other hand, if we change $ \eta(\tau)^{\nu_\sL} I^\text{2nd}_\sN$ by elements of $ \MF_{\nu_\sL/2}\otimes \bQ $, then $ [f(q) \eta(\tau)^4 I^\text{2nd}_\sN]_{q^0}$ does not change by the following reason. 
If $\nu_\sL \notin 4\bZ$, then $ \MF_{\nu_\sL/2} =\{0\}$. 
If $\nu_\sL =4m \in 4\bZ$, then take $\tilde g \in \MF_{2m}\otimes \bQ $ and set $g = \eta(\tau)^{-4m} \tilde g$. 
The prefactor $f$ was chosen precisely under the condition that $f(q) \eta(\tau)^4 g(q) $ is a modular form of weight 2, so its $q^0$-term is zero. 
This was what we wanted to show.

With a little more effort, we can show that $\eta_\sN(f)$ for various $f$ captures all the information of $I^\text{2nd}_\sN$ by varying $f$. If $\nu_\sN \notin -1+4\bZ$, this is obvious since there is no constraint on $f$ (other than that its coefficients are integers), and we can freely adjust it to detect any coefficient of $I^\text{2nd}_\sN$. Therefore, let us focus on the case that $\nu_\sN = 4m-1$. 
In this case, the prefactor $f$ needs to be chosen from the possibilities of the form 
\beq
f = \eta(\tau)^{-4+ 4m} \tilde f, \qquad \tilde f \in \MF_{2-2m}.
\eeq
Then, $\eta_\sN(f)$ would capture all the information of $I^\text{2nd}_\sN$
if the pairing
\beq
\MF_{2-2m}   \times \frac{\bZ((q))}{ \MF_{2m}    } \ni (\tilde f, \tilde h) \mapsto [\tilde f(q) \tilde h(q)]_{q^0} \in \bZ \label{eq:perfect}
\eeq
would be perfect.
Luckily, this perfectness is actually a known fact, see e.g.~\cite{Bunke,Tachikawa:2023lwf}. 
For self-contained-ness, a sketch of the derivation is given here.
Let $\Delta = \eta(\tau)^{24} \in \MF_{12}$ be the modular discriminant, and let $E_4 \in \MF_{4}$ and $E_6 \in \MF_{6}$ be the Eisenstein series of weight 4 and 6, respectively, normalized such that
\beq
E_4= 1 +240 \sum_{n \geq 1} \frac{n^3 q^n}{1-q^n}, \qquad E_6= 1-504 \sum_{n \geq 1} \frac{n^5 q^n}{1-q^n} \label{eq:Eisenstein}
\eeq
 They are related by $(E_4)^3-(E_6)^2=1728 \Delta$. Any element of $\MF_{2m}$ is a linear combination of elements of the form
 \beq
 E_4^a E_6^b \Delta^c, \qquad   (a, b,c) \in (\bZ_{\geq 0}, \{0,1\}, \bZ), ~~ m = 2a +3b +6c.
 \eeq
Notice that $\Delta=q+\cdots$, $E_4=1+\cdots$, and $E_6=1+\cdots$. Let $c_\text{max}^{2m} $ be the maximal possible value of $c$ in $\MF_{2m}$. Then, in $\MF_{2m}$, one can realize any value of the coefficients of $q^n$ in the range $n \leq c_\text{max}^{2m}$. Therefore, it is possible to take a representative of any given $\tilde h \in \bZ((q))/\MF_{2m}$ by using the freedom of adding elements of $\MF_{2m}$ so that
\beq
\tilde h = A q^{d  } + \cdots, \qquad d  \geq c_\text{max}^{2m} +1.
\eeq
If $ -d \leq c_\text{max}^{2-2m} $, it is possible to find $\tilde f \in \MF_{2-2m}$ such that
\beq
\tilde f  = q^{-d} + \cdots 
\eeq
to detect the coefficient $ [\tilde f(q) \tilde h(q)]_{q^0} =A$. 
It is straightforward to check that $ c_\text{max}^{2m} + c_\text{max}^{2-2m} =-1$ by using the fact that 
\beq
c_\text{max}^{12\ell}=c_\text{max}^{12\ell+4}=c_\text{max}^{12\ell+6}=c_\text{max}^{12\ell+8}=c_\text{max}^{12\ell+10}=\ell, \qquad c_\text{max}^{12\ell+2} = \ell-1.
\eeq
Therefore, $ d  \geq c_\text{max}^{2m} +1 \Longleftrightarrow  -d \leq c_\text{max}^{2-2m} $ and hence we can detect any information of $\tilde h$ by appropriately choosing $\tilde f$.
Conversely, any information of $\tilde f$ can be detected by appropriately choosing $\tilde h$. Thus the pairing \eqref{eq:perfect} is perfect. 
From the perfectness of \eqref{eq:perfect}, we find that we can detect any information of $I^\text{2nd}_\sN$ by $\eta_\sN(f)$ for varying $f$. 

A basic example (with $f=1$) is the case that $\sN$ is the sigma model on $S^3$ with the unit $H$-flux, denoted by $S^3_{H=1}$. In this case, we take $\sL$ to be a K3 surface with 24 holes as in Section~\ref{subsec:6}. 
The equation $\d H = -\lambda(R)$ (where $\lambda(R)$ is one-half of the first Pontryagin class) and the fact that $\int \lambda(R)=-24$ on K3 implies that 24 copies of $S^3_{H=1}$ can arise as the boundary of a K3 with 24 holes. 
The index of the Dirac operator on K3 is $2$, and because of the factor $\kappa=1/2$, we get  $[\eta(\tau)^4 I_\sL]_{q^0} = 2\kappa =1$. Therefore, we get
\beq
\eta_{\sN}= \frac{1}{24} \quad \text{when~} \sN = S^3_{H=1}. \label{eq:S3value}
\eeq

\subsection{The spectral pairing}\label{sec:the-pairing}
We now define a spectral pairing between elements of $\SQFT_{d}$ and $\SQ_{-21+d}$, where $\SQ_{-21+d}$ is the set of formal Laurent series of SQMs whose boundary is an SQFT, defined in \eqref{eq:SQM/SQFT}. Consider
\beq
&\sX \in \SQFT_d, \nonumber \\
&\sS \in \SQ_{-21-d} \text{~with~} \partial \sS = \sT \in \SQFT_{-22-d}.
\eeq
We will define a quantity $\eta(\sX,\sS)$ from the spectral data of the associated systems. 
The assumptions we will use here are basically the same as in Section~\ref{sec:new}:
\begin{enumerate}
\item $\SQFT_{-21}({\rm pt}) =0$.
\item For $\sT =\partial \sS$,  we have $[\sX][\sT] =0 \in \SQFT_{-22}({\rm pt}) $.
\end{enumerate}
The second assumption may need explanation.
Notice that the primary invariants (i.e. the usual and mod-2 elliptic genera) have straightforward counterparts in SQMs with time-reversal symmetry. (The time-reversal symmetry is necessary to define the mod-2 indices.) Since $\sT$ is a boundary, its bordism class in SQMs is trivial and hence its usual and mod-2 indices are zero. 
Under the same assumptions as in Section~\ref{sec:new} (i.e. elements of $\SQFT_{-22}({\rm pt}) $ are captured entirely by the primary invariants), the product $\sX \times \sT$ is a boundary of some SQFT. 

\paragraph{Definition.}
Choose an $\sN$ such that
\beq
\partial \sN = \sX \times \sT.
\eeq
Notice also that the product $\sX \times \sS$ has the boundary $\partial(\sX \times \sS) = (-1)^d \sX \times \sT$ where the sign $(-1)^d$ comes from the orientation. 
We glue $\sX \times \sS$ and $\sN$ to get 
$
\sN \cup (-1)^{d-1} \sX \times \sS.
$
This has no boundary, $\partial   (\sN \cup (-1)^{d-1} \sX \times \sS) = \varnothing$.

We define the pairing of $\sX$ and $\sS$ as
\beq
\eta(\sX, \sS) = \eta_{ \sN \cup (-1)^{d-1} \sX \times \sS } \in \bR/\bZ. \label{eq:etaXS}
\eeq
This value is considered only modulo $\bZ$. This is because the $\eta$-invariant is a smooth function of the parameters of the theories only after the reduction modulo $\bZ$ as discussed before. 
We emphasize that this is a spectral (rather than topological) quantity in general. 

\paragraph{Well-definedness.}
We need to show that this definition does not depend on the choice of $\sN$. For this purpose, we use the gluing law of the $\eta$-invariant. Let $\sN'$ be another theory with $\partial \sN' = \partial \sN$. We can consider a compact theory $\sN' \cup \overline{\sN}$. If either $\sN$ or $\sN'$ is generic enough, we expect that there is no zero mode of $Q$ because $\SQFT_{-21}({\rm pt}) =0$ and hence all indices are zero. Then we need not care about the corrections by zero modes in the definition of the $\eta$-invariant \eqref{eq:eta-inv}.
In this case, we claim that the gluing law is given by
\beq
\eta_{\sN' \cup \overline{\sN} } = \eta_{\sN' } -\eta_{{\sN} } , \label{eq: gluing-eta}
\eeq
where $\eta_\sN$ and $\eta_{\sN'}$ are defined as follows. For $\sN$ which has boundary, the definition of $\eta_{{\sN} }$ is given by introducing a certain regularization as in Appendix~\ref{sec:A}. 
In more detail, recall that the $\eta$-invariant is defined by using $\vev{Q}_\sN$. To define it, we may insert a regularization $F(X-a)$ in the path integral, where $X$ is the position operator discussed around \eqref{eq:position}, $F$ is a function such that $F(x) =1$ for $x<-1$ and $F(x) = 0$ for $x >1$, and $a (\to \infty)$ is a large enough number for regularization. 
Now, consider the combined theory $\sN'\cup\overline{\sN}$ as in Figure~\ref{fig:6},
and suppose that $\sN'$ is located around $X \sim -a$ and $\overline{\sN}$ is located around $X \sim +a$. 
Then, we use $1= F(X) + (1-F(X))$ to get \eqref{eq: gluing-eta}. 

At any rate, we assume \eqref{eq: gluing-eta} which is an analog of (a part of) the Dai-Freed theorem \eqref{eq:DaiFreed}. 
Now, notice that $\sN' \cup \overline{\sN}$ is an element of $\SQFT_{-21}$. 
Because $-21 \in 3+24\bZ$, the spectral invariant $\eta $ modulo $ \bZ$ gives a bordism invariant as shown around \eqref{eq:geometricsecondary}, where we take $f=1$ and $\nu=-21$. 
Because of the assumption that $\SQFT_{-21}({\rm pt}) =0$, it must be trivial and hence $ \eta_{\sN' \cup \overline{\sN}} \in \bZ$.

Now, by the same gluing law applied to the definition \eqref{eq:etaXS},
\beq
\eta(\sX, \sS) = \eta_{{\sN} } +(-1)^{d-1} \eta_{\sX \times \sS}.
\eeq
If we use $\sN'$ instead of $\sN$ in the definition of $\eta(\sX, \sS) $, the difference of the two definitions is given by
$
\eta_{ \sN' }  - \eta_{\sN }  = \eta_{\sN' \cup \overline{\sN}},
$
which we already argued to be an integer.
This establishes the well-definedness of $\eta(\sX, \sS)  \in \bR/\bZ$ under our assumptions.

\subsection{General structure of the spectral pairing}\label{sec:generalstr}

Now we want to discuss properties of $\eta(\sX, \sS)$. Before doing so, 
we first would like to put it in a wider context.

\subsubsection{The case of invertible field theories}
Invertible field theories~\cite{Freed:2004yc} form an important subclass of quantum field theory (see e.g. \cite{Kapustin:2014tfa, Kapustin:2014dxa,Freed:2016rqq,Yonekura:2018ufj}).
The partition function $Z_{S}(X)$ of an invertible field theory $S$ on a manifold $X$ of dimension $d$ with some geometric structure has the following properties:
\begin{enumerate}
\item $Z_{S}(X)$ is a pure phase, $Z_{S}(X) \in \U(1)$.\footnote{More precisely, we add topologically less interesting counterterms to make $|Z_S(X)|=1$. } 
In other words, 
\beq
\frac{1}{2\pi \i} \log Z_{S}(X) \in \bR/\bZ. \label{eq:pro1}
\eeq
\item If $X$ is a boundary $X=\partial Y$ of a manifold $Y$ of dimension $d+1$ with the same geometric structure, then 
\beq
\frac{1}{2\pi \i} \log Z_{S}(X) \equiv  \int_Y \cI_S \mod \bZ , \label{eq:pro2}
\eeq
where $\cI_S$ is some characteristic class. 
In particular, $\int_Y \cI_S$ is a smooth function,
\beq
\int_Y \cI_S \in \bR \text{~:~smooth function of $Y$ and $S$.} \label{eq:pro3}
\eeq

\item When $Y$ has no boundary or in other words $X$ is empty, $\partial Y = \varnothing$, then
\beq
\int_Y \cI_S \in \bZ.  \label{eq:pro4}
\eeq
This is required by \eqref{eq:pro1}, \eqref{eq:pro2} and $Z_S(\varnothing) =1$.
\end{enumerate}
We will show that $\eta(\sX, \sS)$ has analogous properties, by replacing $S \to \sS$, $X \to \sX$, $\frac{1}{2\pi \i} \log Z_{S}(X) \to \eta(\sX, \sS)$, and $\int_Y \cI_S \to \hat{A}(\sY,\sS)$ which will be introduced later.

In the case of invertible field theories, the  structures above are related to a differential extension of the Anderson dual of bordism groups~\cite{Yamashita:2021cao}. (The discussion until the end of the subsubsection may be skipped by readers who are not familiar with mathematics related to invertible field theories and TMF.) 
Let $\Omega^H$ be the bordism homology of manifolds with geometric structure under consideration (e.g.~$H=\spin$). The classification of invertible field theories  up to continuous deformation is given by its Anderson dual cohomology theory $I_\bZ \Omega^H$,
as originally put forward in \cite{Freed:2016rqq}.
In more detail, if we have some sigma model target space $W$, then the relevant group is $(I_\bZ \Omega^{H})^{d+1}(W)$. An invertible field theory $S$ itself (rather than its deformation class) gives an element of its differential extension $\widehat {(I_\bZ \Omega^{H}) }{}^{d+1}(W)$ \cite{Yamashita:2021cao}. 
In our context, the theory $\sS$ plays the role of an element of a differential cohomology group.
In the case of TMF, there is a result on the Anderson dual of ${\rm TMF}$, which says
$
 {\rm KO}((q))/{\rm TMF}  \simeq \Sigma^{-20} I_{\bZ}{\rm TMF}
$~\cite{Tachikawa:2023lwf}.
In particular, this relation implies\footnote{For a generalized cohomology theory $E$, we have $(\Sigma^m E)_n({\rm pt}) =\pi_n (\Sigma^m E) =\pi_{n-m}(E)= E_{n-m}({\rm pt}) = E^{-n+m}({\rm pt})$. }
\beq
(I_{\bZ}{\rm TMF})^{d+1}({\rm pt}) = ( {\rm KO}((q))/{\rm TMF})_{-21-d}({\rm pt}).  \label{eq:Andersondual}
\eeq
In the context of SQFT, $ ( {\rm KO}((q))/{\rm TMF})_{-21-d}({\rm pt})$ corresponds to $\SQ_{-21-d}({\rm pt})$, where $\SQwp$ was introduced in \eqref{eq:SQM/SQFT} and the notation $({\rm pt})$ means that we take bordism classes of theories. 
Our (somewhat physically artificial) definition of $\SQwp$ is motivated by the relation to ${\rm KO}((q))/{\rm TMF}$. 
Then, we regard the bordism class $[\sS]$ of $\sS$  as an element 
\beq
[\sS] \in \SQ_{-21-d}({\rm pt}) . 
\eeq
Now, the Anderson dual $ I_{\bZ} E$ of  any spectrum $E$ may be regarded as the universal place 
containing all the things which have nice pairings with $E$. 
For example, a physically concrete realization of this idea in the case $E=\Omega^H$ is the pairing between manifolds with geometric structure and invertible field theories as elements of differential cohomology groups $\widehat {(I_\bZ \Omega^{H}) }{}^{d+1}$. (See also \cite{Yamashita:2021fkd} for its relation to more general cohomology theories.)
Therefore, the spectral pairing 
\beq
\eta(-,-) \colon  \SQFT_d \times \SQ_{-21-d}   \ni (\sX, \sS) \mapsto \eta(\sX,\sS) \in \bR/\bZ.
\eeq
with properties analogous to \eqref{eq:pro1}--\eqref{eq:pro4} might give a map
\beq
[\eta] \colon \SQ_{-21-d}({\rm pt}) \xrightarrow{}  \text{``}(I_{\bZ} \SQFT)^{d+1}({\rm pt}) \text{''},
\eeq
although the rightmost group is not yet defined.\footnote{As remarked in footnote~\ref{footnote:spectrum}, we are avoiding to assume that $\SQFT$, as naively and loosely defined in the present paper, forms an $\Omega$-spectrum.}
The relation of $\SQFT$ to $\TMF$ suggests that this map is an isomorphism,
but we do not have a direct way to confirm it yet.

\subsubsection{The case of our spectral pairing}
Now we are going to show some properties related to the discussions above. Let us recall the basic objects involved in the definition of $\eta(\sX, \sS)$:
\beq
&\sX \in \SQFT_d, \quad \sS \in \SQ_{-21-d}, \quad \sT = \partial \sS \in \SQFT_{-22-d}, \nonumber \\
& \sN~\text{ such that }~ \partial \sN = \sX \times \sT.
\eeq
We are going to discuss the statements which are analogues to \eqref{eq:pro1}--\eqref{eq:pro4} in turn.

We have already shown that $\eta(\sX,\sS) = \eta_{ \sN \cup (-1)^{d-1} \sX \times \sS}$ is well-defined as an element of $\bR/\bZ$.
 This property should be compared with \eqref{eq:pro1}.

\begin{figure}
\centering

\begin{tikzpicture}[scale=2]
  \def\r{1.5}    
  \def\gap{0.65}  

  \draw[thick] (-\r,0) arc (180:270:\r);
  \draw[red,very thick]    (-\r,0) -- (0,0);
  \draw[blue,very thick]   (0,0)   -- (0,-\r);

  \draw[thick] ({\gap+\r},0) arc (0:-90:\r);
  \draw[green!70!black,very thick] (\gap,0)     -- ({\gap+\r},0);
  \draw[blue,very thick]   (\gap,0) -- (\gap,-\r);

  \node at (-0.5*\r,   -0.5*\r)        {$ {\mathsf L}$};
  \node at ({\gap+0.5*\r}, -0.5*\r)    {$ {\mathsf L}_1 = \mathsf{Y} \times \mathsf{S}$};
  \node at (-0.5*\r,   0.15)           {$\mathsf N $};
  \node at ({\gap+0.5*\r}, 0.15)       {$\mathsf{X} \times \mathsf{S}$};
  \node at ({\gap/2},   -0.5*\r)       {$ \mathsf{Y} \times \mathsf{T}$};
\end{tikzpicture}

\caption{Theories with corners and their boundaries. This figure may be compared with Figure~~\ref{fig:1}. 
Here we did not carefully specified the orientations of various components, to reduce clutter.\label{fig:7}}
\end{figure}
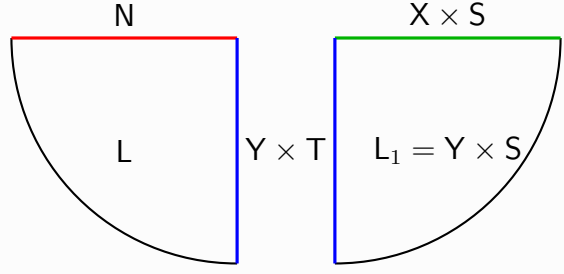

Next, suppose that $\sX$ is a boundary of some SQFT $\sY$, 
\beq
\partial \sY = \sX. 
\eeq
We can glue $\sN$ (whose boundary is $\sX \times \sT$) and the orientation reversal of $\sY \times \sT$ along their common boundary. By $\SQFT_{-21}({\rm pt}) =0$, we can take an $ \mathsf{L} $ such that
\beq
\partial  \mathsf{L} = \sN \cup (\overline{\sY \times \sT} ).
\eeq
We can also consider $\sY \times \sS$, whose boundary is 
\beq
\partial (\sY \times \sS) = \left( \sX \times \sS \right) \cup \left( (-1)^{d+1} \sY \times \sT \right).
\eeq
We can think of $\sY \times \sS$ as a theory with a corner. We can further glue $ {\sL}$ and $(-1)^{d+1}\sY \times \sS$ along $\sY \times \sT$. We denote
\beq
& {\sL}_1 = (-1)^{d+1}\sY \times \sS, \nonumber \\
& {\sL} \cup_0  {\sL}_1 ~\text{ such that }~ \partial( {\sL} \cup_0  {\sL}_1 ) =  \sN \cup  \left( (-1)^{d+1} \sX \times \sS \right)  .
\eeq
See Figure~\ref{fig:7} for the current setup. (This figure may be compared with Figure~~\ref{fig:1}.)

By using the index theorem \eqref{eq:indextheorem}, we get
\beq
 \eta_{ \sN \cup (-1)^{d-1} \sX \times \sS} \equiv -\hat{A}_{ {\sL} \cup_0  {\sL}_1 } \mod \bZ.
\eeq
Let us study $\hat{A}_{ {\sL} \cup_0  {\sL}_1 }$ in more detail. Its definition is
\beq
\hat{A}_{ {\sL} \cup_0  {\sL}_1 } = \frac12  \int_{C_1} \eta(\tau)^4 Z_{ {\sL} \cup_0  {\sL}_1} \d \tau
\eeq
where $C_1$ is defined in \eqref{eq:contours}, and $\frac12 =\kappa$ for the theory $ {\sL} \cup_0  {\sL}_1 $. The regularization of partition functions in Appendix~\ref{sec:A} was such that 
\beq
Z_{ {\sL} \cup_0  {\sL}_1} = Z_{ {\sL} } + Z_{  {\sL}_1}  = Z_{ {\sL} } +(-1)^{d+1}  Z_{  \sY}   Z_{\sS}.
\eeq
The theory $ {\sL}$ is an SQFT and hence by \eqref{eq:vanishingAhat} its contribution to $\hat{A}_{ {\sL} \cup_0  {\sL}_1 }$ vanishes. Let us define
\beq
\hat{A}(\sY, \sS) := (-1)^{d+1} \frac12   \int_{C_1}  \eta(\tau)^4  Z_{  \sY}   Z_{\sS} \d\tau \in \bR.  
\eeq
It is valued in $\bR$, and is expected to be a smooth function of parameters of $\sY$ and $\sS$. This property should be compared with \eqref{eq:pro3}.
We get
\beq
\eta(\sX, \sS) \equiv  -\hat{A}(\sY, \sS)  \mod \bZ \quad (\sX = \partial \sY). 
\eeq
This equation should be compared with \eqref{eq:pro2}.

Finally, when $\partial \sY=\varnothing$, the quantity $Z_{  \sY} $ is the elliptic genus up to a factor $\kappa_\sY$, $Z_\sY = \kappa_Y^{-1} I_{\sY}$. 
Then we get
\beq
\hat{A}(\sY, \sS) = (-1)^{d+1} \frac{1}{2\kappa_\sY}   \int_{C_1}  \eta(\tau)^4  I_{  \sY}   Z_{\sS} \d\tau  
 =  (-1)^{d+1} \frac{1}{2\kappa_\sY \kappa_\sS}  \hat{A}_\sS(I_\sY)
\eeq
where the notation $\hat{A}_{\sL}(f)$ was introduced in \eqref{eq:Ahatgenus}, and $\kappa_\sS$ is the value of $\kappa$ for $\sS$. 
By the index theorem, 
\beq
 \hat{A}_\sS(I_\sY) =  -   {\eta}_\sT(I_\sY)  + [ \eta(\tau)^4 I_\sY I_{\sS}]_{q^0} 
 = -   \frac{\kappa_{\sY} \kappa_{\sS}}{\kappa_{\sY\times \sS}}  \eta_{\sY \times \sT}  + [ \eta(\tau)^4 I_\sY I_{\sS}]_{q^0}.
\eeq
Notice that $ \eta_{\sY \times \sT}$ modulo $\bZ$ gives a bordism invariant of the theory $\sY \times \sT  \in \SQFT_{-21}$. This must be trivial since $\SQFT_{-21}({\rm pt})=0$, and hence
\beq
 \eta_{\sY \times \sT}  \in \bZ.
\eeq
Whenever $I_{\sY}$ is nonzero, one can check that $d+1 \in 4\bZ$ and $\kappa_\sY \kappa_\sS=\kappa_{\sY\times \sS}$.
Combining these results, we get
\beq
\hat{A}(\sY, \sS) =   -\eta_{\sY \times \sT} +  [ \eta(\tau)^4 I_\sY I_{\sS}]_{q^0}   \in \bZ \qquad (\partial \sY = \varnothing). 
\eeq
This result should be compared with \eqref{eq:pro4}. Of course, this is as it should be since $\hat{A}(\sY, \sS) \equiv -\eta(\sX=\varnothing, \sS)=0 \mod \bZ$.

In summary, we have found the following:
\begin{enumerate}
\item 
Analogous to \eqref{eq:pro1}, we have
\beq
\eta(\sX,\sS) \in \bR/\bZ \label{eq:prop1}.
\eeq

\item
Analogous to \eqref{eq:pro2}, we have
\beq
\eta(\sX, \sS) \equiv  -\hat{A}(\sY, \sS)  \mod \bZ \quad (\sX = \partial \sY), \label{eq:prop2}
\eeq
where, analogous to \eqref{eq:pro3},
\beq
\hat{A}(\sY, \sS) = (-1)^{d+1} \frac12   \int_{C_1}  \eta(\tau)^4  Z_{  \sY}   Z_{\sS} \d\tau   \in \bR. \label{eq:prop3}
\eeq

\item
Analogous to \eqref{eq:pro4}, we have
\beq
\hat{A}(\sY, \sS) =   -\eta_{\sY \times \sT} +  [ \eta(\tau)^4 I_\sY I_{\sS}]_{q^0}   \in \bZ \qquad (\partial \sY = \varnothing). \label{eq:prop4}
\eeq

\end{enumerate}

There is one more general comment related to the result of TMF given in \eqref{eq:Andersondual}. When $\sS$ is an SQFT (rather than a formal power series of SQMs), the invariant $\eta(\sX, \sS)$ is trivial. The reason is that in this case we have $ \sN \cup (-1)^{d-1} \sX \times \sS \in \SQFT_{-21}$ and the assumption $\SQFT_{-21}({\rm pt})=0$ implies that $\eta_{ \sN \cup (-1)^{d-1} \sX \times \sS} \equiv 0 \mod \bZ$. This is an expected result from the mathematical properties of $ {\rm KO}((q))/{\rm TMF}$.

\subsection{Descriptions of other invariants in term of the spectral pairing}\label{sec:inv1}

A subset of elements $\sS\in \SQ_\nu$ is given by theories such that $\partial \sS = \sT = \varnothing$. For a while, let us focus our attention to this case. 
Our interest in this subsection is to study what invariants of $\sX$ can be detected by varying $\sS$ such that $\partial \sS = \varnothing$.
In the following discussions, it is convenient to notice the following facts about SQMs with time-reversal symmetry:
\begin{itemize}
\item The index $I_\sS$ can be nonzero only if $\nu_\sS \in 4\bZ$.
\item The mod-2 index $I^{\rm mod\,2}_\sS$ can be nonzero only if $\nu_\sS \in 1+8\bZ$ or $\in 2+8\bZ$. It is given by
\beq
  I_\sS^{\rm mod\,2} \equiv \kappa_{\sS}\dim \ker_{\sS} Q  \mod 2, \label{eq:formulaformod2}
\eeq
where $\kappa_\sS=1$ for $\nu_\sS \in 1+8\bZ$ and $\kappa_\sS=1/2$ for  $\nu_\sS \in 2+8\bZ$.
\item $\vev{Q}_\sS$ can be nonzero only if $\nu_\sS \in 3+4\bZ$.
\item Under the above restrictions, any values of $I_\sS$, $I^{\rm mod\,2}_\sS$, and $\eta_\sS$ can be realized in SQMs.\footnote{For $I_\sS$ and $I^{\rm mod\,2}_\sS$, we can just take finite-dimensional Hilbert spaces on which $Q$ is zero. For $\eta_\sS$, we may freely vary the eigenvalue spectrum of $Q$ to realize any value for $\eta_\sS$. }
\end{itemize}

\paragraph{Elliptic genus.}
Consider the case $d+1 \in 4\bZ$ or equivalently $\nu_\sS \in 4\bZ$.
When $\partial \sY = \varnothing$ as well as $\partial \sS =\varnothing$, Eq.~\eqref{eq:prop4} gives 
\beq
\hat{A}(\sY, \sS) =    [ \eta(\tau)^4 I_\sY I_{\sS}]_{q^0} . \label{eq:integerpairing}
\eeq
In SQMs, there is no restriction on $I_{\sS}$; it can take arbitrary values when $\nu_{\sS}  \in 4\bZ$. Therefore, by varying $\sS$, we can detect the elliptic genus of $\sY \in \SQFT_{d+1}$ completely. 

We remark that the above pairing automatically vanishes when $I_{\sS}$ is given in terms of modular forms as
\beq
I_{\sS} \in \eta(\tau)^{21+d} {\rm MF}_{-(21+d)/2}, \label{eq:Sinmodular}
\eeq
where ${\rm MF}_k$ is the set of (weakly holomorphic) modular forms of weight $k$. In this case, $ \eta(\tau)^4 I_\sY I_{\sS}$ is a modular form of weight $2$ and its $q^0$ component is zero. 
Therefore, what matters in extracting the elliptic genus of $\sX$
is $I_\sS \in \eta(\tau)^{21+d}\bZ((q))$ modulo $\eta(\tau)^{21+d}\MF_{-(21+d)/2}$.

\paragraph{Secondary invariant.} 
Let us still consider the case $d+1 \in 4\bZ$ (or $\nu_\sS \in 4\bZ$), but now suppose that $\partial \sY = \sX \neq \varnothing$ and also that $\sS$ satisfies \eqref{eq:Sinmodular}. We want to see what invariant of $\sX$ (rather than $\sY$) can be detected by such an $\sS$.

The condition $\partial \sS = \varnothing $ means that we can take $\sN = \varnothing$ in the definition of $\eta(\sX, \sS)$. Thus, $ \sN \cup (-1)^{d-1} \sX \times \sS =  \sX \times \sS$ and we get
\beq
\eta(\sX, \sS) =   \eta_{\sX \times \sS} = \eta_{\sX}(I_\sS) =  [ I_\sS \eta(\tau)^4 I^\text{2nd}_\sX]_{q^0}, \label{eq:secondaryinvariant5}
\eeq
where the second equality can be checked from the definition of the $\eta$, and the last equality comes from \eqref{eq:eta-to-2nd}.
The condition \eqref{eq:Sinmodular} implies that the map $\eta(I_\sS) : \sX \mapsto \eta_{\sX}(I_\sS)$ gives the bordism invariant \eqref{eq:geometricsecondary}. Therefore, it can detect the secondary invariant of $\sX$ as explained at the end of Section~\ref{sec:secondrevisit}.

Another interesting point is that if $\sS$ is an SQFT (rather than a formal Laurent series of SQMs), the invariant must vanish as remarked in the previous section. 
This property was already checked directly when we discussed the secondary invariant
in Sec.~\ref{app:BN}, see  Properties~\ref{prop:2} and \ref{prop:3} there.

For example, consider $\sS \in \SQFT_{-24}$ whose elliptic genus $I_\sS$ is given by 
\begin{equation}
I_\sS = a + \cdots. \label{-24}
\end{equation}
By taking $\sX=S^3_{H=1}$, we get  $\eta_{\sX}(I_\sS)=a/24$. This must be trivial and hence $a \in 24\bZ$. 
This consideration will be extensively generalized in the next subsection, Section~\ref{sec:const}.

\paragraph{Continuous theta terms.}
Next let us consider the case that $d \in 4\bZ$ and hence $\nu_\sS \in -1+4\bZ$. In this case, we have
\beq
\eta(\sX, \sS) =   \eta_{\sS}(I_\sX). \label{eq:thetaterm}
\eeq
It depends only on $I_\sX$. 
This is an analog of continuous $\theta$-terms. For instance, on a 4-manifold $X$ we can introduce the usual $\theta$-term $ \i \theta \int_X \frac{1}{48} p_1(R)$,
and a more trivial example is to consider a 0-manifold and introduce a $\theta$-term $\i \theta$.
The role of the $\theta$ parameter is played by $\eta_\sS$ here.

In the case of invertible field theories, such $\theta$-terms are elements of differential cohomology groups which are trivial after taking the deformation class. 
However, we remark that if we reverse the roles of $\sX$ and $\sS$, and consider a theory $\sU$ such that $\partial \sU=\sS$, then it leads to a nontrivial pairing between $\sU$ and $\sX$ (when $\sS =\varnothing$) that is analogous to \eqref{eq:integerpairing}.

\paragraph{Mod-2 elliptic genus.} 

Finally, suppose that $d $ is neither in $-1+4\bZ$ nor $4\bZ$. Then $\sS$ is such that $I_\sS$ and $\vev{Q}_\sS$ are zero. In this case,  the definition of our $\eta$ invariant \eqref{eq:eta-inv} gives
\beq
\eta(\sX, \sS) =  \frac{1}{4}  [\eta(\tau)^3 (\dim \ker_{\sX} Q )(  \dim \ker_{\sS} Q)]_{q^0}
\eeq
where we have used $\dim \ker_{\sX \times \sS} Q =  (\dim \ker_{\sX} Q )(  \dim \ker_{\sS} Q)$ and $\kappa=1/2$ for $\sX \times \sS$.
The only cases in which we expect both $\dim \ker_{\sX} Q$ and $\dim \ker_{\sS} Q$ to be nontrivial are 
\beq
& d \in 1+8\bZ \qquad (\text{i.e.}\quad  \nu_\sX \in  1+8\bZ, \quad  \nu_\sS \in  2+8\bZ) \quad \text{or} \nonumber \\
& d \in 2+8\bZ \qquad (\text{i.e.}\quad   \nu_\sX \in  2+8\bZ, \quad  \nu_\sS \in  1+8\bZ) .
\eeq
In both cases, we have
\beq
\eta(\sX, \sS) =  \frac{1}{2}  [\eta(\tau)^3  I_\sX^{\rm mod\,2}  I_\sS^{\rm mod\,2}]_{q^0}, \label{eq:mod2paring}
\eeq
where we have used $\kappa_\sX\kappa_\sS=1/2$.
This gives a pairing between mod-2 indices. 
There is no restriction on $ I_\sS^{\rm mod\,2}$, so by varying $\sS$ we can detect any mod-2 elliptic genus of $\sX$. 

This is a generalization of the following mod-2 pairing of mod-2 indices of Dirac operators using the $\eta$-invariant.
Let $X$ and $S$ be two spin manifolds such that $\dim X\in 1+8\bZ$ and $\dim S\in 2+8\bZ$.
Then we can consider mod-2 indices of Dirac operators \begin{equation}
I_X^\text{mod 2}:=\dim \cD(X) \in \bZ_2, \qquad
I_S^\text{mod 2}:=\frac12\dim \cD(S) \in \bZ_2.
\end{equation} The product manifold $X\times S$ has dimension $\dim X\times S\in 3+8\bZ$,
and the $\eta$-invariant in this dimension continuously varies in general, but if restricted to the product form $X\times S$, we have the formula 
\begin{equation}
\eta(X\times S)=\frac12 I_X^\text{mod 2} I_S^\text{mod 2} \in \bR/\bZ.
\end{equation}
For example, when $X=S^1$ and $S=T^2$ are both given periodic spin structure,
then $X\times S$ is simply $T^3$ with periodic spin structure.
$S^1$ and $S^2$ both have nontrivial mod-2 indices, and $\eta(T^3)=1/2$.
Our formula above is a generalization of this more elementary pairing.

Coming back to our pairing \eqref{eq:mod2paring},
we note that it vanishes when $\sS$ is an SQFT, rather than a formal Laurent series of SQMs.
Indeed, if otherwise, $\eta(\sX,\sS)$ being nonzero means that there is a nontrivial bordism invariant
detecting $\SQFT_{-21}(\pt)$, which we assumed to be trivial.
The corresponding statement in $\TMF$ can be shown by using detailed properties of $\TMF$~\cite{BrunerRognes,Tachikawa:2023nne} and more precise results will be mentioned in Section~\ref{sec:const}.

For example, let $\sS$ be the sigma model on $T^2$ with periodic spin structure.
Then \begin{equation}
I_\sS^\text{mod 2} = \eta(\tau)^{-2}(1+O(q)).
\end{equation} This means that, if $\sX\in \SQFT_{-23}$  has the mod-2 elliptic genus \begin{equation}
I_\sX^\text{mod 2} = \eta(\tau)^{23}(aq^{-1}+O(1)),
\end{equation}
we have $a\equiv 0$  mod 2.
This restriction is the mod-2 analogue of the restriction of $a$ in \eqref{-24} for $\sX\in \SQFT_{-24}$
which arose from the pairing with $\sS=S^3_{H=1}\in \SQFT_3$.

\subsection{Constraints on the primary and secondary invariants}\label{sec:const}

It is worth emphasizing that the pairing $\eta(\sX, \sS)$ gives interesting constraints on SQFTs \cite{Johnson-Freyd:2024rxr}.
Our basic assumptions are the same as the ones used in the definition of $\eta(\sX, \sS)$ in Section~\ref{sec:the-pairing}, and we derive the constraints under those assumptions. 

As already mentioned, the pairing vanishes if $\sS$ is an SQFT (rather than a formal power series of SQMs),
\beq
\eta(\sX, \sS) =0 \in \bR/\bZ \qquad \text{if $\sS$ is an SQFT}. \label{eq:constraintonX}
\eeq
This equation can be seen as a constraint on (say) the invariants of $\sX$ by appropriately choosing $\sS$ whose properties are well-understood. 

One class of theories whose properties are well-understood are sigma models whose target spaces are string manifolds. The reason that their properties are well-understood is because string manifolds have been studied in mathematics. It is not easy to compare general elements of $\SQFT_\bullet(\pt) $ with general elements of $\TMF_\bullet(\pt)$, but such a direct comparison is possible for string manifolds. In particular, the definitions of the primary and secondary invariants for string manifolds in physics and mathematics coincide in a straightforward way.

Moreover, it is known that string manifolds can realize all elements of $\TMF_\bullet(\pt)$ up to multiplication by a power of the periodicity element, in the following sense.\footnote{A precise statement is that $M \text{String}_\bullet(\pt) \to \text{tmf}_\bullet(\pt)$ is surjective~\cite{Hopkins2002,DevaSurj}. } There is an element $P \in \TMF_{-576}(\pt)$ whose elliptic genus (or the mathematical Witten genus divided by $\eta(\tau)^{\nu_P}$) is just given by $I_P=1$.
For a string manifold $S$, let $[S]$ be its image in $\TMF_\bullet(\pt)$. Then, any element of $\TMF_\bullet(\pt)$ can be realized as $P^{n} [S]$ for nonnegative integer $n\geq 0$. 

Physically, an element $\sP \in \SQFT_{-576}$ has been constructed which has at least the property $I_{\sP}=1$~\cite{Albert:2022gcs}.\footnote{
It would be nice if one can also show that there exists a theory $\sP^{-1}$ such that $\sP \times \sP^{-1}$ is bordant to a theory with only a single vacuum and nothing else (i.e. infinite mass gap), or in other words $[\sP \times \sP^{-1}] =1 \in \SQFT_0(\pt)$, where $1 \in \SQFT_0(\pt)$ is the multiplicative unit of the ring $\SQFT_\bullet(\pt)$. This is the defining property of the periodicity element, and the existence of such $\sP$ and $\sP^{-1}$ is expected from the $\TMF$ side.
} This property is enough for the purpose of the present section. 

We can consider a class of theories $\sP^{\times n} \times \sS$, where $\sS$ is taken to be a sigma model whose target space is a string manifold. This class can realize all primary and secondary invariants on the $\TMF$ side. By using it, the primary and secondary invariants of arbitrary SQFTs $\sX$ are constrained by the condition \eqref{eq:constraintonX}. 

Let us study the situation in more detail. For notational simplicity, let $\bI_\nu$ be the abelian group which is 
either\footnote{In the following, the change from $\bQ$ to $\bR$ in the secondary invariant is not essential. } 
\beq
\bI_\nu =
\left\{ \begin{array}{ll@{}l} 
 \eta(\tau)^{-\nu} \MF_{\nu/2} &  (\nu \equiv 0,&\mod 4),\vspace{0.2cm} \\ 
 \eta(\tau)^{-\nu}  \bZ_2((q)) &  (\nu \equiv  1,2 &\mod 8), \\
\displaystyle \eta(\tau)^{-(\nu+1)} \frac{\bR((q))}{ \bZ((q)) +  \MF_{(\nu+1)/2}\otimes \bR}& (\nu \equiv  3  &\mod 4).
\end{array}
\right.
\eeq
Then there is a natural pairing \begin{equation}
\bI_{\nu}\times \bI_{-21-\nu} \to \bR/\bZ
\end{equation} defined by the formulas \eqref{eq:secondaryinvariant5}, \eqref{eq:thetaterm} and \eqref{eq:mod2paring}.
When $\nu\equiv 0,3$ mod $4$, this is the pairing between the elliptic genus and the secondary invariant,
and when $\nu\equiv 1,2$ mod $8$, this is the pairing between the mod-2 elliptic genera.

The primary or secondary invariant takes values in $\bI_\nu$, but the actual values of invariants form a subgroup of it. We denote this subgroup in $\SQFT$ and $\TMF$ as $\bI^\SQFT_\nu$ and $\bI^\TMF_\nu$, respectively. As mentioned before, the class of elements  $P^n [S]$ for string manifolds $S$ realizes all elements of $\bI^\TMF_\nu$ (and actually the entire $\TMF_\bullet(\pt)$),
and  $[\sP]^n [\sS]$ have at least the same primary and secondary invariants as $P^n [S]$. Therefore, we have 
\beq
\bI^\TMF_\nu \subset \bI^\SQFT_\nu. \label{eq:cons1}
\eeq
On the other hand, the pairing \eqref{eq:constraintonX} which vanishes is given by,
\beq
\bI^\SQFT_\nu \times \bI^\SQFT_{-21-\nu} \to \{0\} \subset \bR/\bZ \label{eq:cons2}.
\eeq 

We claim that the consistency of \eqref{eq:cons1} and  \eqref{eq:cons2} implies that
\beq
  \bI^\SQFT_\nu = \bI^\TMF_\nu.\label{eq:SQFTTMF}
\eeq
The reason for this claim is as follows. Consider the pairing in mathematics,
\beq
\frac{\bI_\nu}{\bI^\TMF_\nu } \times \bI^\TMF_{-21-\nu} \to \bR/\bZ,  \label{eq:perfect2}
\eeq
that corresponds  in physics to \eqref{eq:secondaryinvariant5} for $\nu \equiv 0$ mod $4$, \eqref{eq:thetaterm} for $\nu \equiv 3$ mod $4$ and \eqref{eq:mod2paring} for $\nu \equiv 1,2$ mod $8$, respectively.\footnote{ When $\nu \equiv 3$, the correspondence is not direct. We simply define $(\bI_\nu/\bI^\TMF_\nu) \times  \bI^\TMF_{-21-\nu} \ni (J,K) \mapsto [\eta(\tau)^4 JK]_{q^0}$. This pairing is a direct sum of pairings of the type $\bR/\bZ \times \bZ \to \bR/\bZ$. Its meaning is ``continuous $\theta$-term'' as in \eqref{eq:thetaterm}.
}
The pairing \eqref{eq:perfect2} turns out to be perfect (or in other words the two groups $\bI_\nu/\bI^\TMF_\nu$ and $\bI^\TMF_{-21-\nu}$ are Pontryagin dual of each other).
A derivation of this fact from the results in \cite{Tachikawa:2023lwf} will be postponed to the end of this subsection,
as it is somewhat technical.
Assuming the perfectness of \eqref{eq:perfect2} for now, we have the following consequence. 
Suppose that $\bI^\TMF_\nu \subsetneq \bI^\SQFT_\nu$. Then we could take $A \in \bI^\SQFT_\nu$ that is nonzero in $\bI/\bI^\TMF_\nu$. The perfectness of \eqref{eq:perfect2} implies that there exists an element $B \in \bI^\TMF_{-21-\nu} \subset \bI^\SQFT_{-21-\nu}$ such that the pairing between $A$ and $B$ is nonzero. This is in contradiction with \eqref{eq:cons2}. Therefore, we conclude that the equality \eqref{eq:SQFTTMF} holds. 

The equality \eqref{eq:SQFTTMF}, combined with known mathematical facts, implies that the set of all the values of invariants in physics, $  \bI^\SQFT_\nu $, are as follows:
\begin{enumerate}
\item Elliptic genera for $\SQFT_\nu$ are linear combinations of $a_{i,j,k} E_4^i E_6^j \Delta^k \eta(\tau)^{-\nu}$ ( $i \geq 0, ~j=0,1$ and $\nu=8i+12j+24k$) where\footnote{See \cite{Hopkins2002}. We remark that when $\nu \equiv 4 \mod 8$, we are multiplying $\kappa=1/2$ in the present paper.}
\beq
a_{i,j,k}=\left\{ \begin{array}{ll}
   {24}/{{\rm gcd}(24, k) } & (i=j=0), \\
1 & (\text{otherwise}).
\end{array}
\right.
\eeq
Here ${\rm gcd}(n,m)$ means the greatest common divisor of $n$ and $m$.

\item Secondary invariants for $\SQFT_{-21-\nu} $ with $\nu \in 4\bZ$ are nontrivial only if $\nu=24k$, and in that case it takes values in 
$\bZ_{   {24}/{{\rm gcd}(24, k) }   }$. It is generated by 
\beq
    \frac{ {\rm gcd}(24, k) }{24} \eta(\tau)^{-4} \frac{E_6}{E_4} \in \eta(\tau)^{20+24k} \frac{\bR((q))}{ \bZ((q)) +  \MF_{-(20+24k)/2}\otimes \bR}
\eeq
where the inverse of $E_4=1+\cdots$ is interpreted in terms of power series expansion by $q$.
It vanishes when it is paired with elements of 
$\bI_{24k}^\SQFT\subset \bI_{24k}=\eta(\tau)^{-24k} \MF_{24k/2}$, including in particular  $a_{0,0,k}\Delta^k \eta(\tau)^{-24k}={24}/{{\rm gcd}(24, k) }$.

\item Mod-2 elliptic genera for $\SQFT_\nu$ with $\nu=8m+1$, $8m+2$ are contained in the mod-2 reduction of $\eta(\tau)^{-\nu}\MF_{4m}$ (i.e. the coefficients of formal power series in $q$ of elements of $\eta(\tau)^{-\nu}\MF_{4m}$ are reduced modulo 2). 
Moreover, when $\nu=24k+1$ or $24k+2$, elements of the form $\Delta^{k}\eta(\tau)^{-\nu}$ (reduced modulo $2$) are absent for 
\beq
\nu=1+24k : & \quad k \equiv 2,3,5,6,7 \mod 8, \nonumber \\
\nu=2+24k : & \quad k \equiv 3,6,7 \mod 8.
\eeq
All other elements are present. 

\end{enumerate}
For the somewhat complicated statement about mod-2 elliptic genera, see Section~4 of \cite{Tachikawa:2023nne} and in particular Remark~4.3 there, which comes from the results in \cite{BrunerRognes}.
The missing elements for mod-2 elliptic genera are precisely so that the pairing \eqref{eq:mod2paring} vanishes between SQFTs. 

For instance, consider the pairing between $\eta(\tau)^{-1} \in \bI^\SQFT_{1}$ 
and $E_4^3 \Delta^{-2} \eta(\tau)^{22} \in \bI^\SQFT_{-22}$. Here, both of modular forms are reduced modulo 2. 
The first element $\eta(\tau)^{-1}$ can be realized by the $S^1$ sigma model with periodic spin structure, and the second element $E_4^3 \Delta^{-2} \eta(\tau)^{22} $ can be realized by the product of the $T^2$ sigma model with periodic spin structure (with mod-2 elliptic genus $\eta(\tau)^{-2}$),
the sigma model on the 8-dimensional Bott manifold with $\int \hat A=1$ (with elliptic genus $E_4(\tau) \eta(\tau)^{-8}$), and
the $(\e_8)_1\times (\e_8)_1$ current algebra (with elliptic genus $(E_4(\tau)\Delta^{-1} \eta(\tau)^{16})^2 $).
The pairing defined by \eqref{eq:mod2paring} gives
\beq
\frac 12 [\eta(\tau)^3 \cdot \eta(\tau)^{-1} \cdot E_4^3 \Delta^{-2} \eta(\tau)^{22} ]_{q^0} =  \frac 12 [E_4^3 \Delta^{-1}  ]_{q^0} ,
\eeq
where we have used $\Delta =\eta(\tau)^{24}$.
Now we use the fact that $E_4 \equiv 1$ and $E_6 \equiv 1$ modulo 2 as can be seen from \eqref{eq:Eisenstein}, and in particular one can replace $E_4$ by $E_6$. Then we get
\beq
 \frac 12 [E_4^3 \Delta^{-1}  ]_{q^0} \equiv  \frac 12 [E_4^2 E_6 \Delta^{-1}  ]_{q^0}=0,
\eeq
where the last equality is from the property of modular forms of weight 2. On the other hand, if $ \Delta^{-1} \eta(\tau)^{22} $ were an element of $ \bI^\SQFT_{-22}$, we would get
\beq
\frac 12 [\eta(\tau)^3 \cdot \eta(\tau)^{-1} \cdot   \Delta^{-1} \eta(\tau)^{22} ]_{q^0} =\frac 12 [1 ]_{q^0} =\frac12,
\eeq
contradicting the vanishing of the pairing between $\bI^\SQFT_{1}$ and $ \bI^\SQFT_{-22}$. 
$ \eta(\tau)^{-1}$ is the mod-2 elliptic genus of the sigma model with target space $S^1$ and hence it really exists in $\bI^\SQFT_{1}$. Therefore, $ \Delta^{-1} \eta(\tau)^{22} $ must be absent in $ \bI^\SQFT_{-22}$.

\paragraph{Perfect pairing.}
Let us finally come back to the derivation of the perfectness of \eqref{eq:perfect2}.
Before proceeding, we recall the basic long exact sequence
\beq
\cdots \to \TMF_{\nu}(\pt) \to \mathrm{KO}((q))_{\nu}(\pt) \to (\mathrm{KO}((q))/\TMF)_\nu(\pt) \to \TMF_{\nu-1}(\pt)  \to \cdots \label{eq:KOTMFlong}
\eeq

What was proved mathematically in \cite{Tachikawa:2023lwf}
was the perfectness of the torsion pairing \begin{equation}
\mathrm{Tors}((\mathrm{KO}((q))/\TMF)_\nu(\pt)) \times \mathrm{Tors}(\TMF_{-21-\nu}(\pt))\to \bR/\bZ
\label{eq:torsion-pairing}
\end{equation}
where $\mathrm{Tors}(\cdots)$ means the torsion part,
and  the perfectness of the non-torsion pairing \begin{equation}
\mathrm{Free}((\mathrm{KO}((q))/\TMF)_{\nu+1}(\pt)) \times \mathrm{Free}(\TMF_{-21-\nu}(\pt))\to \bZ,
\label{eq:nontorsion-pairing}
\end{equation}
where $\mathrm{Free}(\cdots)$ means the free part.

The perfectness of \eqref{eq:perfect2} for $\nu\equiv 0$ mod $4$ and $\nu\equiv 1,2$ mod $8$ is simply
the perfectness of the torsion pairing \eqref{eq:torsion-pairing} by the following reason.
Let $\bA_\nu^{\TMF} \subset \TMF_{\nu}(\pt)$ be the TMF version of the subgroup \eqref{eq:nontrivialgroup} characterized by the vanishing of the primary and secondary invariants.
Let $\tilde \iota$ be the restriction of the map 
\beq
\iota : (\mathrm{KO}((q))/\TMF)_\nu(\pt)) \to \TMF_{\nu-1}(\pt) \label{eq:connectinghom}
\eeq
to the torsion part $\mathrm{Tors}((\mathrm{KO}((q))/\TMF)_\nu(\pt))$. The image of $\tilde \iota$  is given by $\bA_{\nu-1}^{\TMF} $.\footnote{$\TMF_{\nu-1}(\pt)$ for $\nu \in 4\bZ$ contains elements that are detected by the secondary invariant. As we will see around \eqref{eq:TMFsecond}, they are images of the free part $\mathrm{Free}((\mathrm{KO}((q))/\TMF)_{\nu}(\pt))$. Therefore, if we restrict to the torsion part $\mathrm{Tors}((\mathrm{KO}((q))/\TMF)_\nu(\pt))$, its image in $\TMF_{\nu-1}(\pt)$ has vanishing secondary invariant. The primary invariants also vanish because it is in the kernel of the map $\TMF_{\nu-1}(\pt) \to \mathrm{KO}((q))_{\nu-1}(\pt)$ due to the exact sequence \eqref{eq:KOTMFlong}. In this way, one can show that the image of $\tilde \iota$ is $\bA_{\nu-1}^{\TMF} $.}
Then, for the above values of $\nu$ we have
\beq
\bI_\nu/\bI_\nu^\TMF &\simeq \tilde \iota^{-1}(0), \nonumber \\
\bI_{-21-\nu}^\TMF &\simeq \mathrm{Tors}(\TMF_{-21-\nu}(\pt))/\bA_{-21-\nu}^{\TMF}.
\eeq
The perfectness of the pairing \eqref{eq:torsion-pairing} is such that the pairing between $\bI_\nu/\bI_\nu^\TMF $ and $\bI_{-21-\nu}^\TMF$ is perfect, and
 the pairing between 
$  \bA_{\nu-1}^{\TMF} \simeq \mathrm{Tors}((\mathrm{KO}((q))/\TMF)_\nu(\pt))/\tilde \iota^{-1}(0) $ and $\bA_{-21-\nu}^{\TMF} $ is also perfect.

The perfectness of \eqref{eq:perfect2} for $\nu\equiv 3$ mod $4$ follows from the perfectness of the non-torsion pairing \eqref{eq:nontorsion-pairing}.
We have $\bI_{-21-\nu}^\TMF\simeq \mathrm{Free}(\TMF_{-21-\nu}(\pt))$, and we need to understand
the relation between $\bI_\nu/\bI_\nu^\TMF$ and $\mathrm{Free}((\mathrm{KO}((q))/\TMF)_{\nu+1}(\pt))$.

Let us first clarify how to describe the free part $\mathrm{Free}((\mathrm{KO}((q))/\TMF)_{\nu+1}(\pt))$. 
Consider the map \eqref{eq:connectinghom} with $\nu$ replaced by $\nu+1$. For $\nu \in 3+4\bZ$, the image is a torsion element. Therefore, for any $A \in  (\mathrm{KO}((q))/\TMF)_{\nu+1}(\pt)$, there exists an integer $k $ such that $\iota(k A)=0$. The exactness of \eqref{eq:KOTMFlong} implies that $kA$ is the image of an element $B \in \mathrm{KO}((q))_{\nu+1}(\pt)$ under the map 
$
\mathrm{KO}((q))_{\nu+1}(\pt) \to  (\mathrm{KO}((q))/\TMF)_{\nu+1}(\pt) . 
$
Let $I_B$ be the primary invariant of $B$. Then we may describe the information of $A$ in the free part $\mathrm{Free}((\mathrm{KO}((q))/\TMF)_{\nu+1}(\pt))$ by defining $I_A=\frac{1}{k} I_B$. 
However, $B$ is ambiguous by the image of the map
$
\TMF_{\nu+1}(\pt) \to \mathrm{KO}((q))_{\nu+1}(\pt).
$
Therefore, $I_A$ is ambiguous by $\eta(\tau)^{-(\nu+1)} \MF_{(\nu+1)/2} \otimes  \bR $. Therefore, we describe the information of $A$ in the free part by
\beq
I_A = \frac{1}{k} I_B \in \eta(\tau)^{-(\nu+1)} \cdot \frac{\bR((q)) }{  \MF_{(\nu+1)/2} \otimes  \bR}. \label{eq:TMFsecond}
\eeq
The further reduction of $I_A$ modulo $\eta(\tau)^{-(\nu+1)} \bZ((q))$,
which comes from the image of $\mathrm{KO}((q))_{\nu+1}(\pt)$,
 gives the secondary invariant $I^\text{2nd}_{\iota(A)}$ of the ``boundary theory'' $\iota(A) \in \TMF_\nu(\pt)$.  
 
Now we consider   
\beq
\bJ_{\nu}^{\mathrm{KO}((q))/\TMF} &:=  \{ I_A \,| \, A \in (\mathrm{KO}((q))/\TMF)_{\nu+1}(\pt)  \}, \nonumber \\
\bI^{\TMF}_{-21-\nu}  &\phantom{:}=   \{ I_A \,| \, A \in \TMF_{-21-\nu}(\pt)  \}.
\eeq
The perfect pairing \eqref{eq:nontorsion-pairing} is 
\beq
 \bJ_{\nu}^{\mathrm{KO}((q))/\TMF} \times    \bI^{\TMF}_{-21-\nu} \to \bZ
\eeq
given by $( f, g) \to [\eta(\tau)^4 fg]_{q^0}$.
Its perfectness implies that the pairing
\beq
\frac{\bJ_{\nu}^{\mathrm{KO}((q))/\TMF}  \otimes \bR}{ \bJ_{\nu}^{\mathrm{KO}((q))/\TMF } } \times    \bI^{\TMF}_{-21-\nu} \to \bR/\bZ, \label{eq:pe1}
\eeq
 with values in $\bR/\bZ$, is also perfect. We have 
\beq
\bJ_{\nu}^{\mathrm{KO}((q))/\TMF}  \otimes \bR = \eta(\tau)^{-(\nu+1)} \cdot \frac{\bR((q)) }{  \MF_{(\nu+1)/2} \otimes  \bR}.
\eeq
In particular, the image of $\mathrm{KO}((q))_{\nu+1}(\pt)$ gives a subgroup $\eta(\tau)^{-(\nu+1)} \cdot (\bZ((q)) / \MF_{(\nu+1)/2) } ) \subset  \bJ_{\nu}^{\mathrm{KO}((q))/\TMF} $ as mentioned above.

The relation between $\bJ_{\nu}^{\mathrm{KO}((q))/\TMF}$ and the secondary invariant mentioned above implies that $\bI_\nu$ and $\bI^\TMF_\nu$ are obtained by reducing $ \bJ_{\nu}^{\mathrm{KO}((q))/\TMF} \otimes \bR$ and $ \bJ_{\nu}^{\mathrm{KO}((q))/\TMF} $ modulo $\eta(\tau)^{-(\nu+1)} \bZ((q))$, respectively, and hence 
\beq
\frac{\bJ_{\nu}^{\mathrm{KO}((q))/\TMF} \otimes \bR}{ \bJ_{\nu}^{\mathrm{KO}((q))/\TMF} } =\frac{\bI_\nu}{\bI^\TMF_\nu } . \label{eq:pe2}
\eeq
Combining \eqref{eq:pe1} and \eqref{eq:pe2} gives the perfectness of \eqref{eq:perfect2}.

\subsection{The new invariant in terms of the spectral pairing} \label{sec:inv2}

The previous subsections have focused on the case that $\partial \sS=\varnothing$. We can ask what invariants of $\sX$ are missed by considering only such $\sS$. In this subsection, we first focus on the case that $\sX$ is such that 
\beq
\eta(\sX, \sS)=0 \quad \text{for all $\sS$ such that}~ \partial \sS=\varnothing, \label{eq:almostvanishing}
\eeq
but we will modify this requirement when $d+1 \in 4\bZ$,
by the reasons which will be discussed later. 

\paragraph{The case when $d+1\notin 4\bZ$.}
Under the above condition on $\sX$, the pairing $\eta(\sX, \sS)$ only depends on $\sT =\partial \sS$ by the following reason. Suppose we have another $\sS'$ such that $\partial \sS'=\sT$. The condition \eqref{eq:almostvanishing} implies that the ordinary and mod-2 elliptic genera of $\sX$ are zero. Then we need not be concerned with zero modes in the definition of our $\eta$-invariant (see \eqref{eq:eta-inv}), and from \eqref{eq: gluing-eta}, we get
\beq
\eta(\sX, \sS') - \eta(\sX,\sS) = \eta(\sX, \sS' \cup \overline{\sS}) =0 
\eeq
where $ \sS' \cup \overline{\sS}$ is obtained by gluing $\sS$ and $\sS'$ and hence $\partial ( \sS' \cup \overline{\sS})=\varnothing$. Therefore, $\eta(\sX, \sS)$ is independent of how  $\sT$ is extended to the bulk $\sS$ and hence we can write
\beq
\GS(\sX,\sT) := \eta(\sX,\sS).\label{5.80}
\eeq
We are going to argue that this is the same invariant as the one defined in Sec.~\ref{sec:def}.

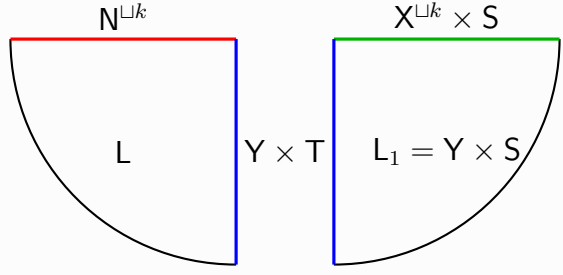
\begin{figure}
\centering

\begin{tikzpicture}[scale=2]
  \def\r{1.5}    
  \def\gap{0.65}  

  \draw[thick] (-\r,0) arc (180:270:\r);
  \draw[red,very thick]    (-\r,0) -- (0,0);
  \draw[blue,very thick]   (0,0)   -- (0,-\r);

  \draw[thick] ({\gap+\r},0) arc (0:-90:\r);
  \draw[green!70!black,very thick] (\gap,0)     -- ({\gap+\r},0);
  \draw[blue,very thick]   (\gap,0) -- (\gap,-\r);

  \node at (-0.5*\r,   -0.5*\r)        {$ {\mathsf L}$};
  \node at ({\gap+0.5*\r}, -0.5*\r)    {$ {\mathsf L}_1 = \mathsf{Y} \times \mathsf{S}$};
  \node at (-0.5*\r,   0.15)           {$\mathsf{N}^{\sqcup k} $};
  \node at ({\gap+0.5*\r}, 0.15)       {$\mathsf{X}^{\sqcup k} \times \mathsf{S}$};
  \node at ({\gap/2},   -0.5*\r)       {$ \mathsf{Y} \times \mathsf{T}$};
\end{tikzpicture}

\caption{The setup used to derive that the quantity as defined in \eqref{5.80}
agrees with the new invariant defined in Sec.~\ref{sec:def}.
 This figure may be compared with Figure~~\ref{fig:7}. 
 To reduce clutter, we did not carefully specify the orientations of the theories involved.
\label{fig:x}}
\end{figure}

Suppose that $k$ copies of $\sX$ is a boundary of some $\sY$, i.e. 
\beq
\partial \sY = \sX^{\sqcup k}.
\eeq
Let us consider the situation shown in Figure~\ref{fig:x}.
Here,
\beq
& \sL \text{ : such that } \partial \sL = \sN^{\sqcup k} \cup \overline{\sY \times \sT}, \nonumber \\
&  {\sL}_1 = (-1)^{d+1}\sY \times \sS , \nonumber \\
& {\sL}  \cup_0  {\sL}_1 ~\text{: such that }~ \partial( {\sL}  \cup_0  {\sL}_1 ) =  \sN^{\sqcup k} \cup  \left( (-1)^{d+1} \sX^{\sqcup k} \times \sS \right) =(\sN \cup (-1)^{d-1} \sX \times \sS)^{\sqcup k}.
\eeq
The index theorem gives
\beq
 k\eta_{\sN \cup (-1)^{d-1} \sX \times \sS} &= -\hat{A}_{ {\sL} \cup_0  {\sL}_1 } +  [\eta(\tau)^4 I_{ {\sL} \cup_0  {\sL}_1 }  ]_{q^0} \nonumber \\
 &= -\hat{A}(\sY, \sS)  + (-1)^{d+1}  [\eta(\tau)^4 I_{ \sY \times \sS }  ]_{q^0} +   [\eta(\tau)^4 I_{ {\sL}   }  ]_{q^0}.
\eeq
Here, we have used the fact that the gluing law of the APS index gives 
\begin{equation}
[\eta(\tau)^4 I_{ {\sL} \cup_0  {\sL}_1 }  ]_{q^0} = [\eta(\tau)^4 
I_{ {\sL}   }  ]_{q^0} +  [\eta(\tau)^4 I_{ {\sL}_1   }  ]_{q^0},
\label{5.83}
\end{equation}
and that the contribution of $\sL$ to $\hat{A}_{ {\sL} \cup_0  {\sL}_1 } $ vanishes as in the discussions of \eqref{eq:prop2} and hence  $\hat{A}_{ {\sL} \cup_0  {\sL}_1 }  = \hat{A}(\sY, \sS) $.
This last quantity  is given by \eqref{eq:prop3},
\beq
\hat{A}(\sY, \sS) = (-1)^{d+1} \frac12   \int_{C_1}  \eta(\tau)^4  Z_{  \sY}   Z_{\sS} \d\tau. \label{eq:prop3-1}
\eeq
Combined, we have 
\beq
\GS(\sX,\sT) &= 
 \eta(\sX,\sS) = \eta_{ \sN \cup (-1)^{d-1} \sX \times \sS} \nonumber \\
 &= \frac1k  \left(  -\hat{A}(\sY, \sS )  + (-1)^{d+1}  [\eta(\tau)^4 I_{ \sY \times \sS }  ]_{q^0} \right)
 +\frac1k  [\eta(\tau)^4 I_{ {\sL}   }  ]_{q^0}.\label{GSGS}
\eeq

Now recall that $\GS(\sX,\sT)$ was defined in Sec.~\ref{sec:def} by the formula 
\beq
\GS(\sX,\sS)=\frac{1}{k}  [\eta(\tau)^4 I_{ {\sL}   }  ]_{q^0} \mod \bZ. 
\eeq
Comparing this with \eqref{GSGS},
what remains to be established is that
the first term on the right hand side of \eqref{GSGS}, i.e.~\begin{equation}
 \frac1k \left(  -\hat{A}(\sY, \sS)  + (-1)^{d+1}  [\eta(\tau)^4 I_{ \sY \times \sS }  ]_{q^0} \right) \label{remainder}
\end{equation}
 does not contribute modulo $\bZ$. 

Whenever boundaries do not have zero modes, the APS index behaves as $I_{ \sY \times \sS } = I_{ \sY } I_{ \sS }$, 
where we have implicitly used the fact that $\kappa_\sY \kappa_\sS =\kappa_{\sY \times \sS} =\frac12$ whenever both $I_{ \sY }$ and $ I_{ \sS } $ are nonzero. Then,
\beq
-\hat{A}(\sY, \sS) +(-1)^{d+1} [\eta(\tau)^4 I_{ \sY \times \sS }  ]_{q^0} = -(-1)^{d+1} \frac12  \int_{F}  \eta(\tau)^4  \frac{\partial}{\partial \bar\tau}( Z_{  \sY}   Z_{\sS}) \d \bar\tau \wedge \d\tau. \label{eq:int-formula}
\eeq
The only case in which either $\vev{Q}_\sY$ or $\vev{Q}_\sS$ can be nonzero is the case that $d+1 \in 4\bZ$. Therefore, by the holomorphic anomaly equation, the derivative  $\frac{\partial}{\partial \bar\tau}( Z_{  \sY}   Z_{\sS})$ is automatically zero unless  $d+1 \in 4\bZ$.

\paragraph{The case when $d+1\in 4\bZ$.}

Now let us consider the case $d+1 \in4\bZ$. In this case, the constraint \eqref{eq:almostvanishing}
is too strong by the following reason. 
When $d+1\in 4\bZ$, the pairing $\eta(\sX,\sS)$ can in general depend on continuous parameters of $\sX$ and $\sS$
and it is not guaranteed to be topological. It can have nonzero values even if $\sS$ or $\sX$ is topologically trivial.

For instance, when $\partial \sS  = \varnothing$, the pairing $\eta(\sX,\sS)$ is given by $ \eta_{\sX}(I_\sS) $ as was seen in the first two equalities of \eqref{eq:secondaryinvariant5}. This may be interpreted as a theta term on $\sS$, where the theta parameter is produced by $\sX$. We expect that such a theta term will no longer be topologically invariant when $ \sS$ has boundary; it is like a Chern-Simons term on $\partial \sS = \sT$ where the Chern-Simons level is the (not-necessarily quantized) theta parameter produced by $\sX$. 

Therefore, the constraint \eqref{eq:almostvanishing} would not be a topological condition. We want to replace it by topological conditions on $\sX$ and $\sS$ to get a nice topological result from the pairing $\eta(\sX,\sS)$.

Let us consider what conditions should be imposed. First, the rightmost quantity $[ I_\sS \eta(\tau)^4 I^\text{2nd}_\sX]_{q^0}$ of the formula \eqref{eq:secondaryinvariant5}, which was obtained under the condition $\partial \sS=\varnothing$ and $I_\sS \in \eta(\tau)^{-\nu_\sS} \MF_{\nu_\sS/2}$,  implies that at least the secondary invariant of $\sX$ should vanish. In fact, we have imposed this condition in Section~\ref{sec:def} when we defined the new invariant.  

The vanishing of the secondary invariant of $\sX$ is not sufficient. The equality $\eta(\sX,\sS) = [ I_\sS \eta(\tau)^4 I^\text{2nd}_\sX]_{q^0}$ was obtained only under the condition $\partial \sS=\varnothing$ and $I_\sS \in \eta(\tau)^{-\nu_\sS} \MF_{\nu_\sS/2}$. We need to generalize this condition to the case $\partial \sS =\sT \neq \varnothing$ to get a nice topological quantity from $\eta(\sX,\sS)$. 

It turns out that the correct condition is that $[\sS]$ is a torsion element of $\SQ_{-21-d}(\pt)$. (Compare the following discussion to the discussion around \eqref{eq:TMFsecond}.) Suppose that 
$[\sS]$ is torsion, which means that there exists an integer $\ell$ and an SQFT $\sU$ such that $\partial \sU = \sT^{\sqcup \ell}$. Then we can glue $\sS^{\sqcup \ell}$ and $\overline{\sU}$ to get a compact theory $\sS^{\sqcup \ell} \cup \overline{\sU}$. The quantity
\beq
\frac{1}{\ell} I_{\sS^{\sqcup \ell} \cup \overline{\sU}} \in \eta(\tau)^{-\nu_\sS} \cdot \frac{\bR((q)) }{  \MF_{\nu_\sS /2} \otimes  \bR} \label{eq:Sprimary}
\eeq
can be regarded as the primary invariant of $[\sS] \in \SQ_{-21-d}(\pt)$. We require that this quantity vanishes. 
This condition does not depend on the choice of $(\ell, \sU)$. 

In fact, the vanishing of \eqref{eq:Sprimary} implies that 
\beq
 I_{\sS^{\sqcup \ell} \cup \overline{\sU}} \in \eta(\tau)^{-\nu_\sS} \MF_{\nu_\sS/2}.
\eeq
Then by \eqref{eq:secondaryinvariant5} we get
\beq
\ell \eta(\sX,\sS) - \eta(\sX, \sU) =  \eta(\sX,\sS^{\sqcup \ell} \cup \overline{\sU})   = [  I_{\sS^{\sqcup \ell} \cup \overline{\sU}} \eta(\tau)^4 I^\text{2nd}_\sX]_{q^0}.
\eeq
Recall that $ \eta(\sX, \sU)$ takes values in integers when $\sU$ is an SQFT, as remarked at the end of Sec.~\ref{sec:generalstr}.
Therefore, $\ell \eta(\sX,\sS)$ is an integer, meaning that $\eta(\sX,\sS)$ takes discrete (rather than continuous) values. 
This is topological as  we wanted.

Recall that the secondary invariant of $\sT$ is given by $I^\text{2nd}_\sT = \frac{1}{\ell} I_\sU$ modulo $\eta(\tau)^{-\nu_\sS} ( \bZ((q))+\MF_{\nu_\sS /2} \otimes  \bR)$.
The gluing law of the APS index gives $\frac{1}{\ell} I_{\sS^{\sqcup \ell} \cup \overline{\sU}} = I_\sS - \frac{1}{\ell} I_\sU$. Therefore, the vanishing of \eqref{eq:Sprimary} implies the vanishing of the secondary invariant of $\sT$. Conversely, when the secondary invariant of $\sT$ vanishes, there exists an $\sS$ with $\partial \sS=\sT$ such that \eqref{eq:Sprimary} vanishes.

Let us summarize our discussion so far.
Recall that  the condition \eqref{eq:almostvanishing} when $d+1\in 4\bZ$ is such that it continuously depends on the parameters of $\sX$,
rather than a topological condition on it. 
Therefore, we may not want to strictly impose it in this case. 
From a topological point of view, it is better to impose only the topological condition on $\sX$ that its secondary invariant is zero. 
Then, we decide to restrict $\sS$ to represent a special class of elements of $\SQ_{-21-d}(\pt)$ defined by the vanishing of the invariant \eqref{eq:Sprimary}.\footnote{In the context of TMF, the vanishing of \eqref{eq:Sprimary} corresponds to the condition that $[\sS]$ is a torsion element of $ ( {\rm KO}((q))/{\rm TMF})_{-21-d}({\rm pt})$. We expect the same for $\SQ_{-21-d}(\pt)$.}
In summary, 
\begin{itemize}
\item The secondary invariant of $\sX$ is zero.
\item $\sS$ (with $\sT=\partial \sS$) is such that the invariant \eqref{eq:Sprimary} vanishes. In particular, the secondary invariant of $\sT$ is zero.
\end{itemize}
We remark that the vanishing of the primary invariants of $\sX$ and $\sT$ is automatic when $d+1 \in 4\bZ$.

Now we show that $\eta(\sX,\sS)$ coincides with $\GS([\sX],[\sT])$.
The secondary invariants of both $\sX$ and $\sT$ vanish.
By Property~\ref{prop:1} of Sec.~\ref{app:BN}, we can (and do) choose $(k,\sY)$ such that $\partial \sY= \sX^{\sqcup k}$ and $I_\sY$ is a multiple of $k$.
By the same argument as in Property~\ref{prop:1} of Sec.~\ref{app:BN}, we can (and do) choose $(\ell, \sU)$ such that $\partial \sU=\sT^{\sqcup \ell}$ and 
$  I_{\sS^{\sqcup \ell} \cup \overline{\sU}}=0$ even without taking modulo $\eta(\tau)^{-\nu_\sS} \MF_{\nu_\sS/2}$. In particular, $I_\sU = \ell I_\sS$ is a multiple of $\ell$.
The gluing law of the partition function and the equality $Z_{\sS^{\sqcup \ell} \cup \overline{\sU}} = I_{\sS^{\sqcup \ell} \cup \overline{\sU}} =0$ implies that
\beq
0=Z_{\sS^{\sqcup \ell} \cup \overline{\sU}} = \ell Z_{\sS} - Z_\sU. \label{eq:SUrel}
\eeq

We apply \eqref{eq:SUrel} to \eqref{eq:int-formula}. 
Recall also that $\hat{A}(\sY, \sU)$ is zero by modular invariance when both $\sY$ and $\sU$ are SQFTs, rather than formal power series of SQMs. Combining these facts, we obtain
that the quantity \eqref{remainder} we would like to control is given by 
\beq
\frac{1}{k} \left( -\hat{A}(\sY, \sS) +(-1)^{d+1} [\eta(\tau)^4 I_{ \sY \times \sS }  ]_{q^0}  \right) = \frac{1}{k\ell} (-1)^{d+1} [\eta(\tau)^4 I_{ \sY } I_{ \sU }  ]_{q^0} .
\eeq 
By assumption, the coefficients of $ \frac{1}{k} I_{ \sY }$ and $\frac{1}{\ell} I_{\sU}$ are integers. Therefore, the right-hand side is an integer. This concludes our argument that 
the part \eqref{remainder}
does not contribute to $ \eta_{ \sN \cup (-1)^{d-1} \sX \times \sS } $ modulo $\bZ $,
thus establishing 
$\GS(\sX,\sT)$ defined in terms of the spectral invariant \eqref{5.80} equals
$\GS([\sX], [\sT])$ defined topologically in Sec.~\ref{sec:def}.

\section{The phase of partition functions in heterotic string theory}\label{sec:string}

Our discussions so far have been mainly about general SQFTs (and SQMs), but at some points we have mentioned relations to heterotic string theory. In this section we would like to discuss the general situation in heterotic string theory. We caution the reader that some of the statements in this section are speculative and require more consistency checks in the future. 

Our interest is in the one loop partition function of the target space theory with Euclidean signature in terms of the worldsheet theory. The standard relation between them is
\beq
Z^\text{1-loop}_\text{target} = \exp (Z^\text{1-loop}_\text{worldsheet} )
\eeq
where $Z^\text{1-loop}_\text{target}$ is the target space partition function and $Z^\text{1-loop}_\text{worldsheet} $ is the worldsheet partition function.
The worldsheet partition function is the sum of four terms because we can take NS and R spin structures in the two directions of $T^2$,
\beq
Z^\text{1-loop}_\text{worldsheet} =Z^\text{1-loop}_\text{NS,NS} +Z^\text{1-loop}_\text{NS,R} +Z^\text{1-loop}_\text{R,NS} +Z^\text{1-loop}_\text{R,R} .\label{eq:1loop}
\eeq
With appropriate care, the first three terms can be computed without problems.\footnote{The one loop partition function is a zero point correlation function at genus one, and one needs care about the treatment of the translation symmetry. See e.g. Section~7.3 of \cite{Polchinski:1998rq} for a textbook account in bosonic string theory. The situation about the first three terms in \eqref{eq:1loop} is analogous to the case of bosonic string theory.} 
However, there is a subtlety about the last term, $Z^\text{1-loop}_\text{R,R} $.

When we try to compute $Z^\text{1-loop}_\text{R,R} $ in the standard approach, we encounter $0/0$ as follows. $T^2$ with $\text{(R,R)}$ spin structure has both
\begin{itemize}
\item the supertranslation symmetry, and
\item a supermodulus
\end{itemize}
The gauge volume $V_\text{supertranslation}$ of the supertranslation symmetry is zero because it is given by an integral over a Grassmann variable. We need to divide by this gauge volume in the computation of $Z^\text{1-loop}_\text{R,R}$. 
On the other hand, the integral over the supermodulus gives supercharge $Q$, and hence $Z^\text{1-loop}_\text{R,R}$ is, roughly speaking, proportional to $\vev{Q}$ of the worldsheet theory.
The total gravitational anomaly of a worldsheet theory $\sM$ is $\nu_\sM=-22$, which is $2 $ modulo $ 8$.
Therefore, the $\vev{Q}$ of $\sM$ is zero as can be shown by the algebra of CPT symmetry given by \eqref{eq:time1}. 
Very roughly, we have $Z^\text{1-loop}_\text{R,R} \sim \vev{Q}/V_\text{supertranslation}$. However, both $V_\text{supertranslation}$ and $\vev{Q}$ are zero, and we get an indeterminate result, $Z^\text{1-loop}_\text{R,R}  \sim 0/0$. 

There is also a problem at the level of the low energy effective theory. It is not straightforward to define the partition function of chiral fermions. A solution to this problem is to go to a higher-dimensional manifold as reviewed in Section~\ref{sec:mfds}. A manifold $M$ on which we want to compute the fermion partition function is realized as the boundary of a higher-dimensional manifold $N$.  Then the phase of the fermion partition function is given by $ \exp ( -2\pi \i \eta(N) )$ as in \eqref{eq:barefermion}, or its modified version $\exp \left( -2\pi \i \eta(N) - 2\pi \i \int_N \cJ_{d+1}\right)$ as in \eqref{eq:totalpf}. 

In the same way, we speculate that the computation of the partition function of a worldvolume theory $\sM$ with $\nu_\sM=-22$ requires a theory $\sN$ such that $\partial \sN = \sM$ and $\nu_\sN=\nu_\sM+1=-21$.  In the standard description in heterotic string theory, the worldvolume theory $\sM$ is required to be a superconformal field theory with central charge $(c_L, c_R)=(26,15)$. However, we will not require $\sN$ to be superconformal, because superconformal invariance plays no role in our discussions of the phase of the partition function.\footnote{ It may also be possible to relax the condition that $\sM$ is superconformal.  Non-conformal SQFTs might be considered as off-shell configurations of the target space. Recall that in the case of sigma models, the vanishing of beta functions of a sigma model corresponds to equations of motion in the target space. Therefore, non-conformal theories do not satisfy equations of motion, suggesting that they are off-shell. When we want to compute quantum gravity partition functions of target space, on-shell configurations may correspond to saddle points of the ``path integral in the target space quantum gravity''. }

Let us compare the partition functions of the worldsheet and low energy effective theory in the limit $\alpha' \to 0$. We are interested in the fermion contributions, and hence we consider R spin structure in the spatial direction of $T^2$. There are two contributions, $Z^\text{1-loop}_\text{R,NS} $ and $Z^\text{1-loop}_\text{R,R}$. By recalling that the worldsheet Hamiltonian is roughly given by $H \sim \alpha'  \cD_d(M)^2+\cdots$, where $\cD_d(M)$ is the Dirac operator on the target space, we can match the part  $Z^\text{1-loop}_\text{R,NS} $ with the absolute value of the fermion partition function,
\beq
\log |{\rm Pf}\, \cD^+_d(M)| \sim -  \int_F  \frac{\d \tau_1 \d\tau_2}{4\tau_2} \left( \frac{|\eta(\tau)|^4 \overline{\eta(\tau)}^2 }{2 \overline{\eta(2\tau)}^2 } \right) \Tr_{\text R} e^{-2\pi \tau_2 H + 2\pi \i \tau_1 P} = Z^\text{1-loop}_\text{R,NS}
 \eeq
 where $\tau = \tau_1 + \i \tau_2$, the trace is in the R-sector of $\sM$, $\rm{Pf}$ means the pfaffian,\footnote{Pfaffians are more appropriate than  determinants in the current context. For instance, gauginos in 10d supersymmetric heterotic string theories are Majorana-Weyl.} and various Dedekind $\eta$'s are ghost contributions.
 
 From the above considerations, it is expected that $Z^\text{1-loop}_\text{R,R} $ should give the phase of the target space partition function. We propose that it is given by
 \beq
 Z^\text{1-loop}_\text{R,R} = -2\pi \i \eta_\sN, \label{eq:conjecture}
 \eeq
 where $\partial \sN = \sM$. 
 
 For simplicity, let us first look at the formula of $ \eta_\sN$ when $\sN$ has no boundary, given by \eqref{eq:eta-inv}, although it is expected to vanish for $\SQFT_{-21}$.
 Neglecting zero mode contributions, it is given by the integral of $\vev{Q}$ over $\tau$. This is the structure that is naturally expected from the existence of the supermodulus and the bosonic modulus on $T^2$. We neglect the gauge volume of the supertranslation symmetry, and instead extend $\sM$ to $\sN$. Ghost contributions are also included in $\eta_\sN$ (see footnote~\ref{foot:ghost}). 
 
 In the presence of a nontrivial boundary $\partial \sN=\sM \neq \varnothing$,  it is not obvious what definition we should use for $ \eta_\sN$. When there are no zero modes, one possibility is to use the regularization as in Appendix~\ref{sec:A}. We have already used this regularization at some points in Section~\ref{sec:differential}. More consistency checks are needed to support this conjecture.
 
As a check, let us consider the dependence of $\eta_\sN$ modulo $\bZ$ on the choice of $\sN$. The difference between two choices $\sN$ and $\sN'$ is given by $\eta_{\sN' \cup \overline{\sN}}$. This is the secondary invariant of $\sN' \cup \overline{\sN} \in \SQFT_{-21}$. The vanishing of the secondary invariant is indeed the condition for the absence of anomalies in heterotic string theory~\cite{Yonekura:2022reu}. 

It is possible that an $\sN$ such that $\partial \sN =\sM$ does not exist. How to treat such a case is also known~\cite{Witten:2016cio,Witten:2019bou}. If $[\sM]$ is an element of order $k$ in the bordism group, i.e. $k[\sM]=0$, we take $\sN$ such that $\partial \sN = \sM^{\sqcup k}$. Then the phase of the partition function of  $\sM^{\sqcup k}$ is determined, but the partition function of $\sM$ itself has an ambiguity by $k$-th roots of unity. Fixing this ambiguity by hand corresponds to choosing a discrete theta angle of the theory. The case of infinite order gives a continuous theta angle. However, in our case, the theta angle might not be so important. We assume that $\SQFT_{-22}(\pt)$ is detected by mod-2 elliptic genus. If $\sM$ is superconformal with central charge $(c_L, c_R)=(26,15)$, the only possibility for a nonzero mod-2 elliptic genus is $\eta(\tau)^{22} E_4^3\Delta^{-2}=\eta(\tau)^{-2}(q^{-1} +0 +\cdots)$. Here the $q^{-1}$ term is a target space supersymmetry generator, and its zero mode might make the sign of the partition function ambiguous.

As another check of \eqref{eq:conjecture}, notice that when $\sM = \sX \times \sT$, $\nu_\sX=d$, $\nu_\sT=-22-d$, $[\sX] \in \bA_d$, $\sT=\partial \sS$ with $\sS$ a Laurant series of SQMs, and $d \notin 3+4\bZ$,  our new invariant is just given by $\eta_\sN$ with the aforementioned regularization. The reason is that $\eta(\sX,\sS)=\eta_{\sN \cup (-1)^{d+1}\sX \times \sS} =\eta_\sN +(-1)^{d+1} \eta_{\sX \times \sS} $, and $\eta_{\sX \times \sS}$ can be shown to vanish when $d \notin 3+4\bZ$. In fact, $\eta_{\sX \times \sS}$ is proportional to either $I_\sX$ (for even $d$) or $\vev{Q}_\sX$ (for odd $d$), and $I_\sX$ is zero because $[\sX] \in \bA_d$ while $\vev{Q}_\sX$ is zero because $d \notin 3+4\bZ$. Therefore, under the conjecture \eqref{eq:conjecture}, our new invariant is precisely the phase of the partition function. This is exactly as we introduced the new invariant in low energy effective theory in \eqref{eq:GScoupling}.

\section*{Acknowledgements}

The work of YT is supported in part  
by WPI Initiative, MEXT, Japan at Kavli IPMU, the University of Tokyo
and by JSPS KAKENHI Grant-in-Aid (Kiban-C), No.24K06883.

 The work of KY is supported in part by JST FOREST Program (Grant Number JPMJFR2030, Japan), 
MEXT-JSPS Grant-in-Aid for Transformative Research Areas (A) ``Extreme Universe'' (No. 21H05188),
and JSPS KAKENHI (21K03546). 

\appendix

\section{The holomorphic anomaly equation}\label{sec:A}

The main purpose of this appendix is to give a derivation of the holomorphic anomaly equation \eqref{eq:holanomeq}. We also give a derivation of the APS index theorem \eqref{eq:APSind} as an application of the holomorphic anomaly equation.

\subsection{Notations and conventions}

First we set some notations and conventions. Let $\sigma=\sigma^1$ and $t=\sigma^0$ be space and time coordinates in two dimensions with periodicity $\sigma \sim \sigma +2\pi$. We also use $\sigma^\pm=t \pm \sigma$ and $\partial_{\pm} = \frac12 ( \partial_t  \pm \partial _\sigma)$. In the Euclidean path integral on $T^2$, we perform Wick rotation $t =\sigma^0 \to -\i \sigma^2$ and use the complex coordinate $z=\sigma^1 + \i \sigma^2$. In particular, under the Wick rotation, $\sigma^{-} \to -z$. It has the periodicity $z \sim z+2\pi$ and $z \sim z +2\pi \tau$ on $T^2$.

Let $(X, \tilde \psi)$ be a supermultiplet for a sigma model with the target space $\bR$. Its action is taken to be 
\beq
S = \frac{1}{4\pi \alpha'}  \int \d^2 \sigma \left( 4\partial_+ X \partial_- X + 2\i \tilde \psi \partial_- \tilde \psi \right)
\eeq
for an arbitrary parameter $\alpha'>0$.
The canonical commutation relations are
\beq
[X(t,\sigma), \partial_t X(t,\sigma')] = 2\pi \i \alpha' \delta(\sigma - \sigma'), \qquad \{\tilde \psi(t,\sigma), \tilde \psi(t,\sigma') \} = 2\pi \alpha'\delta(\sigma - \sigma'). \label{eq:comm}
\eeq
The Hamiltonian is
\beq
H &= \frac{1}{4\pi \alpha'}  \int_0^{2\pi} \d \sigma \left( (\partial_t X)^2+(\partial_\sigma X)^2 +  \i \tilde \psi \partial_\sigma \tilde \psi \right) \nonumber \\
&=H_L  + H_R,
\eeq
where
\beq
H_L =  \frac{1}{2\pi \alpha'}  \int_0^{2\pi} \d \sigma  (\partial_- X)^2  , \qquad H_R =  \frac{1}{2\pi \alpha'}  \int_0^{2\pi} \d \sigma \left( (\partial_+ X)^2 + \frac{\i}{2}  \tilde \psi \partial_\sigma \tilde \psi \right) .
\eeq
The supercharge is\footnote{The minus sign in the definition of the supercharge is taken so that it becomes a Dirac operator $\cD=\i \gamma^i D_i$ for a sigma model, as seen in Section~\ref{sec:SQMAPS}.}
\beq
Q = - \frac{\sqrt{2}}{2\pi \alpha'} \int_0^{2\pi} \d \sigma (  \partial_+ X \tilde \psi)  . \label{eq:supe}
\eeq
By using the canonical commutation relations, one can check that 
\beq
Q^2=H_R, \qquad [X,Q] = - \frac{\i }{\sqrt{2}} \tilde \psi, \qquad \{ \tilde \psi, Q\}= -\sqrt{2}  \partial_+ X.  \label{eq:Xpsicomm}
\eeq

\subsection{A derivation of the holomorphic anomaly equation}

Consider a noncompact SQFT $\sL$ with boundary $\sN$. We assume the existence of a ``position operator'' $X$ as discussed around \eqref{eq:position}. More precisely, $X$ is the bottom component of an $\cN=(0,1)$ chiral operator $(X,\tilde \psi)$. Consider eigenstates $\ket{x}$ of (the center-of-mass mode of) $X$. When $x$ is very large, the theory $\sL$ looks like $\bR \times \sN$ and the pair $(X,\tilde \psi)$ is the multiplet for the sigma model $\bR$. For more general values of $x$, the pair $(X,\tilde \psi)$ is a complicated chiral operator.

\paragraph{Definition of the partition function.} 
We want to define the partition function $Z_\sL$ which is roughly given by $\Tr_\sL (-1)^F q^{H_L} \bar q^{H_R}$ where $q=e^{2\pi \i \tau}$. The problem in defining this partition function is the existence of infinitely many states with finite energies in the region $x \to \infty$. To remedy this problem, we define the partition function as follows. Let $F(x) \geq 0$ be a smooth function such that
\beq
F(x) \to \left\{ \begin{array}{ll} 
1&   x \to -\infty \\
0 & x \to \infty
\end{array} \right. \label{eq:regulatorfunction}
\eeq
(We require that these limits are achieved rapidly enough so that the following discussions will be valid.)
Then we define the partition function $Z_\sL$ by
\beq
Z_\sL &= \lim_{a \to +\infty} Z_\sL(a), \nonumber \\
Z_\sL(a) &:= \Tr_\sL (-1)^F q^{H_L} \bar q^{H_R} F(X_{z=0}-a), \label{eq:regularizedpartition}
\eeq
where $F(X_{z=0}-a)$ is the operator constructed from $X$ at $z=\sigma^1+ \i \sigma^2=0$ by using the function $F$. In this definition, we are using $F(X_{z=0}-a)$ as a regulator of the region $x \to \infty$. 

\paragraph{Well-definedness.} 
We first need to check that the partition function is well-defined.
For a fixed $a$, we expect that $Z_{\sL}(a)$ is well-defined by the following reason. The only problem was the infinite states with finite energies from the center-of-mass mode of $X$ in the region $x \to \infty$. The insertion of $F(X_{z=0}-a)$ gives effectively a potential energy $V(x) = - \log F(x-a)$ in the path integral computation of $Z_\sL(a)$. We have $V(x) \to \infty$ at $x \to \infty$ from the property of the function $F$. This effectively makes the theory compact (in the sense of the energy spectrum) if $F(x) $ in the limit $x\to \infty$ vanishes rapidly enough.\footnote{More precisely, we need to impose that $\int^\infty \d x e^{-V(x)} \sim \int^\infty \d x F(x-a)$ is finite because this integral appears in the path integral as a contribution from the zero mode of $X$.}
Then we expect $Z_\sL(a)$ to be finite.

Let us next consider whether the limit $a \to \infty$ exists. To argue for the existence, we change $a$ by an infinitesimal amount. The change of $Z_\sL(a)$ is
\beq
\frac{\partial Z_\sL(a)}{\partial a} =-\Tr_\sL (-1)^F q^{H_L} \bar q^{H_R} F'(X_{z=0}-a) \label{eq:delofZbya}
\eeq
where $F'$ is the derivative of $F$. From the property of $F$, we see that $F'(X_{z=0}-a)$ is localized in the region $X_{z=0} \sim a$. If $a$ is sufficiently large, the region in which $F'(X_{z=0}-a)$ is significant can be contained in the region where the free sigma model description of $(X, \tilde \psi)$ is valid. 

Suppose that $X_{z=0}$ is at $a'$, where $a'$ is of the order of $a$.
In the path integral, one may have a configuration that extends from $X = a'$ at $z=0$ to $X \sim 0$ at $z \neq 0$. Such a configuration (i.e., a ``stretched string'') has the exponentiated action of order $\exp(- c a'^2)$, where $c$ is a constant that depends on $\tau$. Neglecting such exponentially small effects, let us consider the sigma model $(X,\tilde \psi)$ for $\bR$. In the sigma model, the zero mode of $\tilde \psi$ in the path integral makes \eqref{eq:delofZbya} vanish. Therefore, we conclude that $\partial Z_\sL(a)/\partial a$ vanishes up to errors of order $\exp(-ca'^2)F'(a'-a)$ (or its integral over $a'$). Then the limit $\lim_{a \to \infty} Z_\sL(a)$ exists if the derivative $F'(x)$ vanishes rapidly enough in the limit $x \to -\infty$ (to avoid contributions from $a' \ll a$). 

In the above definition, we have chosen a function $F$. Let us consider the dependence of $Z_\sL$ on the choice of $F$. Suppose we change $F$ by $\delta F$. By the properties of $F$ and $F+\delta F$, we need $\delta F(x) \to 0$ for $x \to \pm \infty$.  $Z_\sL(a)$ changes by
\beq
\Tr_\sL (-1)^F q^{H_L} \bar q^{H_R} \delta F(X_{z=0}-a).
\eeq
This receives contributions only from the region $X_{z=0} \sim a$. By the same reason as above (i.e., the zero mode of $\tilde \psi$), this vanishes up to small errors. Thus, in the limit $a\to \infty$, the partition function is independent of the choice of $F$. 

\paragraph{Properties.}
Let us discuss some properties of the partition function.

First we consider the modular transformation property. 
In the path integral on $T^2$, the operator $F(X_{z=0}-a)$ is inserted at a single point $z=0$ on $T^2$. Thus the partition function is invariant under modular (i.e., $\SL(2,\bZ)$) transformations of $T^2$, up to gravitational anomalies of the theory $\sL$ and the change of the area of $T^2$.

Next consider the relation to the APS index $I_\sL$.
For simplicity, we consider the case that the boundary theory $\sN$ does not have zero modes in the sense of the supercharge $Q=Q_\sN$.
When we take a limit $\bar \tau \to -\i \infty$ with $\tau$ fixed, the only states of $\sL$ that contribute to the partition functions are the ones that are annihilated by $Q=Q_\sL$. In the absence of the zero modes of $Q_\sN$, the zero modes of $Q_\sL$ are localized inside $\sL$. For these localized modes, the regulator $F(X_{z=0}-a)$ has no effect in the limit $a \to \infty$. Therefore, we get $\kappa Z_\sL \to I_\sL$ in the limit $\bar \tau \to -\i \infty$.

\paragraph{The holomorphic anomaly equation.}
We have established the properties of $Z_\sL$ mentioned in Section~\ref{sec:pt} except for the holomorphic anomaly equation. Our remaining task is to derive the holomorphic anomaly equation.

We take derivative of $Z_\sL(a) $ with respect to $\bar \tau$ and use $H_R=Q^2$ to get
\beq
\frac{\partial Z_\sL(a)}{\partial \bar\tau} &=- 2\pi \i \Tr_\sL (-1)^F q^{H_L} \bar q^{H_R} F(X_{z=0}-a) Q^2 \nonumber \\
&=- 2\pi \i \Tr_\sL (-1)^F q^{H_L} \bar q^{H_R} \left( \frac12 \{F(X_{z=0}-a) Q, Q\} + \frac12 [F(X_{z=0}-a),Q]Q   \right).
\eeq
The first term on the right hand side is zero due to the cyclicity of trace and the anticommutation between $Q$ and $(-1)^F q^{H_L} \bar q^{H_R} $. In the second term, we use \eqref{eq:Xpsicomm} to get
\beq
\frac{\partial Z_\sL(a)}{\partial \bar\tau} =- 2\pi \i \Tr_\sL (-1)^F q^{H_L} \bar q^{H_R} \left( - \frac{\i}{2\sqrt{2}} \tilde\psi_{z=0} F'(X_{z=0}-a)Q   \right) \label{eq:bardel}
\eeq
It contains $F'(X_{z=0}-a)$ which is localized around $X_{z=0} \sim a$. Thus, in the limit $a\to \infty$, we can replace the theory $\sL$ by $\bR \times \sN$. 

The supercharge $Q=Q_\sL$ of $\sL$ is given in terms of the supercharges $Q_{\bR}$ and $Q_{\sN}$ of $\bR$ and $\sN$ as $Q_\sL = Q_{\bR} + Q_{\sN}$.
The term $Q_{\bR}$ does not contribute by the following reason. For the path integral computing \eqref{eq:bardel} to be nonzero, we need to insert a single zero mode of $\tilde \psi$ on $T^2$. One can check that the supercharge $Q$ is independent of the zero mode, and hence the zero mode should be provided by the factor $\tilde\psi_{z=0} $. Now, we need even numbers of nonzero modes of $\tilde \psi$. However, each term of $Q_{\bR}$ contains odd number of nonzero modes of $\tilde \psi$, so its contribution vanishes. Therefore we get
\beq
\frac{\d Z_\sL}{\d \bar\tau} = - \frac{\pi}{\sqrt{2}}  \Tr_{\bR \times \sN} (-1)^F q^{H_L} \bar q^{H_R} \left(  \tilde\psi_{z=0} F'(X_{z=0}-a)Q_{\sN}   \right).\label{eq:fo1}
\eeq

Let us first perform integration over $\tilde \psi$ and $\sN$. We use the operator formalism (rather than path integral) for them. In the mode expansion
\beq
\tilde \psi = \alpha'^{1/2} \sum_{n \in \bZ} \tilde \psi_n e^{-  \i n \sigma^+}, \qquad \{\tilde \psi_n, \tilde \psi_m\} = \delta_{n+m,0},
\eeq
we set $\gamma=\sqrt{2} \tilde \psi_0$ so that 
\beq
\gamma^2=1, \quad \{\gamma, (-1)^F\}=0 .
\eeq
We use the following convention for the definition of the trace of $Q_\sN$.\footnote{This convention is chosen so that the APS index theorem discussed later works nicely. Indeed, when we consider the APS index theorem, an analogous definition is necessary where a Dirac operator plays the role of the supercharge. For a brief discussion on the case of the APS index theorem, see e.g. Section~2.1 of \cite{Tachikawa:2018njr}.}  The trace of $Q_\sN$ is nonzero only if $\nu_\sN$ is odd, so we restrict to this case. Consider the Hilbert space for the theory $\{\gamma\} \times \sN$ in which $(-1)^F$ is well-defined. We consider a new supercharge 
\beq
Q'_{\sN}=-\i \gamma Q_{ \sN}
\eeq 
 $Q'_{\sN}$ commutes with $(-1)^F$ and is self-adjoint, 
\beq
[(-1)^F, Q'_{\sN}]=0 ,\qquad (Q'_{\sN})^\dagger = Q'_{\sN}. 
\eeq
Then, we define the trace of $Q_{\sN}$ in the theory $\sN$ to be the trace of $Q'_{\sN}$ in the theory $\{ \gamma\} \times \sN$ in the eigenspace $(-1)^F=1$, that is,
\beq
  \Tr_\sN  q^{H_L} \bar q^{H_R} Q_{\sN} &: = \Tr_{\{\gamma\} \times \sN}q^{H_L} \bar q^{H_R} \left(\frac{1+(-1)^F}{2}\right) Q'_{\sN} \nonumber \\
&= \frac{1}{2} \Tr_{\{\gamma\} \times \sN}q^{H_L} \bar q^{H_R} (-1)^F (-\i \gamma Q_{\sN} ),
\eeq
where in the second equality we have used the fact that the trace without $(-1)^F$ vanishes due to the cyclicity of trace and $ \gamma Q_{\sN} =- Q_{\sN}\gamma$. 

Notice that the eigenspace $(-1)^F=1$ in $\{ \gamma\} \times \sN$ is just isomorphic to the (small) Hilbert space of $\sN$. Then the above definition is just a prescription to fix the sign of $\vev{Q}_\sN$, starting from the supercharge $Q_\sL=Q_\sN +Q_\bR$ of $\sL$. Unless some prescription is given, the sign of $\vev{Q}_\sN$ (without specifying its relation to $\sL$) would be ambiguous in general.

In the theory $\{\tilde \psi \} \times \sN$, we get
\beq
 \Tr_{ \{\tilde \psi \} \times \sN} (-1)^F q^{H_L} \bar q^{H_R}   \tilde\psi_{z=0} Q_{\sN}  
 =\sqrt{2} \i  \alpha'^{1/2}   \overline{\eta( \tau) } \Tr_\sN  q^{H_L} \bar q^{H_R} Q_{\sN}, \label{eq:fo2}
\eeq
where we have used the fact that $ \tilde\psi_{z=0}$ can be replaced by $\alpha'^{1/2} \tilde \psi_0 = \alpha'^{1/2} \gamma/\sqrt{2}$, and the contributions from the modes $\tilde \psi_n$ with $n \neq 0$ give the Dedekind $\eta$-function $\overline{ \eta( \tau)}$. 

The integration over $X$ is done as follows. In the path integral on $T^2$, we decompose $X$ into the zero mode $X_0$ and nonzero modes $X_\text{nonzero}$ as $X=X_0+X_\text{nonzero}$. 
The integration over $X_{0}$ in the path integral gives
\beq
\int [\d X_0] F'(X_0 + X_\text{nonzero}|_{z=0} -a) = F(\infty) - F(-\infty) = -1.
\eeq
On the other hand, as is standard in string theory, the integration over nonzero modes $X_\text{nonzero}$ in the path integral gives a contribution\footnote{
For example, the contribution to $ \Tr_{\bR} q^{H_L} \bar q^{H_R}$ from the center-of-mass of $X$ is computed in momentum space $k$ as $\int \frac{ \d k}{2\pi} \langle k| e^{-\pi   \Im \tau  \alpha' k^2} | k \rangle =(2\pi)^{-1} (\alpha' \Im \tau)^{-1/2} \langle k|k \rangle$, where formally $ \langle k|k \rangle= 2\pi \delta(k-k) =  \text{Vol}(\bR)$.}
\beq
\int [\d X_\text{nonzero}] e^{-S} &= \frac{1}{\text{Vol}(\bR)}\int [\d X ] e^{-S} =\frac{1}{\text{Vol}(\bR)}  \Tr_{\bR} q^{H_L} \bar q^{H_R} \nonumber \\
&= \frac{1}{2\pi (\alpha' \Im \tau )^{1/2} |\eta(\tau)|^2} ,
\eeq
where $\text{Vol}(\bR)$ is the volume of the target space $\bR$.
Therefore, in the theory $\{X\}$ we get
\beq
 \Tr_{\{X\}}   q^{H_L} \bar q^{H_R}      F'(X_{z=0}-a)   = -\frac{1}{2\pi (\alpha' \Im \tau )^{1/2} |\eta(\tau)|^2}. \label{eq:fo3}
\eeq

Combining \eqref{eq:fo1},  \eqref{eq:fo2} and  \eqref{eq:fo3}, we finally get
\beq
\frac{\partial Z_\sL}{\partial \bar\tau} =\frac{\i}{2 (\Im \tau)^{1/2} \eta(\tau)}  \Tr_\sN  q^{H_L} \bar q^{H_R} Q_{\sN}. \label{eq:hformula}
\eeq
This is the formula we wanted.

\subsection{Supersymmetric quantum mechanics and the APS index theorem}\label{sec:SQMAPS}

We can also derive an analogous result in supersymmetric quantum mechanics. Rather than repeating the computations, it is easier to reuse the above result. We just restrict our attention to center-of-mass modes (i.e., modes independent of $\sigma$) and also forget the Casimir energy coming from central charges. Thus we set $P=0$. 

For instance, an $\cN=1$ sigma model on a manifold $L$ with a metric $G_{ij}=G_{ij}(X)$ has the action
\beq
S = \frac{1}{2\alpha'} \int \d t G_{ij} \left( \partial_t X^i  \partial_t X^j  + \i \tilde \psi^i D_t \tilde \psi^j \right),
\eeq
where $D_t$ is the covariant derivative of the sigma model pulled back to the worldline.
Let $P_i$ and $\gamma^i$ be 
\beq
P_i=\alpha'^{-1}G_{ij}\partial_t X^j, \qquad \gamma^i=\frac{\sqrt{2}}{\alpha'^{1/2} }\tilde \psi^i.
\eeq
which satisfies
\beq
[X^i, P_j]= \i \delta^i_j, \qquad \{\gamma^i, \gamma^j\} = 2G^{ij}.
\eeq
The Hamiltonian and  the supercharge are
\beq
Q = -\frac{\alpha'^{1/2}}{2} \gamma^i P_i , \qquad H=2Q^2 . \label{eq:SQMHQ}
\eeq
Here the relation $Q^2=\frac12 H$ is the reduction of $Q^2=H_R=\frac12 (H-P)$ in two dimensions to $P=0$. In this case, we also have $H_L=\frac12 H= Q^2$.

The partition function is
\beq
Z_\sL &= \lim_{a \to +\infty}  \Tr_\sL (-1)^F e^{-\beta H} F(X -a).
\eeq
The parameter $\beta$ here is related to $\tau$ by $\beta = 2\pi \Im \tau=-\pi \i (\tau-\bar\tau)$. 
When we discard contributions from nonzero $P$, the partition function depends only on $\beta$ and we have 
\beq
\frac{\partial}{\partial\bar\tau} = \frac{\partial \beta}{\partial \bar\tau} \frac{\partial}{\partial\beta} = \pi \i \frac{\partial}{\partial\beta}.
\eeq
Then from \eqref{eq:hformula}, we get
\beq
\frac{\d Z_\sL}{\d \beta} =  \frac{1}{(2\pi \beta)^{1/2}}  \Tr_\sN  e^{- 2\beta Q^2} Q. \label{eq:SQMh}
\eeq
This is the supersymmetric quantum mechanical version of the holomorphic anomaly equation.

An interesting application of \eqref{eq:SQMh} is the derivation of the APS index theorem by supersymmetric quantum mechanics~\cite{Dabholkar:2019nnc}. We integrate both sides of the equation to get
\beq
Z_\sL(\beta=\infty) -  Z_\sL(\beta=+0) &= \int_{+0}^{\infty} \d \beta   \frac{1}{(2\pi \beta)^{1/2}}  \Tr_\sN  e^{- 2\beta Q^2} Q \nonumber \\
&= \frac{1}{2} \Tr_\sN \frac{Q}{\sqrt{Q^2}}. \label{eq:Q/|Q|}
\eeq
Let us focus on sigma models (possibly equipped with an internal theory to provide gauge symmetries). Then, from \eqref{eq:SQMHQ}, we see that $Q$ is given by the Dirac operator $\cD$ because $P_i$ is given in terms of the covariant derivatives $D_i$ on the target space manifold as $P_i = -\i D_i$, and $\gamma^i$ are gamma matrices, and hence
\beq
Q = \frac{\alpha'^{1/2}}{2} \cD \quad \text{where} \quad \cD= \i \gamma^i D_i.
\eeq
 Then, the APS $\eta$-invariant (assuming there is no zero mode on the boundary) is defined by
\beq
\eta_\sN =  \frac{1}{2} \Tr_\sN \frac{\cD}{\sqrt{\cD^2}}.
\eeq
On the other hand, the partition function $Z_\sL$ in the limit $\beta \to \infty$ receives contributions only from states with $Q^2=0$ which are localized in the interior region (assuming there is no zero mode on the boundary). It is exactly the APS index $I_\sL$,
\beq
I_\sL = Z_\sL(\beta=\infty).
\eeq
Finally, the partition function at $\beta \to +0$ is computed by the standard heat kernel method (used also in \cite{Alvarez-Gaume:1983zxc} in the context of supersymmetric quantum mechanics) to give
\beq
 Z_\sL(\beta=+0) = \int \cI_{\sL}
\eeq
where $\cI_{\sL}$ is the characteristic class that appears in the Atiyah-Singer index theorem. Therefore, we get
\beq
I_\sL =  \int \cI_{\sL} + \eta_\sN.
\eeq
This is the APS index theorem.


\bibliographystyle{ytphys}
\def\url#1{\href{#1}{#1}}
\bibliography{ref}

\end{document}